\documentclass[prd,12pt,aps,showpacs,superscriptaddress,preprintnumbers,floatfix,amssymb,nofootinbib,tightenlines]{revtex4}
\usepackage{graphicx}

\newcommand{\Tr}{{\mathrm Tr}}
\newcommand{\Z}{{Z \!\!\! Z}}

\newcommand{\eq}[1]{(\ref{#1})}
\newcommand{\beqn}{\begin{eqnarray}}
\newcommand{\eeqn}{\end{eqnarray}}
\newcommand{\be}{\begin{eqnarray}}
\newcommand{\ee}{\end{eqnarray}}
\newcommand{\dual}{\mbox{}^{\ast}}
\newcommand{\cD}{{\cal D}}
\newcommand{\cC}{{\cal C}}

\newcommand{\cL}{{\cal L}}
\newcommand{\cF}{{\cal F}}
\newcommand{\cA}{{\cal A}}

\def\bbbone{{\mathchoice {\rm 1\mskip-4mu l} {\rm 1\mskip-4mu l}
{\rm 1\mskip-4.5mu l} {\rm 1\mskip-5mu l}}}
\newcommand{\latt}{{\mathrm{latt}}}
\newcommand{\phys}{{\mathrm{phys}}}

\newcommand{\Uppsala}{\affiliation{Department of Theoretical Physics, Uppsala University,
P.O. Box 803, S-75108, Uppsala, Sweden}}
\newcommand{\Moscow}{\affiliation{
ITEP, B.Cheremushkinskaya 25, Moscow, 117259, Russia}}

\begin{document}

\title{Embedded monopoles in quark eigenmodes \\ in SU(2) Yang-Mills Theory}

\author{M.\,N.~Chernodub}\Moscow\Uppsala
\author{S.\,M.~Morozov}\Moscow

\begin{abstract}
We study the embedded QCD monopoles (``quark monopoles'') using
low-lying eigenmodes of the overlap Dirac operator in zero- and
finite-temperature SU(2) Yang-Mills theory on the lattice.
These monopoles correspond to the gauge-invariant hedgehogs in the
quark-antiquark condensates. The monopoles were suggested to be
agents of the chiral symmetry restoration since their cores should
suppress the chiral condensate. We study numerically the scalar,
axial and chirally invariant definitions of the embedded monopoles
and show that the monopole densities are in fact globally
anti-correlated with the density of the Dirac eigenmodes. We
observe, that the embedded monopoles corresponding to low-lying
Dirac eigenvalues are dense in the chirally invariant (high
temperature) phase and dilute in the chirally broken (low
temperature) phase. We find that the scaling of the scalar and
axial monopole densities towards the continuum limit
is similar to the scaling of the string-like objects while the
chirally invariant monopoles scale as membranes. The excess of gluon
energy at monopole positions reveals that the embedded QCD monopole
possesses a gluonic core, which is, however, empty at the very
center of the monopole.
\end{abstract}

\preprint{UUITP-28/05}
\preprint{ITEP-LAT/2005-29}


\maketitle

\section{Introduction}

It is generally
believed~\cite{ref:general:reviews:QCD,ref:crossover:lattice:facts}
that the low-temperature (confinement) and the high-temperature
(deconfinement) phases in QCD with realistic quark masses and
vanishing chemical potential $\mu$ are separated by a smooth
crossover which takes place at temperature $T_c \approx 170$~MeV.
As the system goes through the crossover all thermodynamic
quantities and their derivatives change smoothly, being
non-singular functions of the temperature $T$. Therefore there
is no {\it local} order parameter which can distinguish between
these two phases at $\mu=0$.

Recently it was suggested~\cite{ref:embedded:QCD} that a
well-defined boundary between the QCD phases  at $\mu=0$ can still
be rigorously defined as a proliferation (percolation)
transition of the so-called ``embedded QCD monopoles'' or, as we also call
them, ``quark monopoles''. These monopoles are (gauge-invariant)
composite objects made of quark and gluon fields. The monopoles
are assumed to be proliferating at infinitely long distances in
the high temperature phase while in the low-temperature phase they
are moderately dilute. Contrary to Abelian monopoles in
compact gauge theories, the embedded QCD monopoles are in general,
{\it not} directly associated with the confining properties of the
vacuum\footnote{There should be
however, an indirect relation between the confining properties and
the embedded monopole dynamics since as it is well known the
confinement phenomena and the chiral symmetry are intimately
related to each other in QCD.}. The embedded monopoles can be considered as
agents of the chiral symmetry restoration: in the low-temperature phase the
chiral condensate should be suppressed in the cores of the
embedded QCD monopoles while outside the monopoles the chiral
condensate is suggested to be non-zero.

The assumption that the chiral phase transition should be driven
by the percolation of such monopoles can intuitively be understood
as follows~\cite{ref:embedded:QCD}. At low temperatures the
density of the embedded monopoles is low and suppression of the
chiral condensate by the monopole cores is negligibly small.
However, as the temperature increases, the density of the embedded
monopoles gets larger and, consequently, the chiral condensate
becomes more suppressed. One can also look at the relation between
the embedded monopole density and the chiral condensate from
another side: with an increase of the temperature the chiral
condensate becomes weaker, and the embedded monopoles -- which are
energetically unfavorable hedgehogs in the quark-antiquark
condensates -- become more populated. At some point the chiral
condensate gets low enough for the embedded QCD monopoles to
become sufficiently dense to start proliferating themselves.

The quark monopole in QCD is conceptually similar to the embedded
defects of the Standard Electroweak (EW)
model~\cite{ref:semilocal:review}. These defects are called as the
Nambu monopoles~\cite{ref:Nambu} and the
$Z$-vortices~\cite{ref:Z:vortex}. The $Z$--vortices (if they are
not closed) begin and end on the Nambu monopoles. In the broken
low-temperature phase the value of the Higgs field is suppressed
inside the embedded EW defects, and is asymptotically non-zero
outside the defects. According to analytical estimates~\cite{Va94}
the $Z$-vortices are proliferating for long distances in the
high-temperature symmetric phase in which they form a dense
percolating network. In the broken phase the $Z$-vortex network is
destroyed and these objects become dilute. The Nambu monopoles
possess similar properties~\cite{ref:chernodub:nambu}. Numerical
simulations~\cite{ref:hot:electroweak,ref:Z:percolation} show that
the percolation transition of the $Z$-vortices takes place both at
the region of the relatively small Higgs
mass~\cite{ref:hot:electroweak}, $M_{H}  \lesssim
72\,\mathrm{GeV}$, where the phase transition of the first order,
and at large Higgs masses~\cite{ref:Z:percolation}, $M_{H} \gtrsim
72\,\mathrm{GeV}$, where the transition is a smooth analytical
crossover~\cite{ref:EW:phase}.

In the condensed matter physics, an onset of percolation realized
in the absence of a thermodynamic phase transition is usually referred
to as the Kert\'esz transition~\cite{ref:Kertesz}. The Ising model
in an external magnetic field provides the best known example of
the Kert\'esz transition which is defined with respect to the so
called Fortuin--Kasteleyn (FK)
clusters~\cite{ref:Fortuin:Kasteleyn}. The FK clusters are sets of
lattice links connecting nearest spins in the same spin states.
These clusters are proliferating in the high temperature
(disordered) phase and they are short-sized and dilute in the low
temperature (ordered) phase. In zero magnetic field, $H=0$, the
ordered and the disordered phases are separated by a phase
transition at the Curie temperature $T_c$, which coincides with
the percolation transition for the FK clusters. In an external
magnetic field the phase transition is known to be absent and the
ordered and disordered phases are connected analytically by a
crossover in the $T$-$H$ plane. Nevertheless, the phases are
still separated by a Kert\'esz transition line $T_K=T_K(H)$ which marks
the proliferation (percolation) transition for the FK clusters.
Obviously, in the zero-field limit the Kert\'esz line meets the
Curie point, $\lim_{H\to 0}T_K(H) \to T_c$.

The Kert\'esz-type transitions often appear in the gauge theories
coupled to fundamental mater fields. Besides mentioned cases of
embedded monopoles in QCD~\cite{ref:embedded:QCD} and the embedded
defects in the Electroweak
model~\cite{ref:chernodub:nambu,ref:hot:electroweak,ref:Z:percolation},
the Kert\'esz line appears, for example, in the compact $U(1)$
gauge theories~\cite{ref:Arwed}. The manifestation of the
Kert\'esz line in the SU(2) Higgs model (which is similar to the
Electroweak model) can be found as the percolation of the center
vortices~\cite{ref:Greensite:Faber}.

The picture of percolating embedded monopoles in QCD is most
probably related to the percolation of the hadron clusters at high
temperature and non-zero density ($\mu \neq 0$) environment, which
may be realized, for example, in the heavy-ion collision
experiments. In these extreme conditions hadrons may overlap and
form clusters within which the quarks are no more confined. The
onset of the quark-gluon plasma phase is thus associated with the
percolation transition of the hadron
clusters~\cite{ref:Satz:theory,ref:Satz:percolation}.

In this paper we study basic properties of the embedded (or,
quark) monopoles using numerical simulations in the SU(2)
gauge theory without dynamical matter field. The monopoles are defined with the help of
$c$-valued eigenmodes of the overlap Dirac operator. In
Section~\ref{sec:construction} we describe the structure of such
monopoles in the continuum space-time. We show that in the SU(2) Yang-Mills theory
there are three types of these monopoles characterized by their behavior
with respect to the global axial transformations. We also discuss the extension
of our construction to the realistic SU(3) case. In the same
Section we provide a lattice construction of the quark monopoles,
which is suitable for utilization in numerical simulations. In
Section~\ref{sec:density} we describe results of our numerical
simulations for the density of the embedded monopoles. In
Section~\ref{sec:spectral} we discuss a relation between the monopole
density and the spectral density of the Dirac fermions. We also
discuss a relation of our results to the Banks-Casher
formula~\cite{ref:Banks:Casher}. Section~\ref{sec:energy} is devoted to
numerical analysis of the structure of the chromomagnetic fields around
the monopoles. Our conclusions are given in the last Section.

\section{Quark monopoles in continuum and on the lattice}
\label{sec:construction}

The quark monopoles in QCD are analogues of the embedded (Nambu)
monopoles~\cite{ref:Nambu,ref:semilocal:review} in the Standard
Electroweak model. Here we briefly outline the definition of the
embedded QCD monopoles following Ref.~\cite{ref:embedded:QCD}. For
the sake of simplicity we consider the gauge theory with the reduced
number of colors, $N_c=2$. An outline of the generalization of our
approach to the bigger number of colors is given at the end of this Section.

\subsection{Quark monopoles in continuum SU(2) gauge theory}

Let consider the SU(2) Yang-Mills theory with one (for simplicity)
species of the fermion field~$\psi$ which transforms in the
fundamental representation of the gauge group. Using $\psi$
one can define the bilinear functions of the fermion field,
\beqn
\vec \xi_\Gamma = \bar\psi(x) \Gamma \vec \tau \psi(x)\,,
\qquad \Gamma = \bbbone\,, i \gamma_5\,,
\label{eq:xi}
\eeqn
where $\vec \tau=(\tau_1,\tau_2,\tau_3)$ are the Pauli matrices
acting in the color space and
$\gamma_\mu$, $\gamma_5$ is the standard set of the
spinor $\gamma$--matrices in the four-dimensional space-time. The
real-valued composite fields
$\vec \xi_S$ and $\vec \xi_A$ (with the subscripts $S$ and $A$
corresponding to the scalar, $\bbbone$,
and axial, $i \gamma_5$, operators, respectively)
are scalar and, respectively, pseudoscalar (axial) fields from the point of view of
space-time transformations. Both these fields transform as adjoint three-component
quantities with respect to the action of the gauge group.

In the EW model the role of the adjoint composite field~\eq{eq:xi}
is played by the scalar triplet $\Phi^\dagger \vec\tau \Phi$, where $\Phi$
is the two-component Higgs field. The EW embedded defects can then
be formulated in terms of the classical or asymptotic configurations
of the gauge and the Higgs fields. To make a tight link between the
embedded defects in both theories we assume from the very
beginning that the fermion field~$\psi$ used in the
definition~\eq{eq:xi} is a $c$-valued function. It is convenient to choose
the field~$\psi$ to be an eigenmode of the Dirac operator $\cD$,
\beqn
\cD[A] \psi_\lambda(x) = \lambda \psi_\lambda(x)\,, \qquad
\cD[A] = \gamma_\mu (\partial_\mu + i \frac{1}{2} \tau^a A^a_\mu) + m\,,
\label{eq:Dirac:equation}
\eeqn
corresponding to a configuration of the gauge fields $A^a_\mu(x)$. In our
numerical analysis we use the massless Dirac operator with $m=0$. The
Dirac eigenmodes are labeled by the eigenvalues $\lambda$ of the Dirac operator.
The label $\lambda$ will be omitted in this Section for the sake of simplicity.

The axial transformations are defined by the global Abelian parameter $\alpha$,
\beqn
U_A(1): \qquad \psi \to e^{i \alpha \gamma_5} \psi\,, \quad
\bar\psi \to \bar\psi e^{i \alpha \gamma_5}\,.
\label{eq:axial:U1}
\eeqn
The color vectors $\vec \xi_S$ and $\vec \xi_A$ are transforming into each other
under the axial transformations~\eq{eq:axial:U1} as follows:
\beqn
\left(\begin{array}{c} \vec \xi_S \\ \vec \xi_A \end{array} \right) \to
{\left(\begin{array}{c} \vec \xi_S \\ \vec \xi_A \end{array} \right)}'
=
\left(\begin{array}{cc}
\cos 2 \alpha & \sin 2 \alpha \\ - \sin 2 \alpha & \cos 2 \alpha
\end{array}
\right)
\left(\begin{array}{c} \vec \xi_S \\ \vec \xi_A \end{array} \right)\,.
\label{eq:axial:SO2}
\eeqn

Using two adjoint fields~\eq{eq:xi} we define three unit color vectors
\beqn
\vec n_S = \frac{\vec \xi_S}{|\vec \xi_S|}\,, \quad
\vec n_A = \frac{\vec \xi_A}{|\vec \xi_A|}\,, \quad
\vec n_I = \frac{\vec \xi_S \times \vec \xi_A}{|\vec \xi_S \times \vec \xi_A|}\,,
\label{eq:ns}
\eeqn
where $(\vec A, \vec B)$ and $[\vec A \times \vec B]^a = \epsilon^{abc} A^b B^c$
are, respectively, the scalar and the vector products in the color space and
$|\vec A| = (\vec A, \vec A)^{1/2}$ is the norm of the color vector $A$. The last term in Eq.~\eq{eq:ns}, $\vec n_I$,
is a (normalized) vector product of the scalar and axial color vectors.
The vector $\vec n_I$ is interesting because it is invariant under the axial
transformations~(\ref{eq:axial:U1}), (\ref{eq:axial:SO2}). The index $I$ in the subscript of $\vec n_I$
stands for "invariant".

The crucial observation is to interpret the unit
vectors~\eq{eq:ns} as directions of the corresponding composite
adjoint Higgs field. Thus we have three Georgi-Glashow multiplets
$(n^a_\Gamma,A^a_\mu)$, $\Gamma=S,A,I$, which can be used to
construct the gauge invariant 't~Hooft tensors~\cite{ref:thooft},
\beqn
\cF^\Gamma_{\mu\nu}(n_\Gamma,A) = F^a_{\mu\nu}(A)\, n_\Gamma^a -
\frac{1}{g} \epsilon^{abc} n_\Gamma^a {(D^{\mathrm{ad}}_\mu n_\Gamma)}^b {(D^{\mathrm{ad}}_\nu n_\Gamma)}^c\,,
\qquad
\Gamma = S\,,A\,,I\,,
\label{eq:thooft:tensor}
\eeqn
where $F^a_{\mu\nu} = \partial_\mu A^a_\nu - \partial_\nu A^a_\mu + g \epsilon^{abc} A^b_\mu A^c_\nu $
is the field strength tensor for the SU(2) gauge field $A^a_\mu$, and
\beqn
{(D^{\mathrm{ad}}_\mu)}^{ab} = \delta^{ab}\, \partial_\mu + g \, \epsilon^{abc} A^c_\mu\,,
\label{eq:D:adjoint}
\eeqn
is the adjoint covariant derivative.
The 't~Hooft tensor~\eq{eq:thooft:tensor} is the gauge-invariant field strength tensor for the
diagonal (with respect to the color direction $\vec n_\Gamma$) component of the gauge
field,
\beqn
\cA^\Gamma_\mu = A^a_\mu n^a_\Gamma\,, \qquad \Gamma = S\,,A\,,I\,.
\label{eq:A:Gamma}
\eeqn

The current of the quark monopole of the $\Gamma^{\mbox{th}}$ type,
\beqn
k^\Gamma_\nu = \frac{g}{4 \pi} \partial_\mu \tilde \cF^\Gamma_{\mu\nu} \equiv
\int_{\cC^\Gamma} d\tau \frac{\partial X^{\cC^\Gamma}_\nu(\tau)}{\partial \tau}\, \delta^{(4)}(x - X^{\cC^\Gamma}(\tau))\,,
\qquad
\tilde \cF^\Gamma_{\mu\nu} = \frac{1}{2} \epsilon_{\mu\nu\alpha\beta} \cF^\Gamma_{\alpha\beta}\,,
\label{eq:k:Nambu}
\eeqn
has a $\delta$--like singularity at the corresponding worldline $\cC^\Gamma$.
The monopole worldline is parameterized by the vector $x_\mu = X^{\cC^\Gamma}_\mu(\tau)$ and the parameter $\tau$.
The quark monopoles defined according to Eq.~\eq{eq:k:Nambu} are quantized and
the corresponding monopole charge is conserved ({\it i.e.}, the worldlines $\cC^\Gamma$ are closed).

The quark monopoles $k^\Gamma_\mu$ carry the magnetic charges with
respect to the ``scalar'', ``axial'' and ``chirally invariant''
components of the gauge field $\cA^\Gamma_\mu$,
Eq.~\eq{eq:A:Gamma}. In the corresponding Unitary gauges
$n^a_\Gamma = \delta^{a3}$, the quark monopoles correspond to
monopoles ``embedded'' into the diagonal
component~\eq{eq:A:Gamma}. In the gauges, where the diagonal
component $\cA^\Gamma_\mu$ is regular, such monopoles are
hedgehogs in the composite quark-antiquark fields. The
corresponding quark condensates are characterized by the typical
hedgehog behavior $n^a_\Gamma \sim x^a$ in the local (transverse)
vicinity of the monopoles. The fact of the existence
of these monopoles in QCD
is not a dynamical fact but rather a simple (kinematical)
consequence of the existence of the adjoint real-valued
fields~(\ref{eq:xi}), (\ref{eq:ns}). Note that there is an infinite
number of equivalent formulations of the embedded monopoles associated with
triplet isovectors which are given by a chiral
rotation~\eq{eq:axial:SO2} of, say, isovector $\vec \xi_A$ with an
arbitrary angle $\alpha$.
The dynamics of these monopoles is studied below.

Let us summarize briefly: if one has the configuration of the
gauge field $A_\mu$ and the configuration of the
(generally massive)
quark $c$-field~$\psi$ then the location of the embedded quark monopoles
of all three types (``scalar'', ``axial'' and ``invariant'') can
be found with the help of
relations~(\ref{eq:xi}), (\ref{eq:ns}), (\ref{eq:thooft:tensor}), (\ref{eq:k:Nambu}).
The quark $c$-fields~$\psi$ can be defined as a set of eigenmodes
of the Dirac operator~\eq{eq:Dirac:equation}, labeled by the
eigenvalue $\lambda$.

\subsection{Quark monopoles in SU(2) gauge theory on the lattice}

The lattice construction of the quark monopoles in Euclidean QCD closely resembles a similar
lattice construction~\cite{ref:sphaleron} of the embedded defects in the
standard model of electroweak interactions. Consider a configuration of the lattice gauge
fields $U(x,\mu)$ and a configuration of the $c$-valued
fermion matter field $\Psi(x)$. Using the fermionic field $\Psi(x)$ one can construct the composite
color fields on the Euclidean lattice,
\beqn
\xi^a_S(x) = \Psi^\dagger(x) \tau^a \Psi(x)\,, \qquad \xi^a_A(x) = \Psi^\dagger(x) \tau^a \gamma_5 \Psi(x)\,,
\label{eq:n}
\eeqn
which are the lattice analogues of the continuum expressions~\eq{eq:xi}. Then the adjoint variables
$\xi^a_S$ and $\xi^a_A$ can be used to construct the lattice unit vectors $n_S$, $n_A$, and $n_I$,
in a manner completely similar to Eq.~\eq{eq:ns}.

The next step is to define the (un-normalized) projections of the gauge field onto the color
directions $n_\Gamma \equiv n^a_\Gamma \tau^a$:
\[
V_\Gamma(x,\mu) = U(x,\mu) + n_\Gamma(x) U(x,\mu) n_\Gamma(x + \hat\mu)\,, \qquad \Gamma = S\,,A\,,I\,.
\]
The field $V_\Gamma$ behaves under the
action of the gauge transformation $\Omega$ as a gauge field,
\[
V_\Gamma(x,\mu) \to \Omega^\dagger(x) V_\Gamma(x,\mu) \Omega(x + \hat \mu)\,.
\]

The lattice analogue of  the 't~Hooft tensor~\eq{eq:thooft:tensor} is given~\cite{ref:sphaleron}
by the compact field ${\bar \theta}(x,\mu\nu) \in (-\pi,\pi]$ defined on the plaquette $P = \{x,\mu\nu\}$:
\beqn
{\bar \theta}^\Gamma(x,\mu\nu) = \arg \Bigl( {\mathrm {Tr}}
 \left\{\Bigl[\bbbone + n_\Gamma(x)\Bigr] V_\Gamma(x,\mu) V_\Gamma(x+\hat\mu,\nu)
V^\dagger_\Gamma(x + \hat\nu,\mu) V^\dagger_\Gamma(x,\nu) \right\}\Bigr)\,.
\label{eq:theta:P}
\eeqn
Due to the property $n_\Gamma(x) V_\Gamma(x,\mu) = V_\Gamma(x,\mu) n_\Gamma(x+\hat\mu)$,
the definition~\eq{eq:theta:P} is independent of the choice of the reference point $x$
on the plaquette $P$. One can show that in the Unitary gauge, $n_\Gamma(x) = \tau^3$, the
gauge invariant plaquette function~\eq{eq:theta:P} coincides with the standard Abelian
plaquette formed out of the compact Abelian
fields~$\theta^u_\Gamma(x,\mu) = {\mathrm{arg}} \, U^{11}_\Gamma(x,\mu)$.

The singularities in the compact fields ${\bar \theta}_\Gamma$ correspond to the quark monopoles
of the scalar, axial and invariant types. The quark monopoles are defined on the links of
dual lattice $\dual \{x,\mu\}$ which are dual to the cubes $c_{x,\mu}$ of the primary lattice:
\beqn
j^\Gamma(x,\mu) = - \frac{1}{2\pi} \sum_{P \in \partial c_{x,\mu}} {\bar \theta}^\Gamma_P\,,
\label{eq:j}
\eeqn
where the sum is taken over all six plaquettes $P$ forming the faces of the cube $c_{x,\mu}$.
Equation~\eq{eq:j} is an analog of the standard definition~\cite{DGT} of the Abelian monopole
in the lattice gauge theory of compact Abelian fields. By construction,
the monopole current~\eq{eq:j} is integer-valued, $j^\Gamma \in \Z$, and conserved,
$\delta \dual j^\Gamma = 0$. Here the operator $\delta$ is the lattice divergence.

\subsection{Generalization to SU(3) gauge group}

The construction of the embedded quark monopoles can be generalized to the realistic case of
the $SU(3)$ Yang-Mills theory. The structure of such monopoles shares similarity with
the non-Abelian monopoles in the $SU(3)$ gauge theory coupled to an octet Higgs field.
A good review of the subject on the monopole configurations in the $SU(3)$ gauge-Higgs models
be found in Refs.~\cite{ref:Yasha}. Below we briefly outline the construction to be
discussed elsewhere in more detail~\cite{ref:chernodub:inpreparation}.

The generators of the $SU(3)$ gauge group are given by eight traceless matrices
$T^a = \lambda^a/2$ normalized as $\Tr T^a T^b = \delta^{ab}/2$. Here $\lambda^a$ are the
standard Gell-Mann matrices. Contrary to the simplest case of two colors, there are two
structure constants in the $SU(3)$ gauge group. The totally symmetric constant $f^{abc}$
is defined via the relation $[T^a,T^b] = i f^{abc} T^c$, while the values of the
totally symmetric structure $d^{abc}$ can be deduced from
the equation $\{T^a,T^b\} = \frac{1}{3} \delta^{ab} + d^{abc} T^c$. These constants can also
be expressed via the single relation, $4 \Tr T^a T^b T^c = d^{abc} + i f^{abc}$.

Similarly to the $SU(2)$ case~\eq{eq:xi} the $SU(3)$ model admits two composite octet vectors
\be
\xi^a_\Gamma = \bar\psi(x) \Gamma \vec T^a \psi(x)\,, \qquad \Gamma = \bbbone\,, i \gamma_5\,,
\label{eq:xi:SU3}
\ee
where we use the same notations as in the $SU(2)$ case. This should not, however, lead to a confusion
since the $SU(3)$ gauge group is discussed in this Section only.

Besides the scalar octet $\xi^a_S$ with $\Gamma = \bbbone$ and the axial octet $\xi^a_A$
with $\Gamma = i \gamma_5$ the structure $SU(3)$ gauge group allows us to build the invariant field
\be
\xi^a_{I} = f^{abc} \xi^b_S \xi^b_A\,,
\label{eq:inv:SU3}
\ee
which transforms in the adjoint representation of the $SU(3)$ gauge group.
The octet $\xi_{I}$ is the $SU(3)$ generalization of the invariant $SU(2)$ triplet
field $\xi_I = \vec \xi_S \times \vec \xi_A$ which is used in Eq.~\eq{eq:ns}. One can explicitly show
that the composite field~\eq{eq:inv:SU3} is invariant under the global axial rotations
which read in terms of the $SU(3)$ octet fields~\eq{eq:xi:SU3} as follows:
\beqn
\xi^a_\Gamma \to R_{\Gamma\Gamma'} \xi^a_{\Gamma'}\,,
\qquad
R =
\left(\begin{array}{cc}
\cos 2 \alpha & \sin 2 \alpha \\ - \sin 2 \alpha & \cos 2 \alpha
\end{array}
\right)
\left(\begin{array}{c} \xi^a_S \\ \vec \xi^a_A \end{array} \right)\,,
\qquad
\Gamma,\Gamma'=S,A\,.
\label{eq:axial:SU3}
\eeqn

In addition to the asymmetric constant $f^{abc}$ the $SU(3)$ gauge group possess the symmetric structure
constant $d^{abc}$. Therefore one can define three additional octet fields,
$\xi^a_{SS}$, $\xi^a_{AS} \equiv \xi^a_{SA}$ and $\xi^a_{AA}$ which form a symmetric rank-2
tensor field,
\be
\xi^a_{\Gamma_1\Gamma_2} = d^{abc} \xi^b_{\Gamma_1} \xi^b_{\Gamma_2}\,,
\qquad
\Gamma_{1,2}=S,A\,,
\label{eq:tensor:SO2}
\ee
with respect to the global axial rotations~\eq{eq:axial:SU3}:
\be
\xi^a_{\Gamma_1\Gamma_2} \to R_{\Gamma_1\Gamma'_1} R_{\Gamma_2\Gamma'_2} \xi^a_{\Gamma'_1\Gamma'_2}\,.
\ee

Summarizing, in the realistic case of three colors
we have six\footnote{In a multiflavor case there are six octet structures per one flavor, as well as
a number of the flavor-mixing structures composed of quarks of different flavors.} independent structures which
are classified with respect to the global
axial rotations~\eq{eq:axial:SU3} as the scalar ($\xi^a_I$), vector ($\xi^a_S$ and $\xi^a_A$),
and rank-2 symmetric tensor ($\xi^a_{SS}$, $\xi^a_{SA}$, and $\xi^a_{AA}$). All these structures behave as
octet fields with respect to the $SU(3)$ gauge transformations.

Note, that from a kinematical point of view any of the six octet fields can equivalently be used to construct
the embedded quark monopoles in the $SU(3)$ gauge theory.
The relevance of one or another octet field is to be figured out with the help of the dynamical
considerations ({\it i.e.}, with the help of numerical simulations). In the rest of this Section we
describe the kinematical construction of the embedded monopoles in the theory with the $SU(3)$
gauge group and therefore use a generic notation $\xi^a$ for any of the six composite fields, or
for a linear combinations of them.

In order to characterize the embedded monopoles in QCD it is convenient to work in the Cartan-Weyl
basis~\cite{ref:Weinberg}. The Cartan subgroup of the $SU(3)$ gauge group is generated by two diagonal
generators
\be
H_1 \equiv T^3 = \frac{1}{2}
\left(
\begin{array}{rrr}
1 & 0 & 0 \\
0 & -1 & 0 \\
0 & 0 & 0 \\
\end{array}
\right)
\,,
\quad
H_2 \equiv T^8 = \frac{1}{2 \sqrt{3}}
\left(
\begin{array}{rrr}
1 & 0 & 0 \\
0 & 1 & 0 \\
0 & 0 & -2 \\
\end{array}
\right)
\,,
\ee
which form the two-component vector $\vec H = (H_1,H_2)$.
In any point of the space-time the octet field $\xi^a$ can be gauge-rotated to
the Cartan subalgebra
\be
\Phi(x) \to |\Phi(x)| \, (\vec h(x), \vec H)\,,\quad \vec h^2 =1\,.
\label{eq:Phi:infty}
\ee
where $(a,b)$ denotes the scalar product in the Cartan space, and $\vec h$ is the unit two-component
vector pointing to the direction of the composite octet field $\xi$ in the Cartan space.

The magnetic charge of the $SU(3)$ monopole~\cite{ref:Weinberg}
\be
{\vec g}_M = \frac{4 \pi}{g} \Bigl(n_1 {\vec \beta}^*_1 + n_2 {\vec \beta}^*_2\Bigr)\,,
\label{eq:g:M}
\ee
is given in terms of the dual simple roots ${\vec \beta}^*_\alpha$ with $\alpha=1,2$.
The dual roots ${\vec \beta}^*_\alpha$ are expressed in terms
of the original simple roots ${\vec \beta}_\alpha$ of the $SU(3)$ group
as ${\vec \beta}^*_\alpha = {\vec \beta}_\alpha/|{\vec \beta}_\alpha|^2$.
These roots are often chosen in the self-dual form,
\be
{\vec \beta}_1 = (1,0)\,,\qquad {\vec \beta}_2 = (- 1/2, \sqrt{3}/2)\,.
\label{eq:roots}
\ee

The generalization of the standard Dirac quantization of the monopole charge to the case of the $SU(3)$ monopole reads as
\be
e^{i g (\vec g_M, \vec H)} = \bbbone\,.
\label{eq:Dirac}
\ee
The Dirac condition~\eq{eq:Dirac} is satisfied by Eq.~\eq{eq:g:M} provided the numbers
$n_1$ and $n_2$ are integer.

The classification of the $SU(3)$ monopoles qualitatively depends on the direction of the local Higgs field
in the Cartan subalgebra. If the vector $\vec h$ in Eq.~\eq{eq:Phi:infty} is not orthogonal to any of the
simple roots $\vec \beta_1$ and $\vec \beta_2$, Eq.~\eq{eq:roots}, then the pattern of the symmetry breaking
is ``maximal''~\cite{ref:Weinberg}
\be
SU(3) \to U(1) \times U(1)\,,
\label{eq:maximal}
\ee
and the corresponding vacuum manifold is characterized by a non-trivial second homotopy group,
\be
\pi_2 \left(\frac{SU(3)}{U(1) \times U(1)}\right) = {\mathbb Z}^2\,.
\ee
Thus, the monopoles are described by the two integer numbers. These numbers are $n_1$ and $n_2$ which enter
the definition of the monopole charge~\eq{eq:g:M}.

If the asymptotic Higgs field is orthogonal either to the simple root $\vec \beta_1$ or to the simple root
$\vec \beta_2$, then the symmetry breaking is ``minimal''
\be
SU(3) \to U(2)\,.
\label{eq:minimal}
\ee
This pattern also possess a non-trivial second homotopy group,
\be
\pi_2 \left(\frac{SU(3)}{U(2)}\right) = {\mathbb Z}\,,
\ee
and the monopoles are characterized by one integer number which is either $n_1$ (if $(\vec \beta_2, \vec h)=0$)
or $n_2$ (if $(\vec \beta_1, \vec h)=0$).

The position of the $SU(3)$ monopole can locally be determined with the help of the 't~Hooft tensors
similarly to the $SU(2)$ case~\eq{eq:k:Nambu}. In the Cartan space the monopole is a point-like source of
the magnetic field which is described locally as
${\mathbf{B}} \simeq ({\vec g}_M, \vec H) {\mathbf{r}}/r^3$ provided ${\vec r}$ is close to the center of the monopole.
The type of the symmetry breaking pattern corresponds to the type of the embedding of the $SU(2)$ HP monopole into the
larger $SU(3)$ group. In the general case of the $SU(N)$ gauge group the embedded monopoles can be formulated
similarly to the case of the $SU(N)$ gauge-Higgs monopoles reviewed in detail in Ref.~\cite{ref:Yasha}.

It is known~\cite{ref:Yasha} that in the case of the monopole-like solutions to the classical equations of motion of the $SU(3)$
gauge-Higgs model, the choice of either maximal \eq{eq:maximal} or minimal \eq{eq:minimal} symmetry breaking patterns depends on
the details of the Higgs potential~\cite{ref:Yasha}. In the case of QCD one can realize both patterns simultaneously.
Let us take any of the six octet fields \eq{eq:xi:SU3}, \eq{eq:inv:SU3} or \eq{eq:tensor:SO2}, and
project it onto the Cartan subgroup . The color direction of this field in any point is usually not perpendicular to
any of the root vectors~\eq{eq:roots} apart from rare degenerate cases. Thus, if we choose
any of the above octets as the effective Higgs field $\xi$, then it is likely that the maximal embedding pattern~\eq{eq:maximal} is realized. On the other
hand, a linear combination of any of the two octets can always be fine-tuned (again, apart from rare degenerate cases) to be
perpendicular to a chosen root vector, so that the minimal embedding pattern~\eq{eq:minimal} can also be realized in
QCD. A discussion on the kinematical construction and well as on the possible dynamical significance of such
monopoles in QCD will be published elsewhere~\cite{ref:chernodub:inpreparation}. Below we discuss numerical signatures
of the embedded quark monopoles in the simpler case with two colors.

\section{Density of quark monopoles at zero and finite temperature}
\label{sec:density}

In order to study basic properties of the embedded QCD monopoles
we perform a simulation of the pure SU(2) Yang-Mills model on the lattice
at zero and finite temperatures. The technical details of
numerical simulations are given in Appendix~\ref{app:details}, and
below we discuss the results of the simulations.

\subsection{Monopole densities and the effect of temperature}

\begin{figure}[!htb]
\begin{tabular}{cc}
\includegraphics[width=74mm,clip=true]{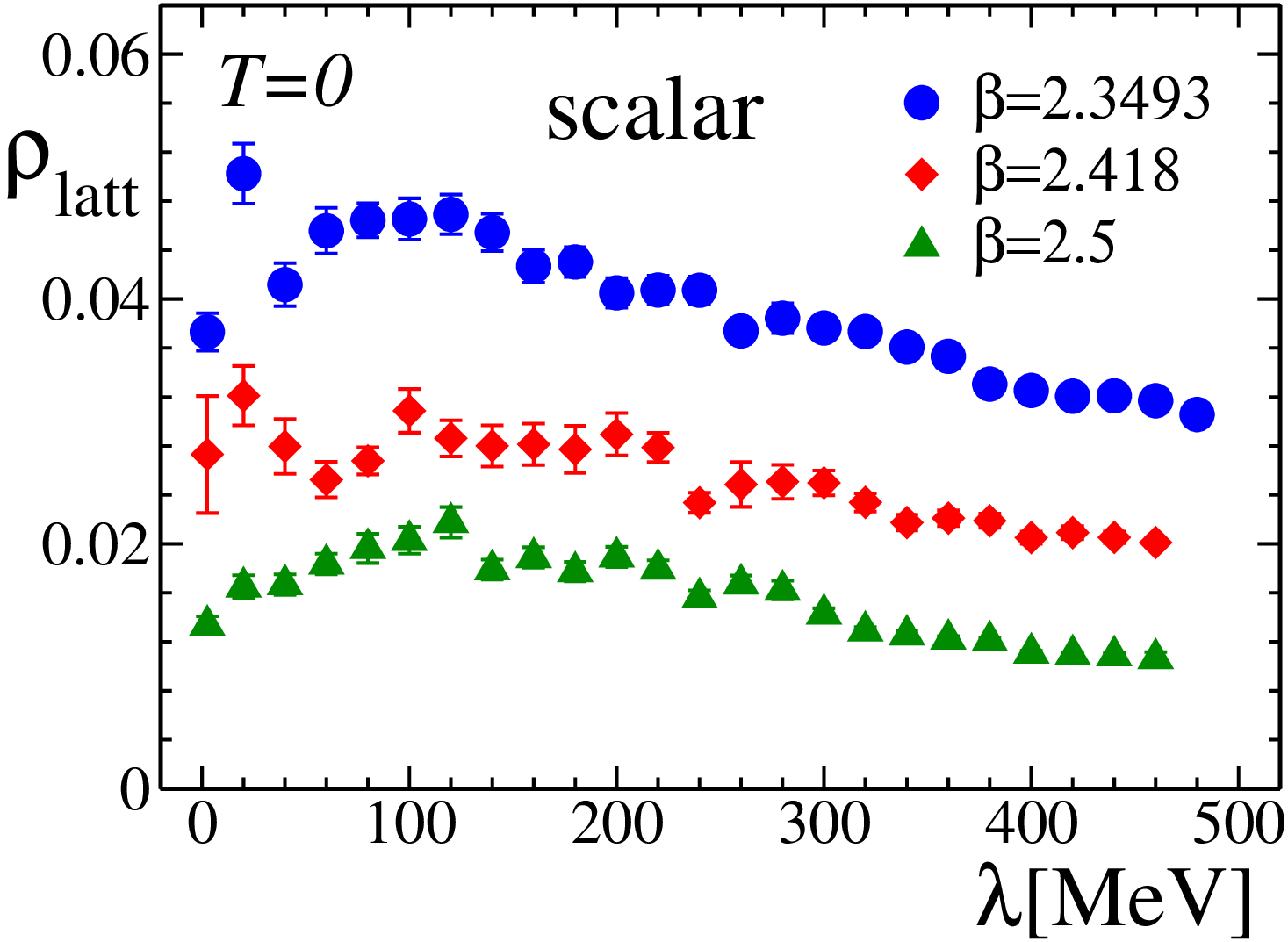} &
\includegraphics[width=74mm,clip=true]{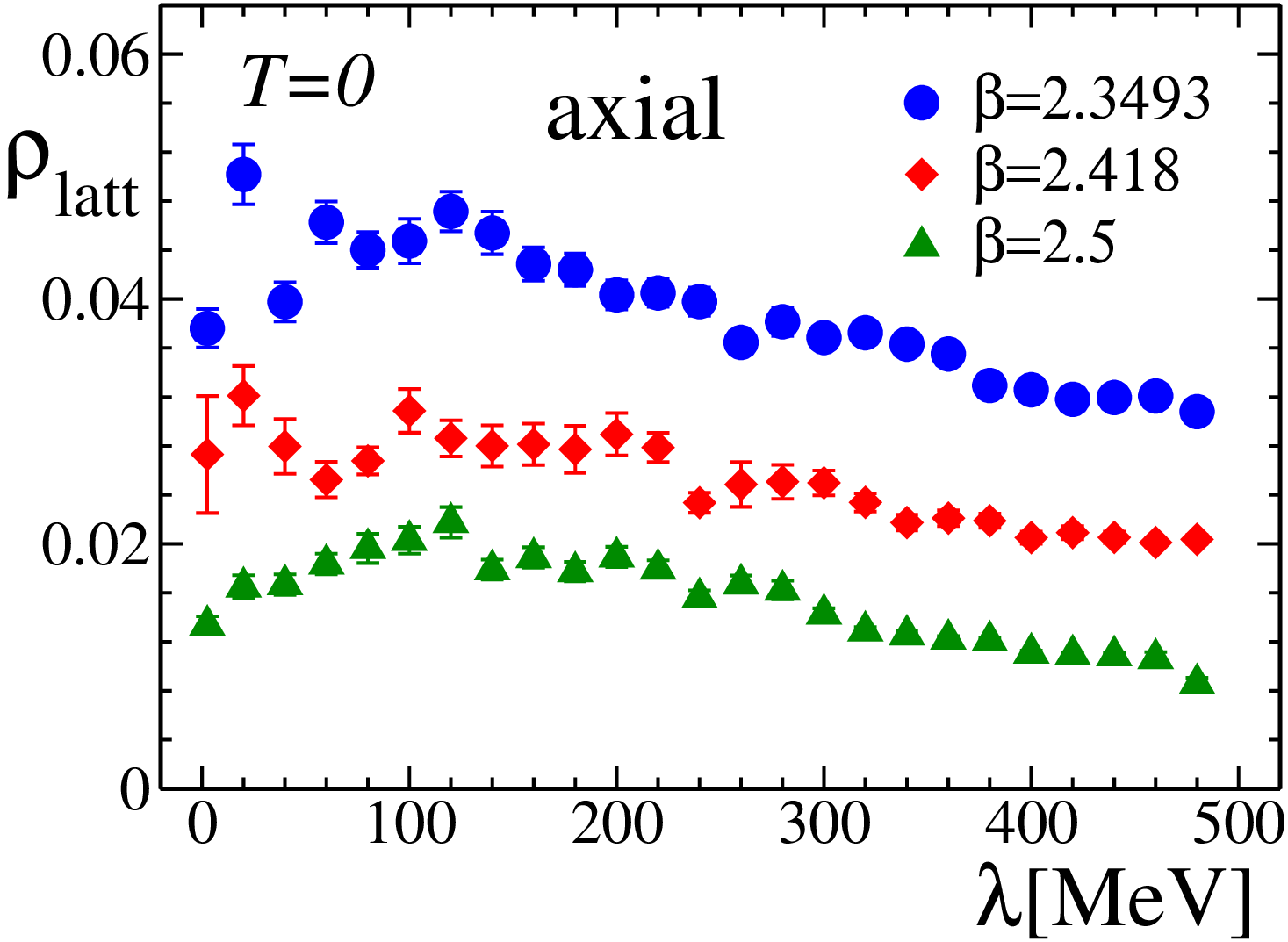} \\
(a) & (b)\\
\includegraphics[width=74mm,clip=true]{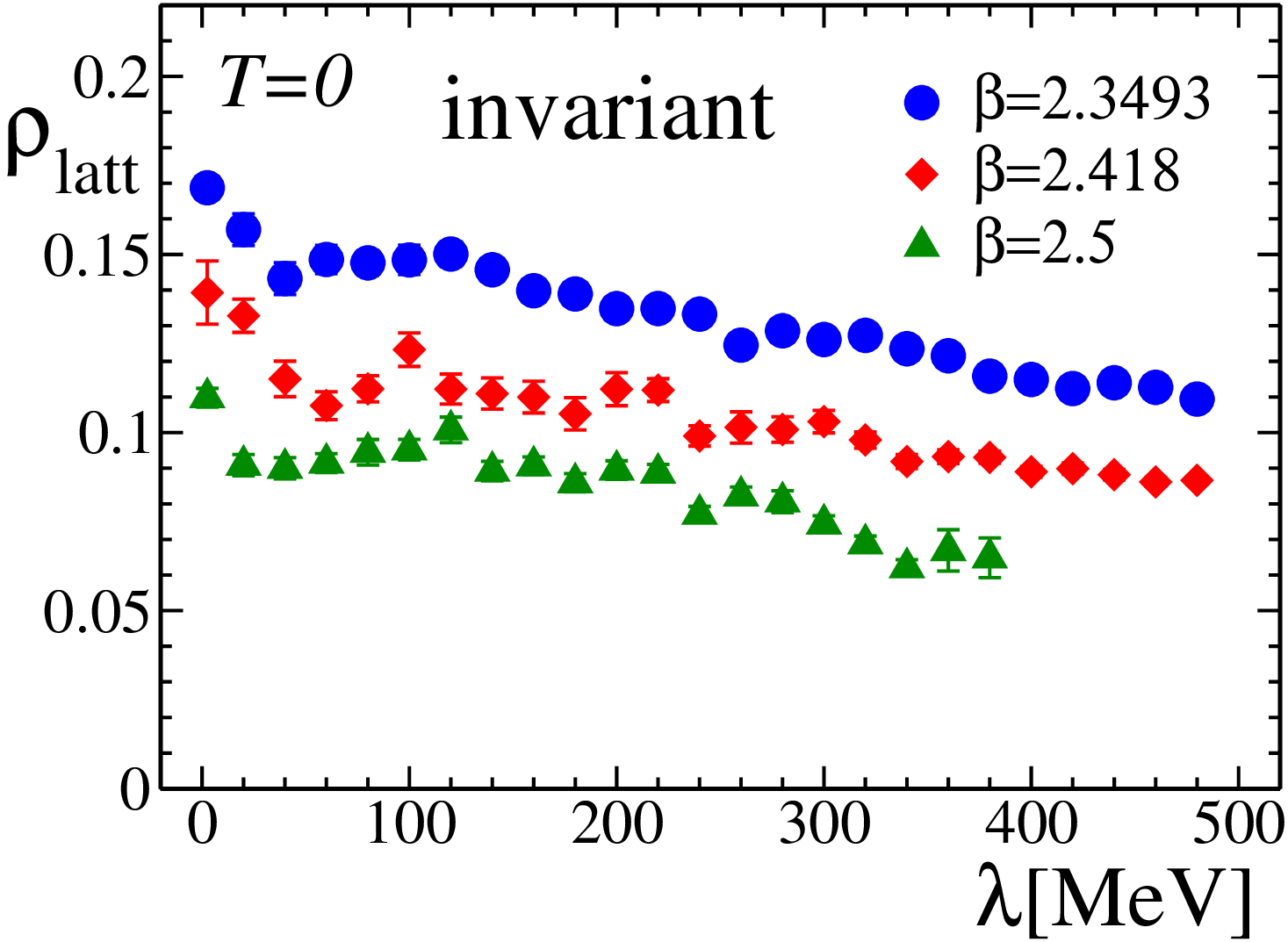} &
\includegraphics[width=74mm,clip=true]{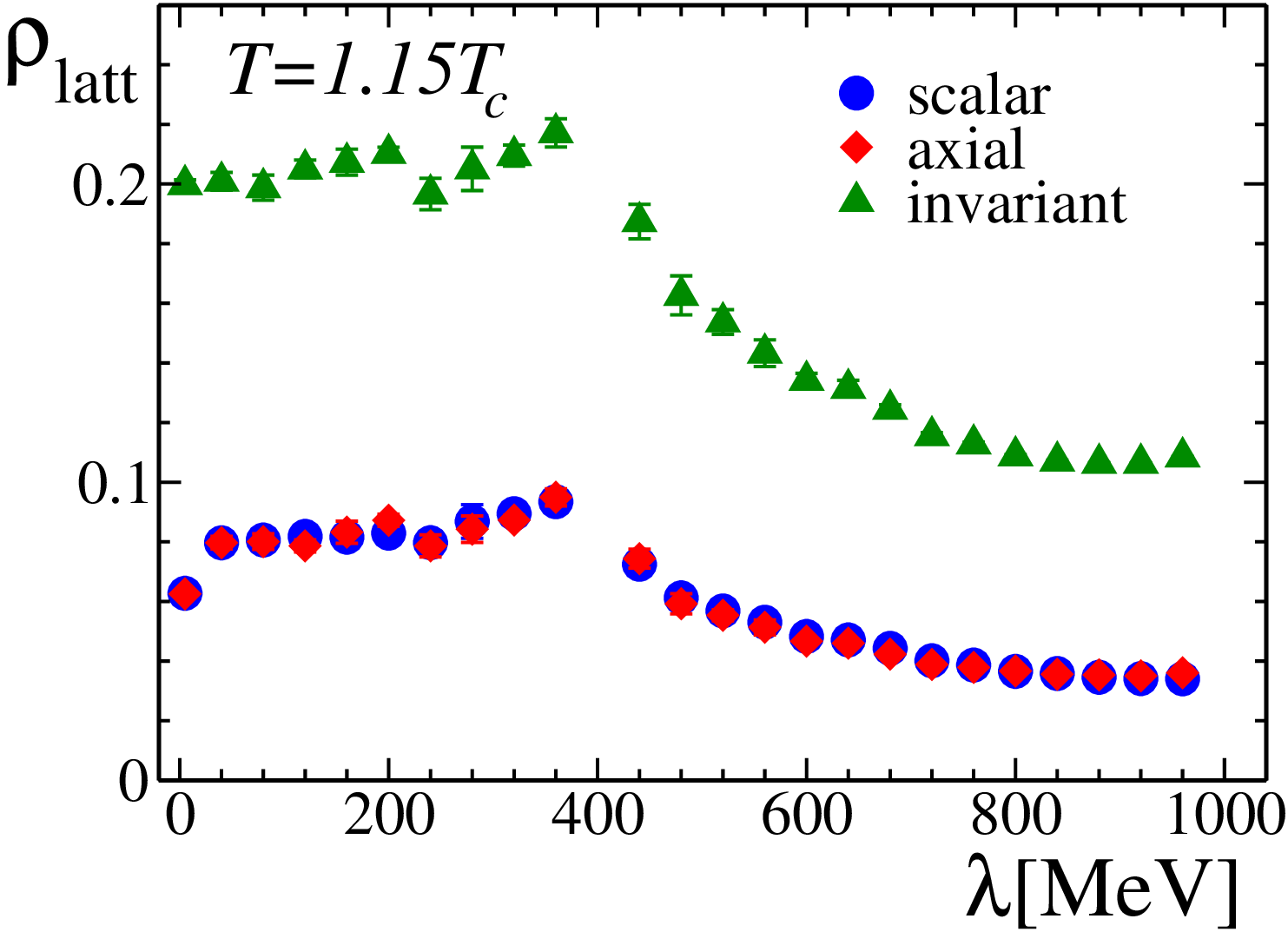} \\
(c) & (d)
\end{tabular}
\caption{(Color online) The densities of the (a) scalar, (b) axial, and (c) chirally invariant embedded monopoles
in confinement phase for $\beta = 2.3493$, $2.418$, and $2.5$ on {\it vs.} the Dirac eigenvalue $\lambda$.
(d) The same but for the deconfinement phase at $T = 1.15T_c$. The densities are given
in the units of the lattice spacing $a$.}
\label{fig:densities:lambda:coarsed}
\end{figure}

In Figures~\ref{fig:densities:lambda:coarsed} (a), (b) and (c) we
show the lattice densities $\rho_{\latt}$ of, respectively,
scalar, axial and invariant embedded monopoles at zero
temperature. The densities, plotted in the units of the lattice
spacing $a$, are shown as functions of the Dirac eigenmode energy
$\lambda$ for three values of the coupling $\beta$. Apart from a
few irregular points (which we ascribe to statistical
fluctuations), the densities are smooth functions of the eigenmode
energy $\lambda$. Moreover the scalar and axial densities at
zero-temperature are decreasing functions of $\lambda$ for $\lambda
\gtrsim 150\,\mbox{MeV}$. The density of the chirally invariant
quark monopole is a decreasing function for all considered values
of $\lambda$. The density of the chirally invariant monopoles is higher
than the scalar and axial monopole densities.

According to Figures~\ref{fig:densities:lambda:coarsed} (a) and
(b) the scalar and axial densities coincide with each other within
the error bars. This observation does not contradict the statement
that the chiral symmetry is broken in the low-temperature phase.
In order to illustrate this fact let us consider, as a toy example,
a lattice model describing the $SO(2)$ global scalar
field $\phi = (\phi_1, \phi_2)^T$ with two real-valued
components $\phi_1$ and $\phi_2$. The global $SO(2)$ symmetry
treats the field $\phi$ as a two-component vector in an internal space.
This symmetry is an analogue of the axial symmetry~\eq{eq:axial:U1}, while the
components $\phi_1$ and $\phi_2$ can be associated with the scalar, $\xi_S$, and
axial, $\xi_A$, triplets, respectively.

Apart from a kinetic term, $(\partial_\mu \phi)^2$, a generic Lagrangian of
the toy model should contain the $SO(2)$-invariant potential
\be
V(\phi^2) = \mu^2 \phi^2 + \lambda (\phi^2)^2\,,
\label{eq:V}
\ee
with $\phi^2 \equiv \phi^2_1 + \phi^2_2$. In the broken phase ($\mu^2 < 0$)
the scalar field develops a non-zero expectation value in the infinite volume system,
$\langle \phi \rangle \neq 0$. As a consequence, the vacuum of
the model becomes non-invariant under the $SO(2)$ transformations. In the
symmetric phase ($\mu^2 \geqslant 0$) the global symmetry is
restored since $\langle \phi \rangle = 0$.

However, in a finite volume the expectation values of the components
of the Higgs fields should vanish in both phases, $\langle \phi_1\rangle_V = 0$ and
$\langle \phi_2\rangle_V = 0$ (here the subscript $V$ in $\langle \dots \rangle_V$
indicates that the averages are taken in a finite volume). This happens
because in the averages the finite number of integrals over the scalar fields $\phi_{1,2}(x)$ includes also
integrations over all orientations of the fields in the internal space. Thus, in a finite-element
system all the internal directions are formally equivalent.
Moreover, one gets $\langle \phi^2_1\rangle_V
\equiv \langle \phi^2_1\rangle_V$ both in broken and unbroken phases because of the
same reason. This is precisely the same what we could expect for the expectation values
of the triplet fields, $\langle \vec \xi^2_S \rangle_V = \langle \vec \xi^2_A\rangle_V$, in
a finite volume system regardless if the chiral symmetry is broken in the thermodynamic
limit or not.

As a consequence, finite volume expectation values of any operator ${\mathcal O}(\xi_S)$
should be equivalent to the expectation value of its axial
counterpart ${\mathcal O}(\xi_A)$. In other words,
the choice of the axial isovector $\vec \xi_A$ in
a role of the adjoint composite Higgs field is as good as the
choice of the scalar isovector $\vec \xi_S$. Thus the densities of
the scalar and axial quark monopoles should coincide with each other in the finite volume.

Apart from the quenched case, in the real QCD with dynamical quarks the breaking
of the chiral symmetry must explicitly be seen in the densities of the embedded
monopoles: the density of the scalar and axial monopoles must in general be
different. For example, it is expected~\cite{ref:embedded:QCD} that
at sufficiently high temperatures the density of the axial quark monopoles
should be higher than the density of the scalar monopoles. We expect that the
effect should also be observable in the finite volume because of an explicit
breaking of the axial symmetry by the fermion determinant.
In fact, the measure of the integration over the fermion fields is axially
anomalous~\cite{ref:Fujikawa}, and therefore the expectation values of the
operator ${\mathcal O}(\xi_S)$ should be different from its axial
analogue ${\mathcal O}(\xi_A)$. In terms of the toy $SO(2)$ model
the effect of non-invariant measure can be emulated by
an addition of an extra $SO(2)$ breaking term (say $\delta V \propto \phi_1$) to the
potential~\eq{eq:V}.

In Figure~\ref{fig:densities:lambda:coarsed} (d) we show the density of the
quark monopoles at the temperature $T=1.15\, T_c$ corresponding to the deconfinement phase.
Similarly to the zero temperature case, the density of the scalar
and axial monopoles coincide with each other. The invariant monopoles are denser
than the scalar/axial monopoles for all values of eigenvalue $\lambda$.
The monopole density is independent of the eigenvalue
in the region $0 \leqslant \lambda \lesssim 400\,\mbox{MeV}$. In the limit
$\lambda \to 0$ the densities of the scalar/axial and invariant monopoles
in physical units are, respectively,
\beqn
\lim_{\lambda \to 0} \rho^\phys_{\Gamma}(\lambda) \to
\left\{
\begin{array}{ll}
\approx {(3\, \mathrm{fm})}^3 & \quad \mbox{[scalar and axial]}\\
\approx {(4\, \mathrm{fm})}^3 & \quad \mbox{[invariant]}
\end{array}
\right.
\quad\qquad\quad T = 1.15\, T_c\,.
\eeqn
In the region $\lambda \gtrsim 400\,\mbox{MeV}$ the density
of the monopoles of all three types quickly drops down.
This observation will be confronted with the fermion spectral function
in Section~\ref{sec:spectral}.

To estimate the effect of temperature on the monopole density
it worth comparing the lattice monopole densities at zero
temperature for $\beta=2.3493$ (shown by filled circles in
Figures~\ref{fig:densities:lambda:coarsed}(a),(b), and (c)) and at
$T=1.15\, T_c$ for $\beta=2.35$ (shown in
Figure~\ref{fig:densities:lambda:coarsed}(d)). The selected values
of the lattice coupling $\beta$ are very close to each other and
therefore they correspond to almost the same value of the lattice
spacing $a$ according to Table~\ref{tbl:simulation:parameters} of the appendix~\ref{app:details}.
In a wide region of the Dirac eigenvalues, $0 < \lambda < 500\,\mbox{MeV}$, the density of the scalar and axial
monopoles at $T=0$ case are noticeably smaller then the density
of the these monopoles at $T=1.15\, T_c$:
\beqn
\frac{\rho^{\latt}_{S,A}(T=1.15\,T_c)}{\rho^{\latt}_{S,A}(T=0)} \sim  2 \dots 3\,.
\label{eq:T:ratio:SA}
\eeqn
The effect of temperature on the invariant quark monopoles is
milder compared to the scalar/axial monopoles:
\beqn
\frac{\rho^{\latt}_{I}(T=1.15\,T_c)}{\rho^{\latt}_{I}(T=0)} \sim  1.5 \dots 2\,.
\label{eq:T:ratio:I}
\eeqn

The difference in ratios~(\ref{eq:T:ratio:SA}) and (\ref{eq:T:ratio:I})
can probably be explained by the fact that the chirally invariant
embedded monopoles are, by definition, explicitly invariant under
the axial transformations~\eq{eq:axial:U1}, while the scalar and
axial monopoles are not. Thus the invariant monopole is less
insensitive to the effects of the chiral symmetry breaking/restoration
compared to the scalar and axial monopoles.

Summarizing, the results of this Section show that the density of
the quark monopoles is an increasing function of the temperature
in agreement with general expectations~\cite{ref:embedded:QCD}.

\subsection{Scaling towards continuum limit}

An extrapolation to the continuum limit of numerically calculated
quantities is one of the most important issues of the
lattice simulations. In general, the monopole densities
can be extrapolated to the continuum with the help of the
following polynomial formula:
\beqn
\rho^{\mathrm{latt}}(a) = C + v \cdot a + s \cdot a^2 + \rho \cdot a^3\,,
\label{eq:extrapolation}
\eeqn
where $C$, $v$, $s$ and $\rho$ are the fitting coefficients. The terms of the order $O(a^4)$
and possible logarithmic corrections are neglected in Eq.~\eq{eq:extrapolation}. Naively,
if the monopoles are physical objects which
form a gas-like ensemble then one could expect that the coefficient $\rho$ -- representing
the physical density of the monopoles -- is to be non-zero while the other coefficients
in Eq.~\eq{eq:extrapolation} are vanishing. Below we show numerically that this is not the case.

We found that the scaling of densities for all
non-zero modes, $\lambda \neq 0$, is universal in a sense that the
form of the scaling function does not depend on $\lambda$ while being
sensitive to the monopole type. For illustrative properties we take
here the eigenvalue $\lambda=235\,\mbox{MeV}$. We show in
Figures~\eq{fig:density:various:fits:high}(a) and (b) the
densities of, respectively, the scalar and invariant embedded
monopoles {\it vs.} the lattice spacing $a$. Since the scalar and
axial monopoles have the same (within error bars) densities we
show the data for the scalar monopoles only. In the same figures
we show the best fitting curves for the (truncated) fitting
function~\eq{eq:extrapolation}.
\begin{figure}[!htb]
\begin{tabular}{cc}
\includegraphics[width=74mm,clip=true]{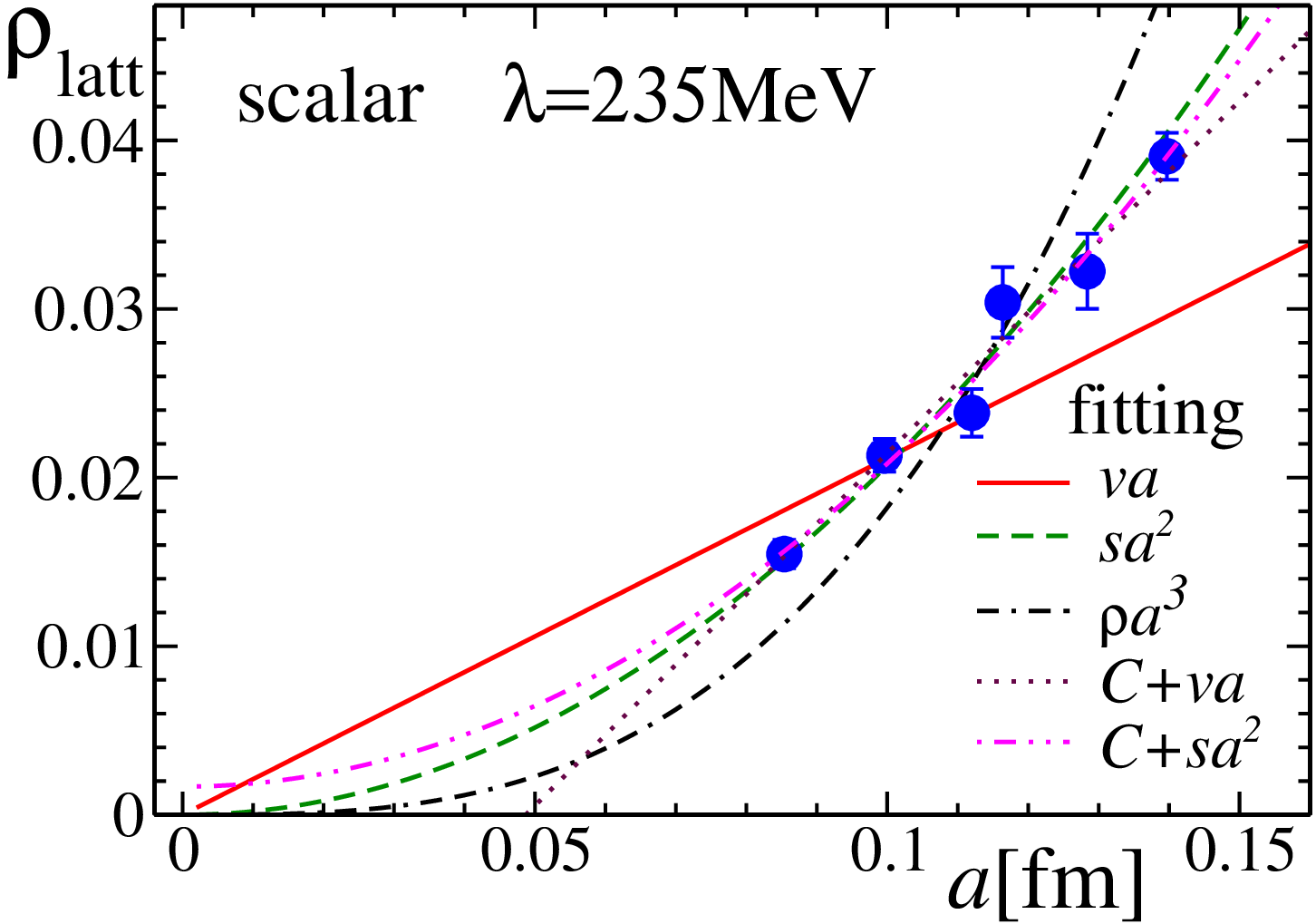} & \hspace{5mm}
\includegraphics[width=74mm,clip=true]{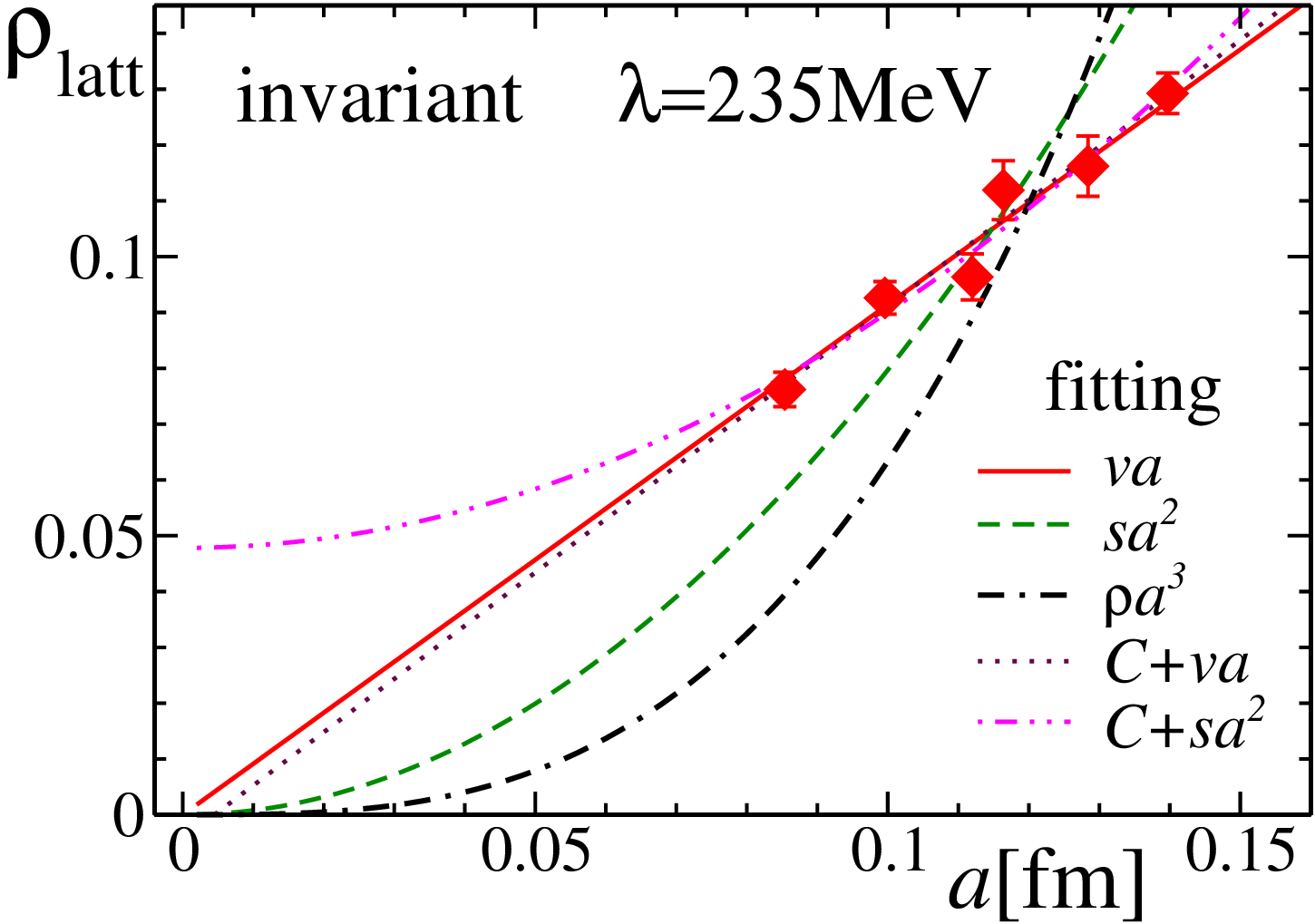} \\
(a) &  \hspace{5mm} (b)
\end{tabular}
\caption{(Color online) The extrapolation of the scaling coefficients for the densities of the (a) scalar and axial, and (b)
chirally invariant quark monopoles using various fits for $\lambda=235\,\mbox{MeV}$.}
\label{fig:density:various:fits:high}
\end{figure}

As it seen from Figures~\ref{fig:density:various:fits:high} the
expected fit
$\rho_\latt \propto a^3$ does not work for all types of monopoles.
The corresponding quality of the
fit is $\chi^2/d.o.f. = 20$ and 82 for scalar/axial and invariant
monopoles, respectively. However, the fits
$\rho^{\mathrm{latt}}(a) = C + v \cdot a$ and
$\rho^{\mathrm{latt}}(a) = C + s \cdot a^2$ give reasonable values for
$\chi^2/d.o.f.$ (of the order of unity), while the coefficient $C$
is consistent with zero within error bars in all our fits. Setting
$C=0$ we obtain that the best fits for the scalar/axial and
invariant monopole densities is achieved, respectively,  by the functions
\beqn
\rho^{\mathrm{latt}}_{S,A}(a,\lambda) = s_{S,A}(\lambda) \cdot a^2\,,\quad
\rho^{\mathrm{latt}}_{I}(a,\lambda) = v_{I}(\lambda) \cdot a\,, \qquad \lambda \neq 0\,.
\label{eq:extrapolation:Lneq0}
\eeqn
In all these cases $\chi^2/d.o.f. \sim 1$. Note that the
density of the scalar/axial and invariant monopoles can not be
well fitted by the linear and, respectively, quadratic functions
of $a$ since in these cases the quality of fits is as large as $10-20$.
All discussed fits are shown in Figures~\ref{fig:density:various:fits:high}
(a) and (b) by lines.

\begin{figure}[!htb]
\begin{tabular}{cc}
\includegraphics[width=74mm,clip=true]{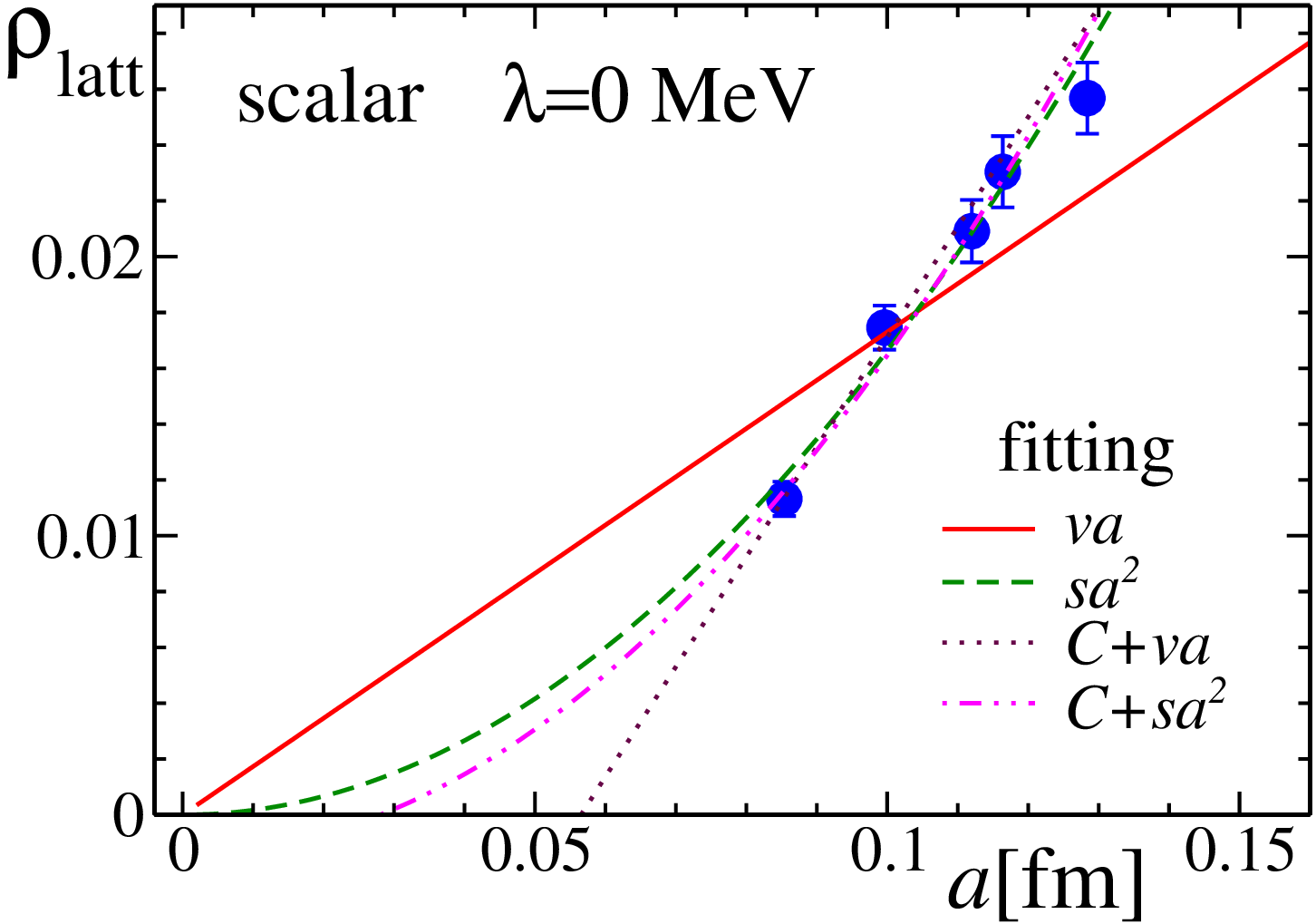} & \hspace{5mm}
\includegraphics[width=74mm,clip=true]{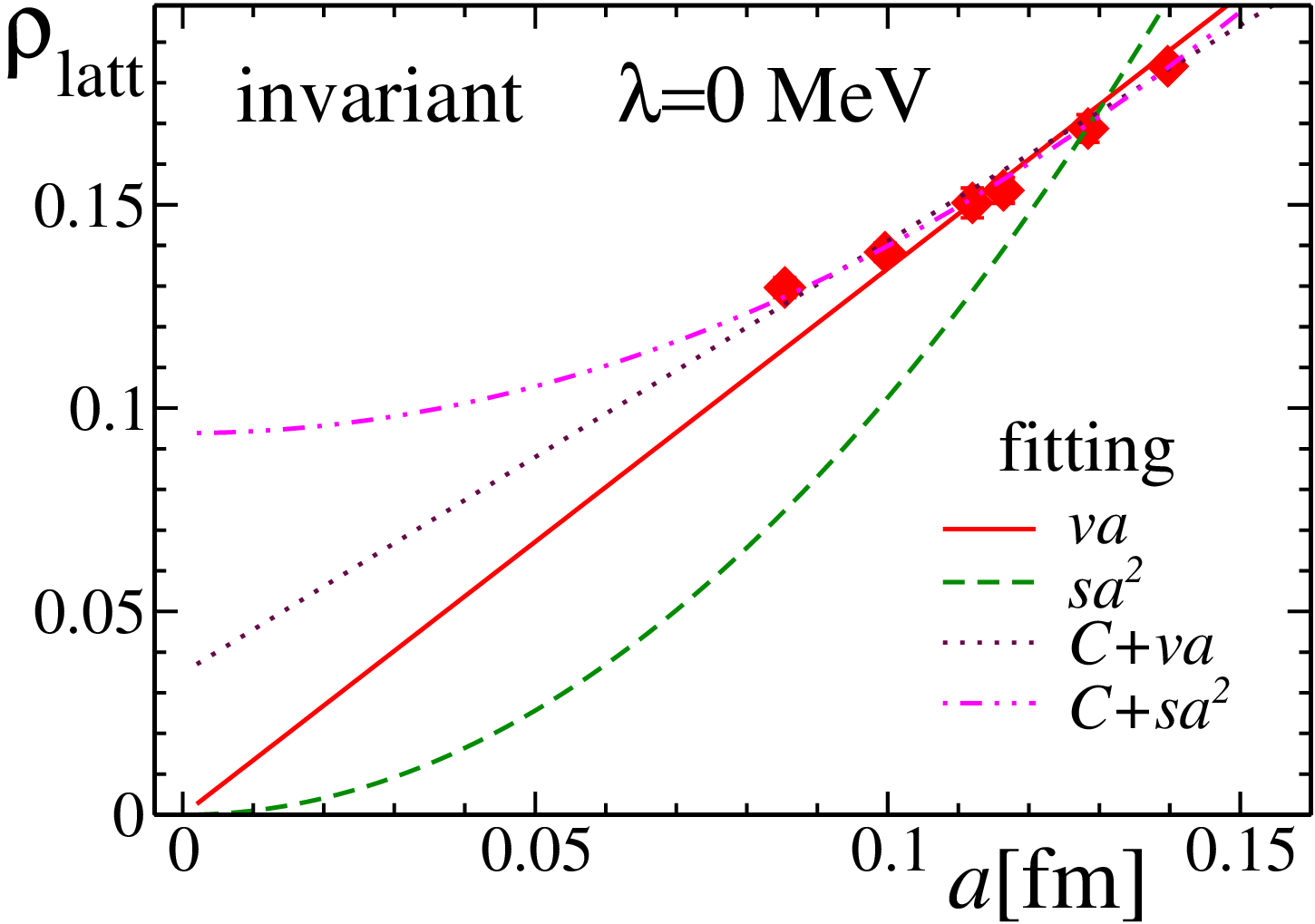} \\
(a) & \hspace{5mm} (b)
\end{tabular}
\caption{(Color online) The same as in Figure~\eq{fig:density:various:fits:high} but for the zero mode, $\lambda=0$.}
\label{fig:density:various:fits:zero}
\end{figure}
The similar analysis can be performed for the zero mode, Figures~\ref{fig:density:various:fits:zero}(a) and (b).
We find that the best fit functions are
\beqn
\rho^{\mathrm{latt}}_{S,A}(a) = s_{S,A}(0) \cdot a^2\,,\quad
\rho^{\mathrm{latt}}_{I}(a) = C_I(0) + s_I(0) \cdot a^2\,, \qquad \lambda = 0\,,
\label{eq:extrapolation:L0}
\eeqn
with $\chi^2/d.o.f. \sim 1.3$ and $0.5$, respectively. The best fit parameters are
\beqn
s_{S,A}(0) = 1.66(3)\,\mbox{fm}^2\,, \qquad C_I(0) = 0.094(2)\,, \quad s_I(0) = 4.6(2)\,\mbox{fm}^2\,.
\label{eq:zero:scaling:res}
\eeqn
Thus, the scaling properties of the density of the invariant monopoles
at $\lambda\neq 0$, Eq.~\eq{eq:extrapolation:Lneq0}, and at $\lambda = 0$,
Eq.~\eq{eq:extrapolation:L0}, are different from each other even on the
qualitative level.

We also attempted to determine the scaling behavior of the
densities using the power fit of the form $C\, a^\alpha$, where
$C$ and $\alpha$ are fitting parameters. For the non-zero modes we
typically get $\alpha_{S,A} \approx 2$ and $\alpha_{I}\approx 1$
which is in agreement with the best fit function used above. For
example, for the case of $\lambda = 235\,\mbox{MeV}$ we get
$\alpha_{S,A} = 1.9(1)$ and $\alpha_{I} = 1.04(7)$. The
scalar/axial monopoles constructed from the of the zero Dirac mode
give $\alpha_{S,A}(0) = 2.1(1)$. The fit by the same dependence of
the invariant monopole with $\lambda = 0$ give $\alpha_{I}(0) =
0.78(6)$ with higher values of $\chi^2/d.o.f. \approx 3$.
Therefore the constant term $C_I \neq 0$ in the corresponding
fitting function (the middle formula in
Eq.~\eq{eq:extrapolation:L0}) is essential.

\begin{figure}[!htb]
\begin{tabular}{cc}
\includegraphics[width=74mm,clip=true]{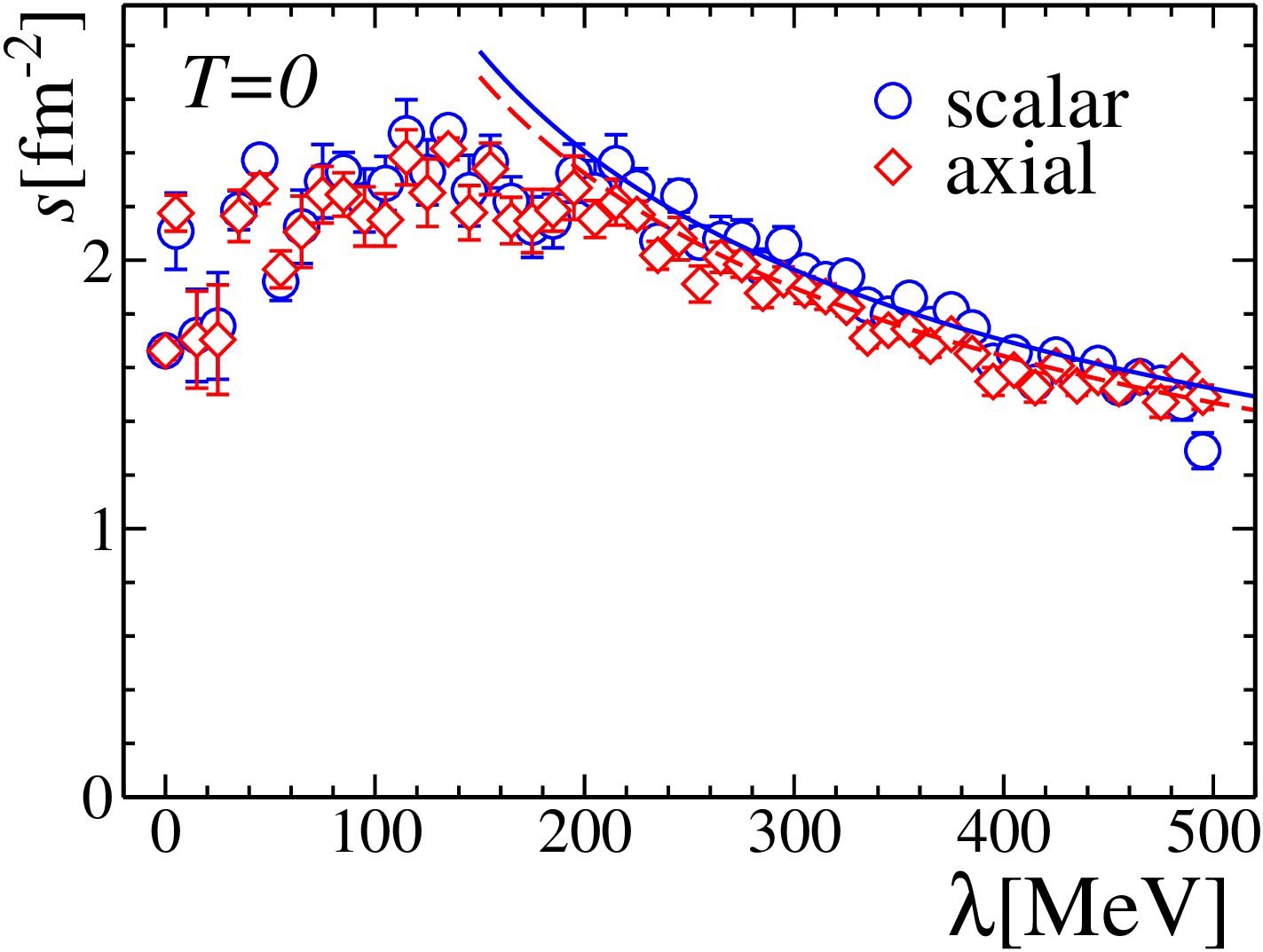}& \hspace{5mm}
\includegraphics[width=74mm,clip=true]{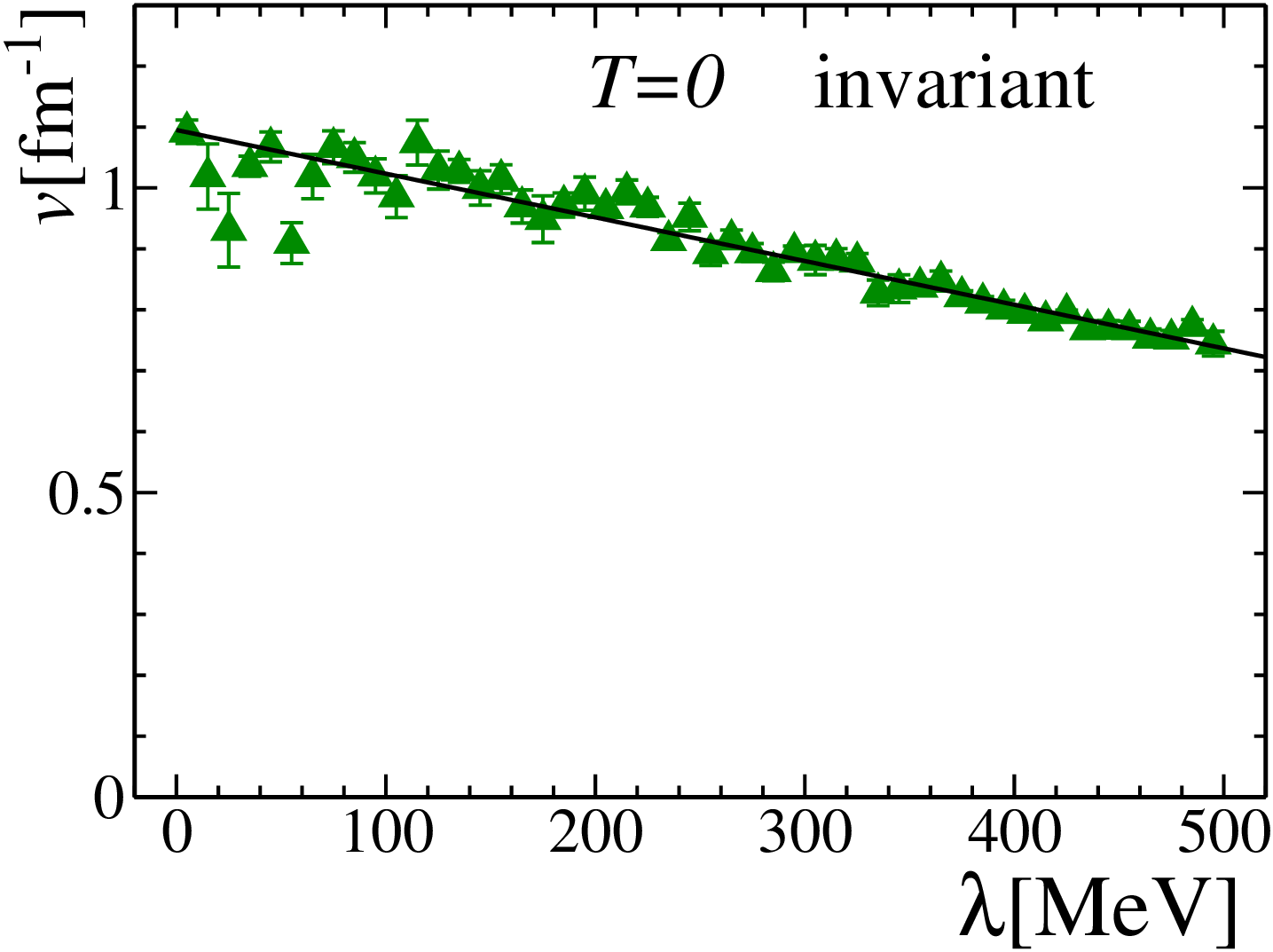} \\
(a) & \hspace{5mm} (b)\\
\end{tabular}
\caption{(Color online) The scaling coefficients for the densities of the (a) scalar and axial, and (b)
chirally invariant quark monopoles. The fits by functions~\eq{eq:s:fit} and \eq{eq:v:fit} are
shown by solid and dashed lines.}
\label{fig:monopole:densities}
\end{figure}

The scaling coefficients $s_{S,A}$ and $v_I$ obtained with the help of extrapolation~(\ref{eq:extrapolation:Lneq0})
to the continuum limit are shown in Figures~\ref{fig:monopole:densities} (a) and (b), respectively,
as functions of the Dirac eigenvalue $\lambda$.
The scaling coefficient $s_{S,A}(\lambda)$ of the scalar and axial monopoles has a
peak around $\lambda \sim 150\,\mbox{MeV}$ while the scaling coefficient $v_{I}(\lambda)$
of the invariant mode is a monotonically decreasing function for all studied eigenvalues $\lambda$.

The behavior of the scaling coefficients $s_{S,A}$ and $v_I$ has some particularities. For example,
we find that these scaling coefficients can be described by the formulae
\beqn
s^{\mathrm{fit}}_{S,A}(\lambda) & = &
l^2_{S,A} \cdot {\Bigl(\frac{\lambda}{1\,\mbox{MeV}}\Bigr)}^{-\gamma_{S,A}}\,,
\qquad \quad \mbox{for} \quad \lambda > 250\,\mbox{MeV}\,,
\label{eq:s:fit}\\
v^{\mathrm{fit}}_{I}(\lambda) & = & B_I \Bigl(1 - \frac{\lambda}{\lambda_I} \Bigr)\,,
\hskip 25.5mm \mbox{for all\ } \lambda \,,
\label{eq:v:fit}
\eeqn
where the data for $s_{S,A}$ is compared with the fitting function~\eq{eq:s:fit}
only in the region of large $\lambda$ with $\lambda > 250\,\mbox{MeV}$ since
in the small $\lambda$-region the behavior of this quantity is statistically unclear (there
is however, a noticeable tend of $s_{S,A}$ to decrease as $\lambda \to 0$).
We get (with $\chi^2/d.o.f. \approx 2 $) the similar values for scalar and axial monopoles:
\beqn
\gamma_S & = & 0.54(4)\,,\qquad l_S = 0.15(2)\,\mbox{fm} \,,\\
\gamma_A & = & 0.44(4)\,,\qquad l_A = 0.21(2)\,\mbox{fm}\,.
\eeqn
This result suggest that the scaling exponent $\gamma$ may be close to $1/2$
for scalar and axial types of the quark monopoles. Setting $\gamma_{S,A} = 1/2$
one gets
\beqn
\begin{array}{rcl}
l_S & = & 0.171(1)\,\mbox{fm} \\
\qquad l_A & = & 0.174(1)\,\mbox{fm}
\end{array}
\qquad \quad \mbox{for} \quad \gamma_{S,A} = 1/2\,.
\eeqn
The last fits are shown in Figure~\ref{fig:monopole:densities}(a)
by the solid and dashed lines for the scalar and axial monopoles,
respectively.

The scaling coefficient $v_I$ is compared to the fitting
function~\eq{eq:v:fit} in the whole available region of the
eigenvalues $\lambda$. We get the following best fit parameters:
\beqn
B_I = 1.09(1)\,\mbox{fm}^{-1}\,,\qquad \lambda_I = 1.53(4)\,\mbox{GeV}\,.
\eeqn
The corresponding fit is shown in Figure~\ref{fig:monopole:densities}(b) by the solid line.
One can see that the scaling of the coefficient $v_I$ for the invariant monopoles towards small values
of the eigenvalues $\lambda$ is a smooth linear function over the whole region of studied eigenvalues $\lambda$.
The $\lambda\to 0$ limit for the coefficients $s_{S,A}$ corresponding to the scalar and axial monopoles are known
less accurately, as it can be seen from Figure~\ref{fig:monopole:densities}(b).
Summarizing, in the $\lambda\to 0$ limit we find:
\beqn
\lim_{\lambda \to 0} s_{S,A}(\lambda) \approx 1.9(3) \,\mbox{fm}^{-1}\,,\qquad
\lim_{\lambda \to 0} v_I(\lambda) \equiv B_I = 1.09(1)\,\mbox{fm}^{-1} \,.
\label{eq:nonzero:scaling:res}
\eeqn
As it is seen from
Eqs.~(\ref{eq:zero:scaling:res}), (\ref{eq:nonzero:scaling:res}) the
scaling coefficients $s_{S,A}$ at $\lambda=0$ for scalar and axial
modes seems to coincide with the corresponding limits,
$\lim_{\lambda \to 0} s_{S,A}(\lambda) \approx s_{S,A}(0)$. On the
other hand, the scaling coefficient $v_I$ for the invariant
monopole has a discontinuity at $\lambda=0$,
$v_I(0) \neq \lim_{\lambda \to 0} v_{I}(\lambda)$, since
the corresponding scaling formulae,
Eqs.~(\ref{eq:extrapolation:Lneq0}), (\ref{eq:extrapolation:L0}), are
different from each other even on the qualitative level.

\subsection{Cluster structure of the monopole ensembles}

The ensembles of the trajectories of the embedded monopoles can be
characterized by percolation properties. As it happens in the case
of the FK clusters in the Ising model, a general ensemble of the
monopole trajectories consists of clusters of different types. If
in the thermodynamic limit at certain physical conditions there
exists a non-zero probability to find a cluster of infinite
length, then the objects are said to be percolating and are often
called as ``condensed''. In the finite volume the role of the
percolating cluster is played by a monopole cluster with the size
of the order of the system volume. Using the standard terminology
we call the percolating clusters as ``infrared'' (IR) and the
short-length clusters are referred to as ``ultraviolet'' (UV).

In our studies we have used the following definitions of the IR and UV clusters
\cite{Bornyakov:2001ux}:
\begin{itemize}
\item The largest cluster is called the IR cluster;
\item The wrapped cluster is also called the IR cluster. More precisely, for each
monopole cluster $C$ we calculate the sum $S_\mu = \sum_{j \in C} j_\mu$.
If this sum is nonzero then the cluster is called the IR cluster;
\item Other clusters are called the UV clusters.
\end{itemize}

\begin{figure}[!htb]
\begin{tabular}{ccc}
\includegraphics[width=52mm,clip=true]{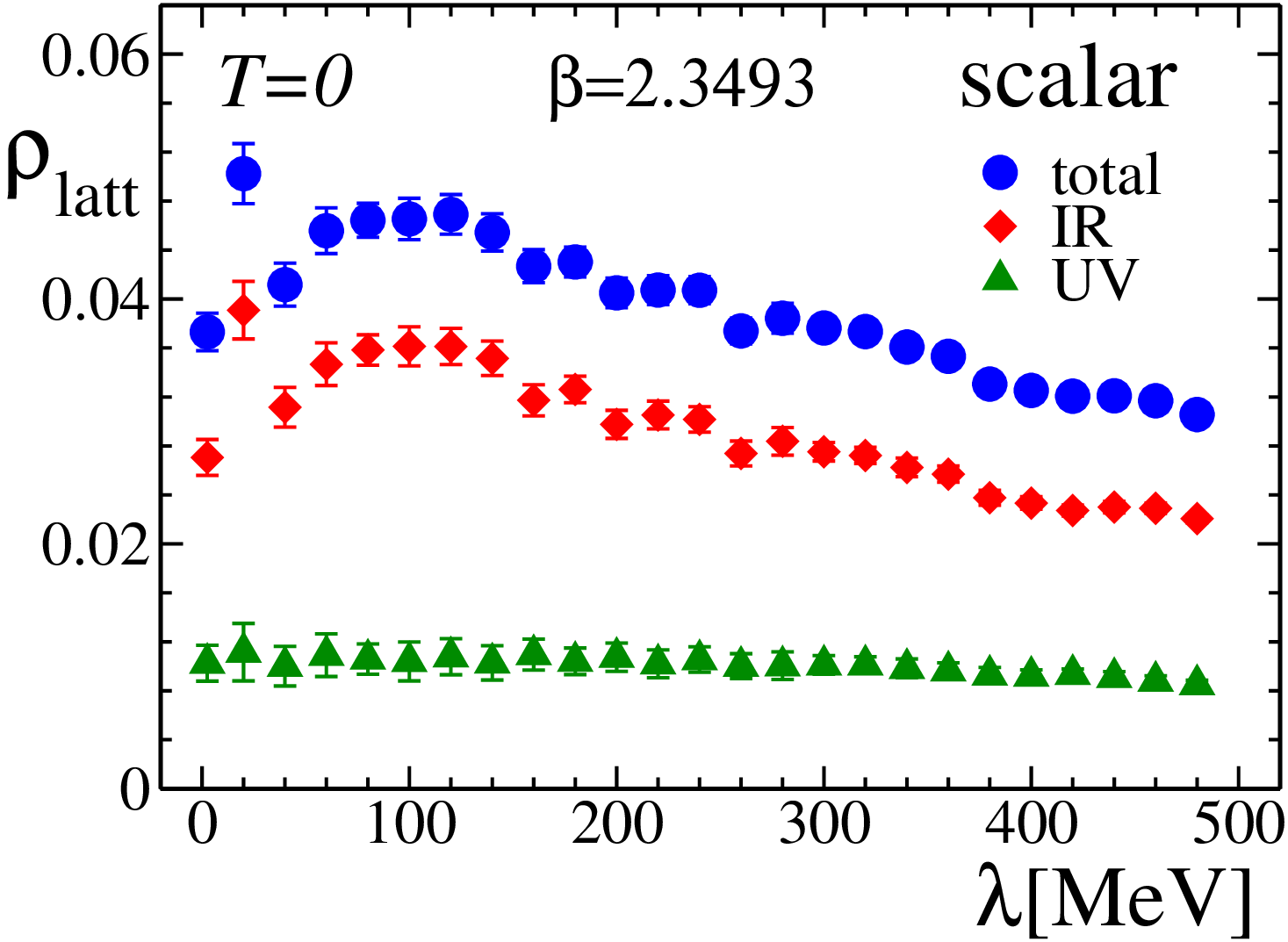}&
\includegraphics[width=52mm,clip=true]{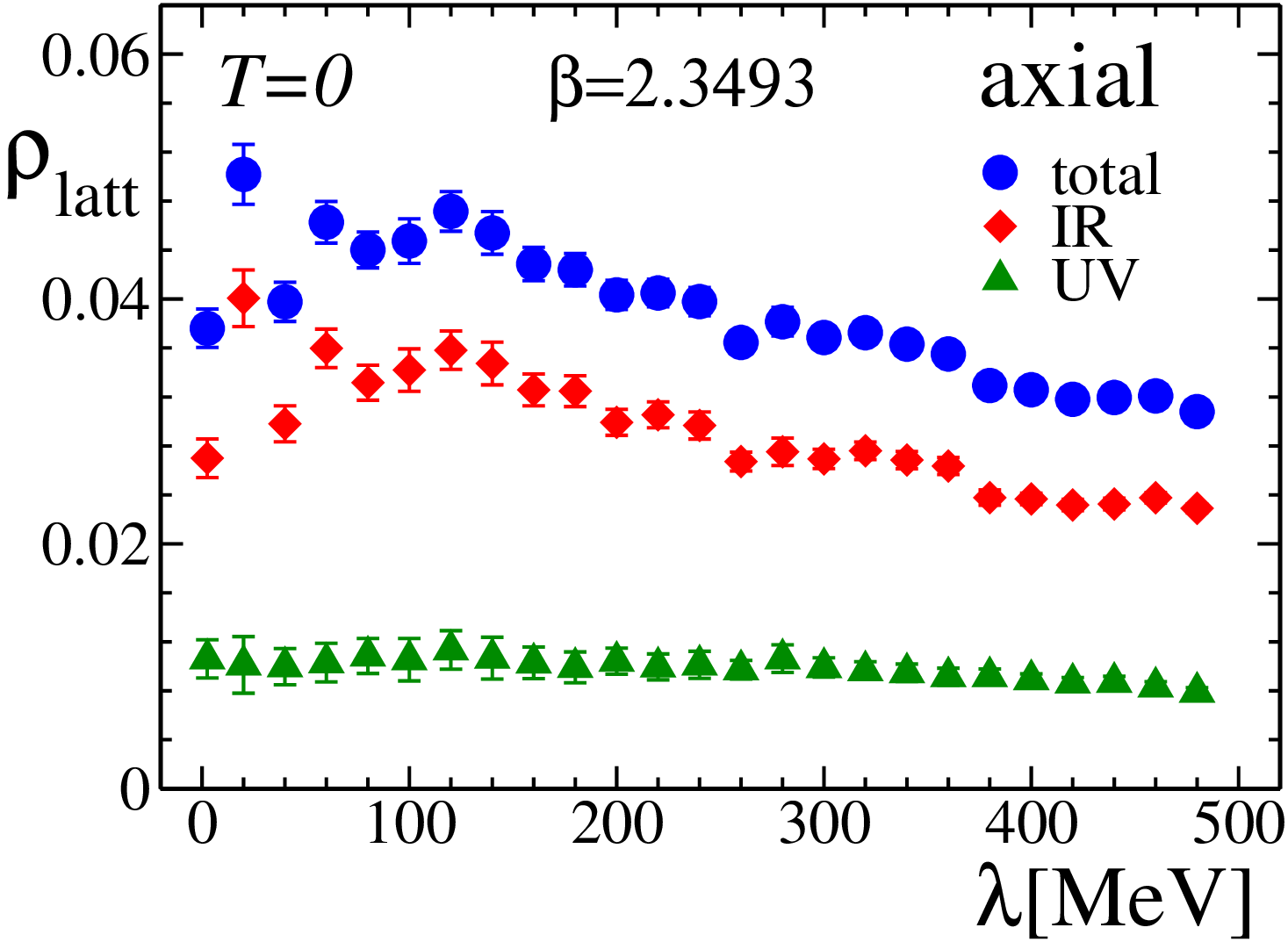} &
\includegraphics[width=52mm,clip=true]{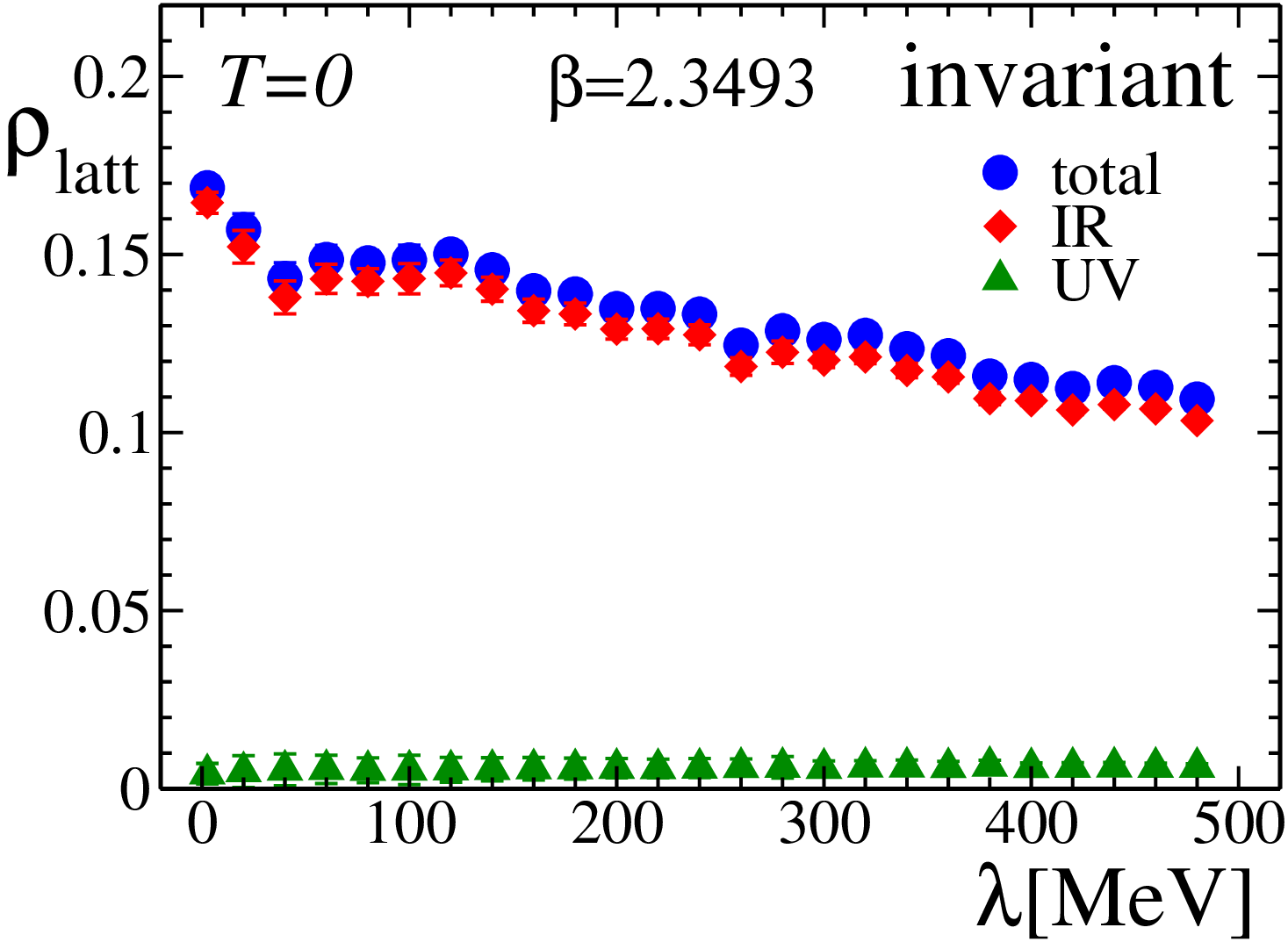} \\
(a) & (b) & (c)
\end{tabular}
\caption{(Color online) The total, infrared and ultraviolet densities of the (a) scalar, (b) axial, and (c)
chirally invariant embedded monopoles
{\it vs.} the Dirac eigenvalue $\lambda$.
}
\label{fig:densities:lambda:parts:scaled}
\end{figure}

As an example, we show in
Figures~\ref{fig:densities:lambda:parts:scaled}(a), (b) and (c)
the total monopole density along with the density of the quark
monopoles in the IR and the UV clusters for scalar, axial and
invariant monopoles, respectively. The densities are shown at zero
temperature (for $\beta=2.3493$) as functions of the eigenvalue
$\lambda$.
One finds that the most part of monopoles of all
types belongs to the IR clusters. The IR monopole density is about
3/4 of the total monopole density in the case of the scalar and
axial monopoles, while in the axially invariant case
almost all (about 95\%) monopoles are residing in the IR clusters.
Another interesting feature of the monopole density spectrum is
that the UV part of the monopole clusters is almost insensitive to
the value of the Dirac eigenvalue~$\lambda$. The UV density of the
invariant monopoles is very small being slightly increasing function
of the eigenvalue $\lambda$.

\begin{figure}[!htb]
\begin{tabular}{ccc}
\includegraphics[width=52mm,clip=true]{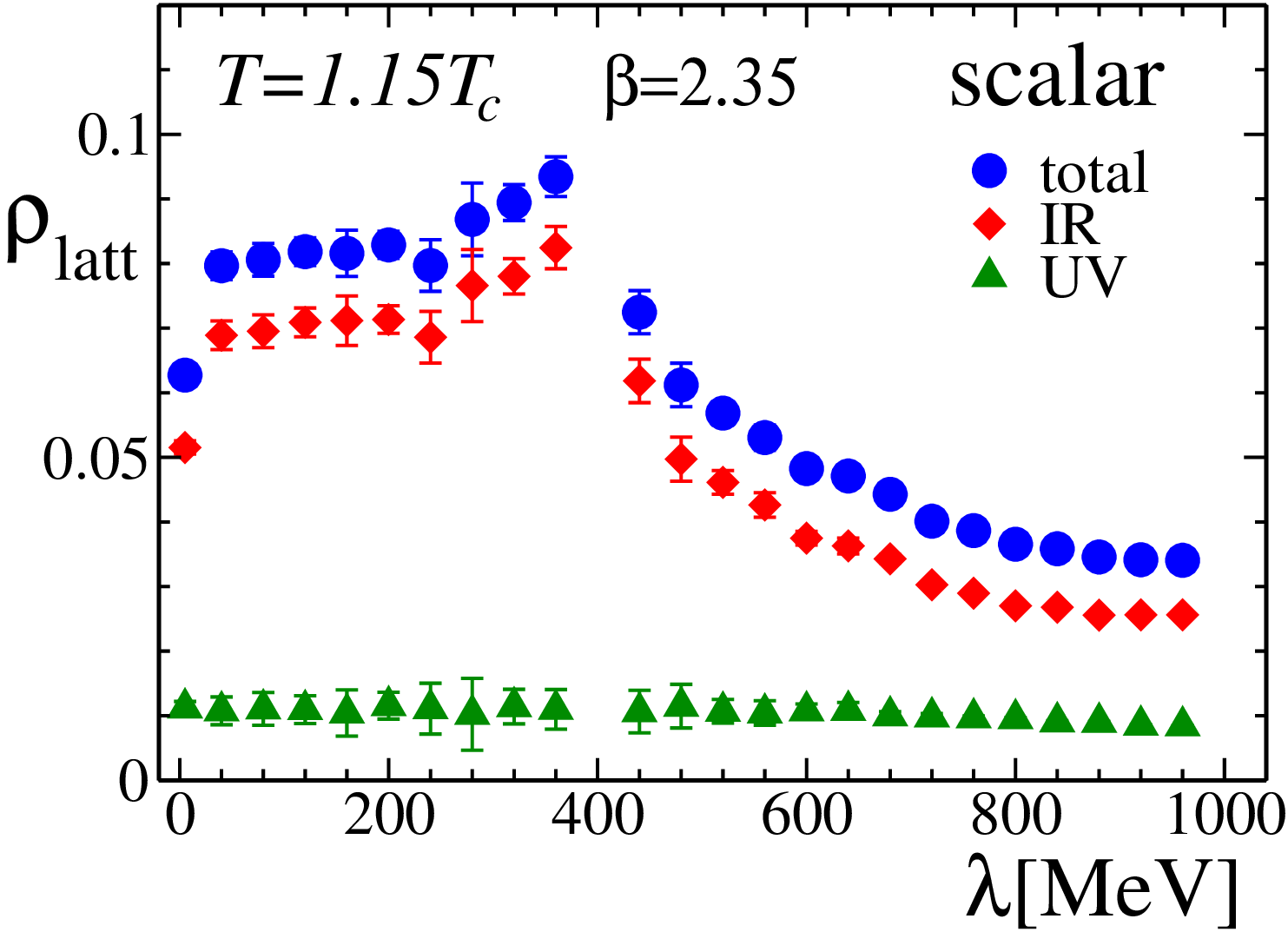}&
\includegraphics[width=52mm,clip=true]{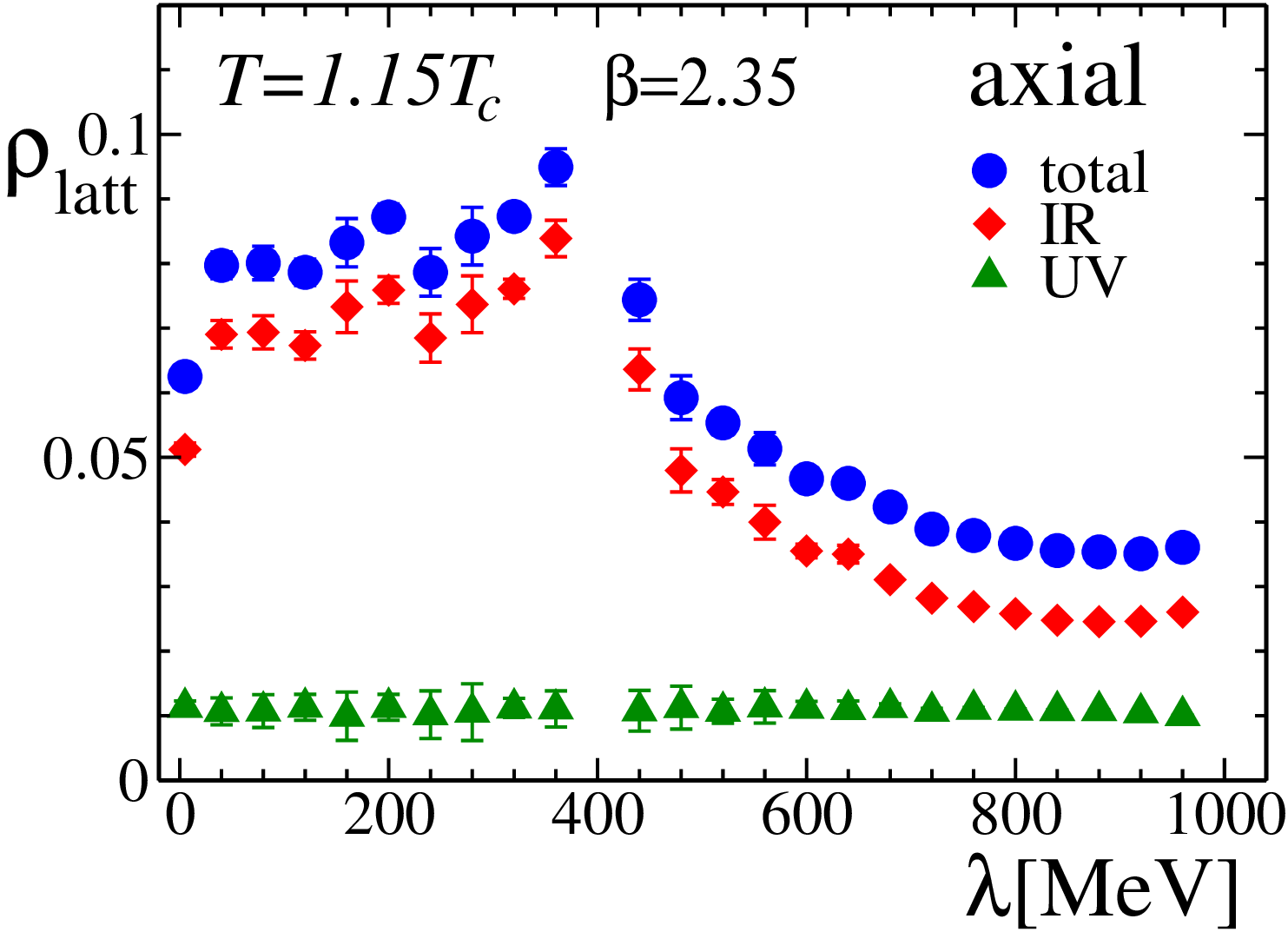} &
\includegraphics[width=52mm,clip=true]{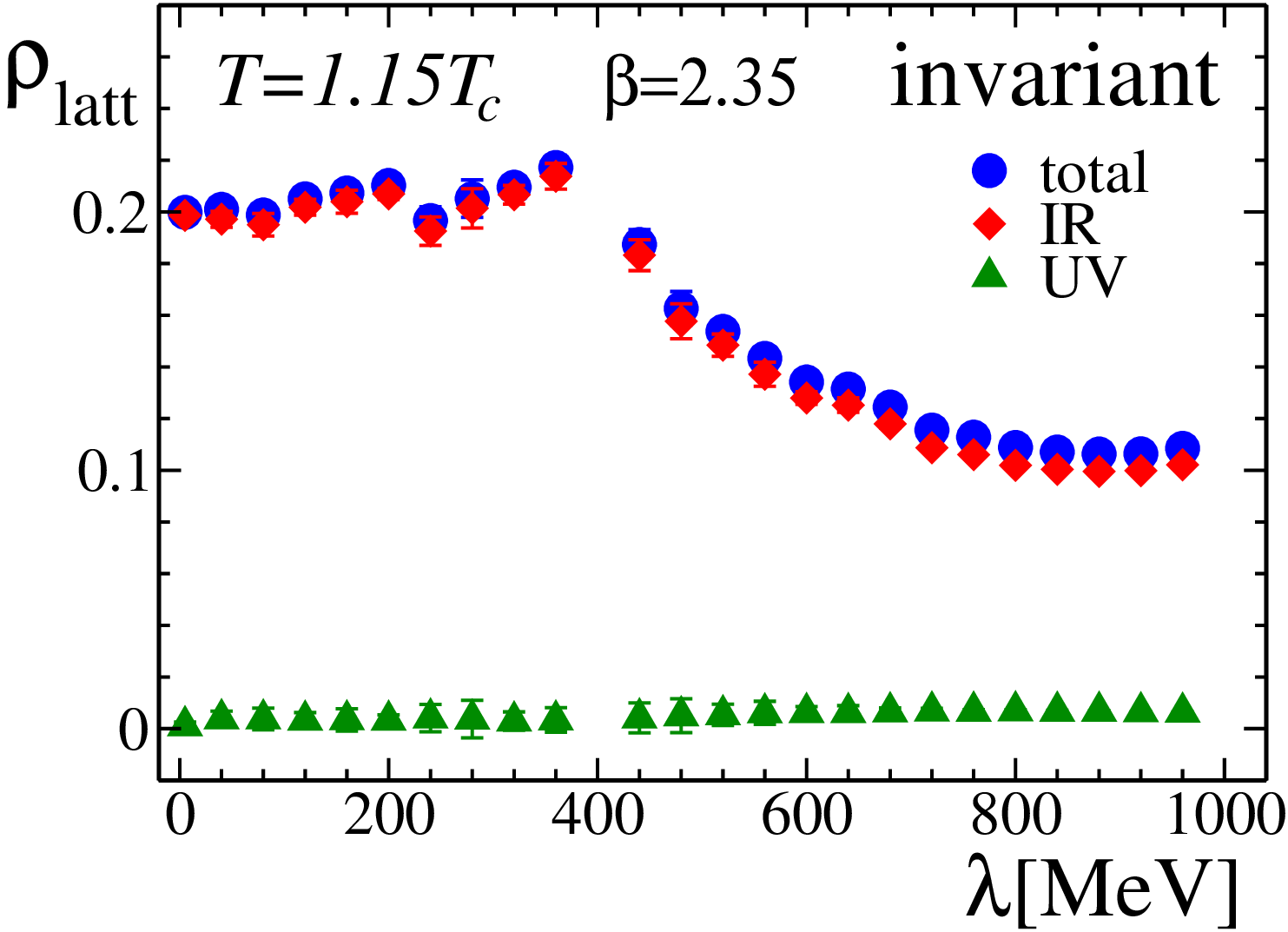} \\
(a) & (b) & (c)
\end{tabular}
\caption{(Color online) The same as in Figure~\eq{fig:densities:lambda:parts:scaled} but for the deconfinement
phase at $T = 1.15T_c$.}
\label{fig:densities:lambda:parts:scaled:temperature}
\end{figure}
In order to estimate how the temperature affects the monopole
densities, we show in
Figure~\eq{fig:densities:lambda:parts:scaled:temperature} the
densities of the monopoles in the IR and UV clusters in the
deconfinement phase at $T = 1.15T_c$. The lattice spacing for the
data shown in Figures~\ref{fig:densities:lambda:parts:scaled} and
\ref{fig:densities:lambda:parts:scaled:temperature} is chosen to
be almost the same. One can clearly see that the basic features of
the cluster structure in the deconfinement phase are similar to
those in the confinement phase except for the quantitative
difference: in the deconfinement phase a bigger
(compared to the confinement phase) fraction of the
monopoles belong to the IR cluster.

The scaling of the individual contributions (total, IR and UV)
towards continuum limit is especially interesting. We found that
the total and the IR parts of the lattice density of the chirally
invariant monopole scale towards the continuum limit proportionally
the lattice spacing $a$ for all non-zero ($\lambda > 0$) modes.
The UV part of the lattice density does not depend on the coupling
$a$ at all, which indicates that this part is a lattice artifact.
Note that the last two observations do not contradict to each
other in the sense of the numerical fitting since the constant UV
part is very small and is almost consistent with zero. For
example, at $\lambda = 235\,\mbox{MeV}$ we have
$\rho^{\mathrm{latt}}_{I,UV} = 0.006(5)$. Therefore the scaling of
the IR and the total parts should numerically be the
indistinguishable from each other, and both should follow the
functional dependence~\eq{eq:extrapolation:Lneq0}.

\begin{figure}[!htb]
\begin{tabular}{cc}
\includegraphics[width=74mm,clip=true]{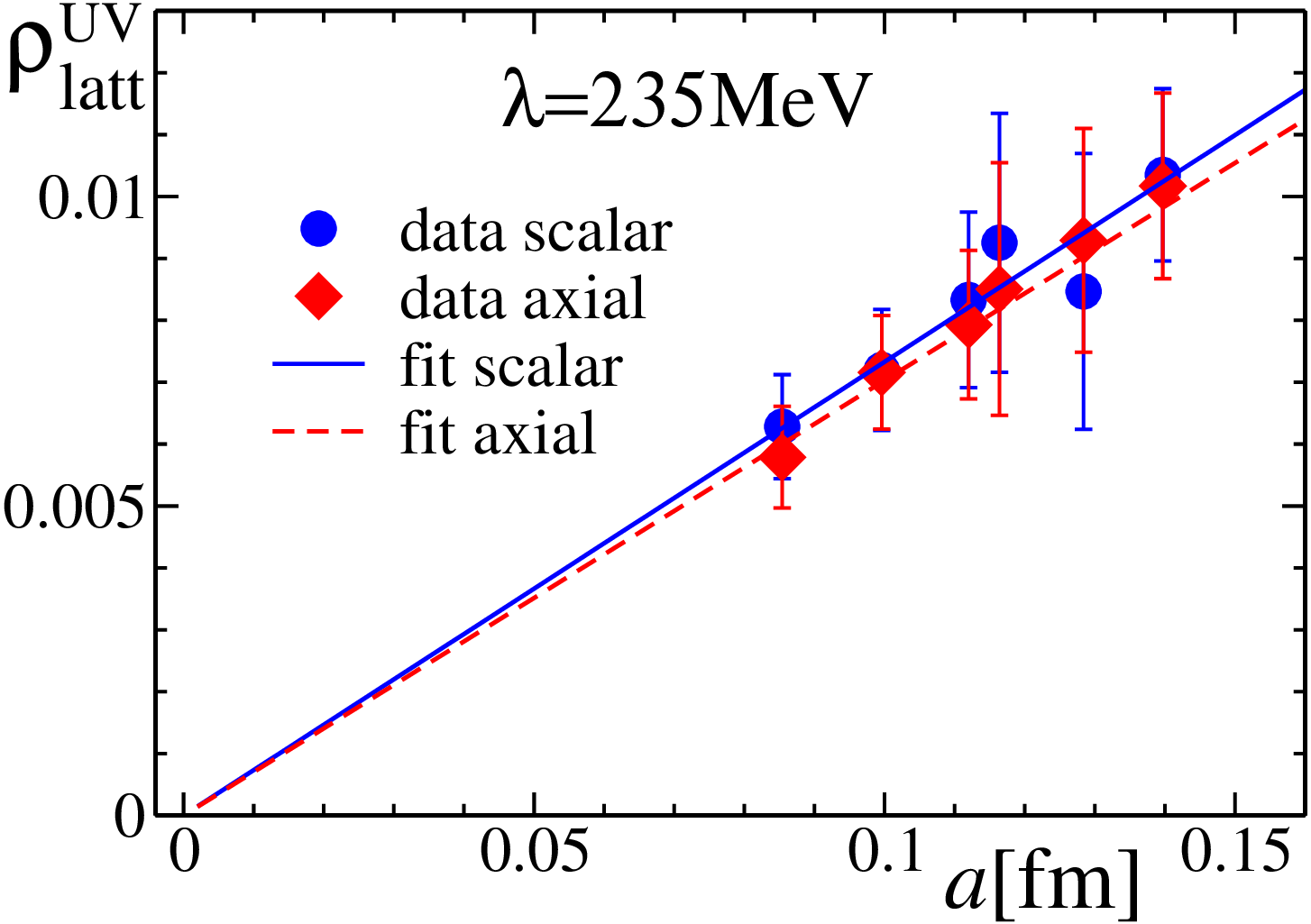}  &
\includegraphics[width=70mm,clip=true]{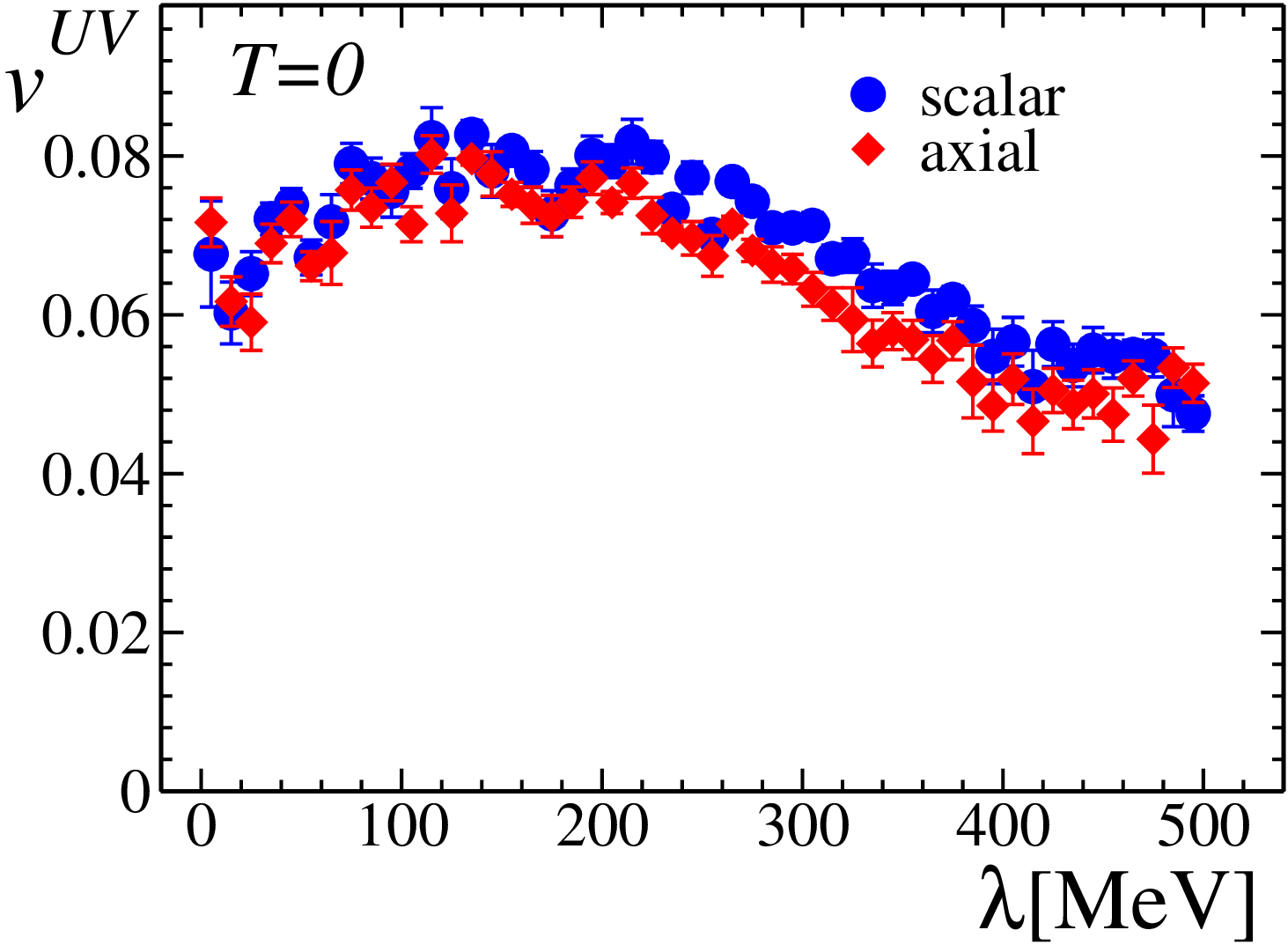} \\
(a) & (b)
\end{tabular}
\caption{(Color online) (a) Example of the extrapolation of the UV fraction of
the scalar and axial monopole densities at the eigenvalue
$\lambda=235\,\mbox{MeV}$. The fit is done by the linear
formula~\eq{eq:UV:fit}. (b) The scaling coefficient $v^{UV}$ {\it
vs.}~$\lambda$ for the UV part of the scalar and axial monopole
densities. The quantity $v^{UV}$ is extrapolated the continuum
limit.}
\label{fig:UV:proportionality}
\end{figure}

As for the scalar and axial monopoles, their IR and UV clusters
also scale differently. The IR part of the scalar and axial
monopole densities (written in the lattice units) scales as $a^2$,
similarly to the total density~\eq{eq:extrapolation:Lneq0}. As for
the UV part, we have found that the scaling of the corresponding
lattice density is proportional to $a$. This is drastically
different from scaling of the total and infrared parts. Unlike the
invariant monopole case, in the case of scalar/axial monopoles
there is a substantial part of the monopoles residing in the UV
clusters. Therefore the different scaling of the UV part cannot in
general be neglected. Unfortunately, the accuracy of our data is
such that the truncated fit~\eq{eq:extrapolation} with two fitting
parameters $v$, $s$ and with $C=\rho=0$ can not give a reliable
estimate of the coefficient $v$. In order to get this coefficient
with a good accuracy, we fit the data for the monopole density in
the ultraviolet clusters using the linear formula \beqn
\rho^{\mathrm{latt}}_{\Gamma,UV} = v^{\mathrm{UV}}_\Gamma a\,,
\qquad \Gamma=S,A\,. \label{eq:UV:fit} \eeqn An example of this
fit is shown in Figure~\ref{fig:UV:proportionality}(a) and the
corresponding coefficient of proportionality $v^{\mathrm{UV}}$ is
plotted in Figure~\ref{fig:UV:proportionality}(b).

Summarizing, the scaling laws for the total, IR and UV monopole densities corresponding to non-zero
Dirac eigenvalues, are
\beqn
\begin{array}{llllll}
\rho^{\mathrm{latt,total}}_{S,A}(a,\lambda) & = & s^{\mathrm{total}}_{S,A}(\lambda) \cdot a^2\,,\qquad
\rho^{\mathrm{latt,total}}_{I}(a,\lambda) & = & v^{\mathrm{total}}_{I}(\lambda) \cdot a\,, \\
\rho^{\mathrm{latt,IR}}_{S,A}(a,\lambda) & = & s^{\mathrm{IR}}_{S,A}(\lambda) \cdot a^2\,,\qquad
\rho^{\mathrm{latt,IR}}_{I}(a,\lambda) & = & v^{\mathrm{IR}}_{I}(\lambda) \cdot a\,,\\
\rho^{\mathrm{latt,UV}}_{S,A}(a,\lambda) & = & v^{\mathrm{UV}}_{S,A}(\lambda) \cdot a^1\,,\qquad
\rho^{\mathrm{latt,UV}}_{I}(a,\lambda) & = & C^{\mathrm{UV}}_{I}(\lambda) \cdot a^0\,,
\end{array}
\qquad \lambda \neq 0\,.
\label{eq:extrapolation:parts}
\eeqn

As for the zero mode, the UV part of the density of the invariant
monopoles is consistent with zero while the IR part coincide with
the total monopole density within error bars. In the case of the
scalar and axial monopoles both total, IR and UV parts satisfy the
quadratic scaling law~\eq{eq:extrapolation:L0}. Moreover,
Eq.~\eq{eq:extrapolation:parts} indicates that in the continuum
limit the most part of the scalar, axial, and invariant monopoles
corresponding to non-zero Dirac eigenvalues ($\lambda \neq 0$)
resides predominantly in UV monopole clusters. This is not the
case for the exact zero mode ($\lambda = 0$) which even in the
continuum limit may possess both IR and UV components of the
densities. So, the exact zero modes and the non-zero modes have,
in fact, different embedded monopole content.

\begin{figure}[!htb]
\begin{tabular}{ccc}
\includegraphics[width=74mm,clip=true]{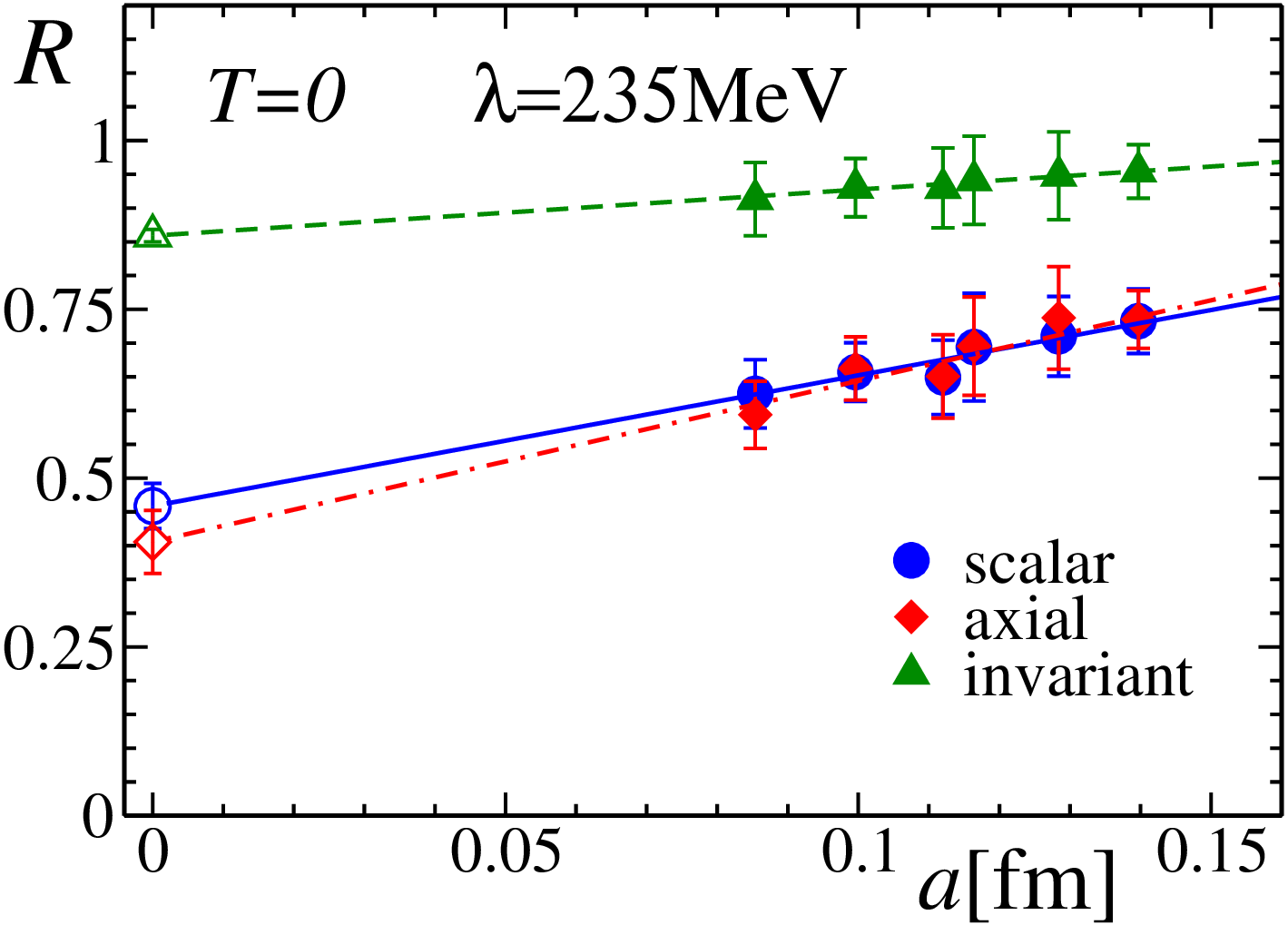} &
\includegraphics[width=70mm,clip=true]{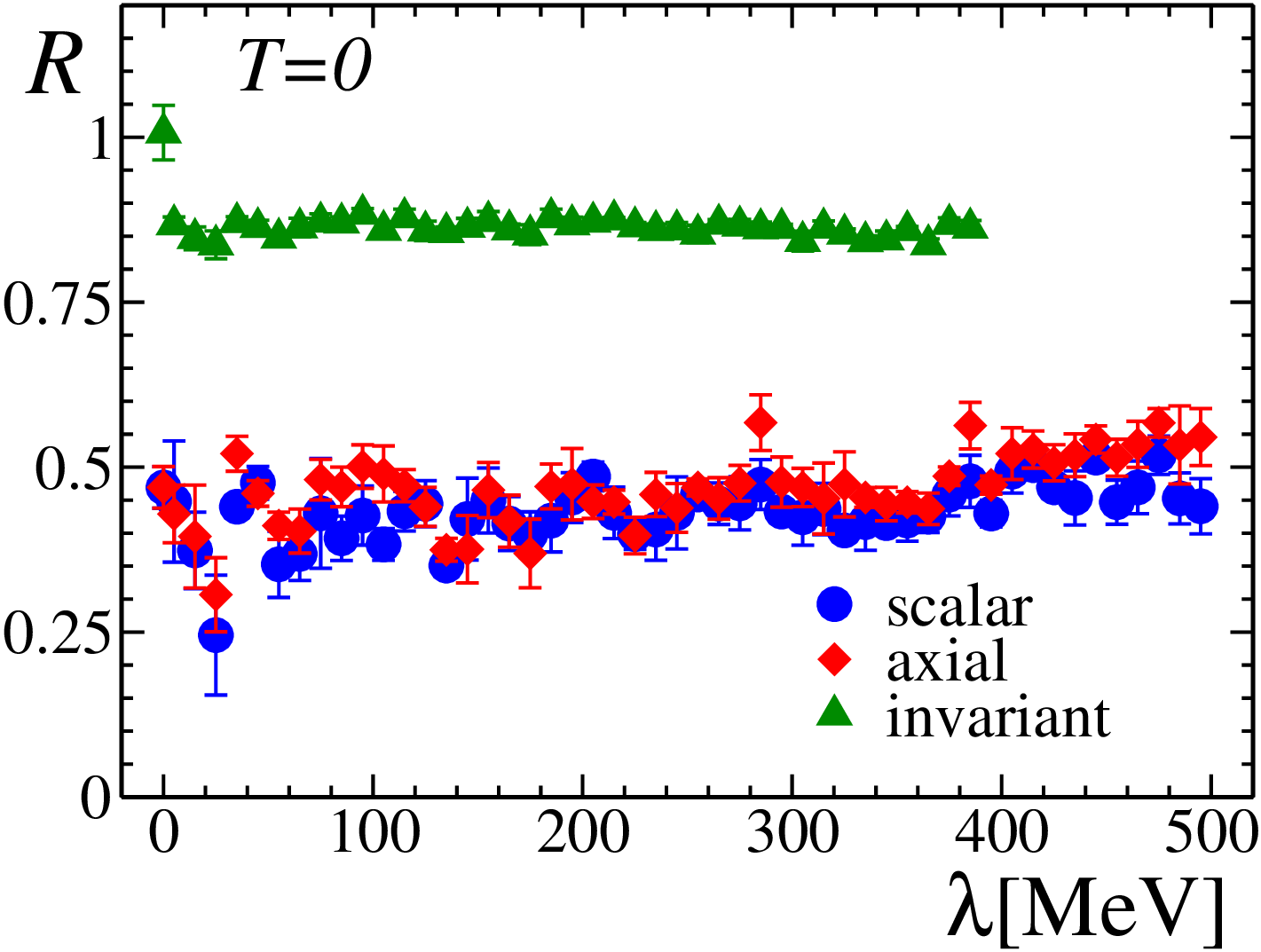} \\
(a) & (b)
\end{tabular}
\caption{(Color online) (a) Example of extrapolation of the infrared-to-total ratios~\eq{R:def} corresponding to
the scalar, axial, and chirally invariant embedded monopoles for the eigenvalue $\lambda=235\,\mbox{MeV}$.
The fit is done by the linear formula~\eq{eq:R}.
(b) The infrared-to-total ratios of the scalar, axial and invariant monopole densities
extrapolated by Eq.~\eq{eq:R} to the continuum limit {\it vs.}~$\lambda$.}
\label{fig:densities:ratio:extrapolation}
\end{figure}

We also study the relative ratio $R$ of the monopole density
in the IR clusters $\rho^\latt_{\mathrm{IR}}$ compared to the total
monopole density $\rho^\latt_{\mathrm{total}}$,
\beqn
R = \frac{\rho^\latt_{\mathrm{IR}}}{\rho^\latt_{\mathrm{total}}}\,.
\label{R:def}
\eeqn
In order to extrapolate this ratio to the continuum limit we use
the linear formula:
\beqn
R^{\mathrm{fit}}(a) = R_0 (1 - K a)\,.
\label{eq:R}
\eeqn
Here $R_0$ and $K$ are the fitting parameters. An example of the extrapolation
and the extrapolated values of $R$ are shown in
Figures~\ref{fig:densities:ratio:extrapolation} (a) and (b), respectively.
Note that according to Eq.~\eq{eq:extrapolation:parts} the formula
for extrapolation~\eq{eq:R} should contain $O(a^2)$ corrections,
which, however, can not be traced out due to limited accuracy
of our data.

\subsection{Discussion on scaling properties}

It is interesting to speculate about the nature of the observed
scaling behavior of the embedded monopole
densities~\eq{eq:extrapolation}, \eq{eq:extrapolation:L0},
\eq{eq:extrapolation:Lneq0}, \eq{eq:UV:fit}, \eq{eq:extrapolation:parts}.
In the zero temperature case the naive physical density,
$\rho^{\mathrm{phys}}(a,\lambda) = a^{-3} \rho^{\mathrm{latt}}(a,\lambda)$,
of the scalar, axial and invariant embedded monopoles diverges in the continuum limit as
\beqn
\rho^{\mathrm{phys}}_{S,A}(a,\lambda) & \sim & a^{-1}\,, \qquad
\rho^{\mathrm{phys}}_{I}(a,\lambda) \sim a^{-2}\,, \quad
\lambda \neq 0\,, \\
\rho^{\mathrm{phys}}_{S,A}(a,0) & \sim & a^{-1}\,, \qquad
\rho^{\mathrm{phys}}_{I}(a,0)  \sim a^{-3}\,, \quad
\lambda = 0\,.
\label{ref:phys:scaling}
\eeqn

Let us suppose for a moment that the general
quantity~\eq{eq:extrapolation} is a density of objects of an
unknown dimension, and that the individual objects are {\it not}
strongly correlated in the lattice
ensembles. Then, on general grounds, the terms in the
formula~\eq{eq:extrapolation} can be interpreted as follows.

If the object is of a pure lattice origin (a lattice artifact),
then its lattice density should  not change with the variation of
the physical scale $a$. Thus, if the first term $C$ is non-zero in
the continuum limit $a \to 0$, then density $\rho$ corresponds to
a purely lattice object. The physics of these objects is
determined by the ultraviolet cutoff $\Lambda_{\mathrm{UV}} \sim
a^{-1}$ only.

Now suppose, that parameter the $C$ vanishes and the leading
behavior of the lattice density is $\rho_{\mathrm{latt}} = v a +
\dots$ as $a\to 0$, where the coefficient $v$ is of the order of
the physical QCD scale $v \sim \Lambda_{\mathrm{QCD}}$. Then the
world-manifolds of the objects are the three-dimensional volumes
distributed in the four-dimensional space time with the physical
density $v$. The corresponding object is a membrane.

The leading scaling behavior in the form $\rho_{\mathrm{latt}} = s
a^2 + \dots$ corresponds to string-like objects, density of which
is given by the quantity $s \sim \Lambda_{\mathrm{QCD}}^2$.
Finally, if one studies pointlike objects ({\it i.e.}, monopoles),
which are not strongly correlated, then the scaling of their
density should be $\rho_{\mathrm{latt}} = \rho a^3 + \dots$, and
the physical density of the objects should be $\rho \sim
\Lambda_{\mathrm{QCD}}^3$.

However, these simple considerations become incorrect if the
objects are (strongly) interacting with each other. As an
illustrative example, it may be useful to consider currents of
Abelian monopoles in an Abelian projection of pure SU(2)
Yang-Mills theory. A general configuration of the gauge fields
typically contain~\cite{ref:clusters,ref:vitaly} two components of
the monopole clusters, one of them is infrared and the other one
is ultraviolet. The physical density of the infrared monopole
currents is finite in the continuum limit $a\to 0$, which means
that the for these monopoles the coefficients $C$, $v$ and $s$ in
Eq.~\eq{eq:extrapolation} are zero. On the other hand, the
ultraviolet component of the Abelian monopole density diverges as
$a^{-1}$ in physical units. One can understand this scaling as a
consequence of a strong correlation between segments of the
monopole loops at the scale of the lattice spacing $a$ because a
typical UV monopole cluster is, in fact, a loop of the length of a
few lattice spacings. One can equivalently say that the monopole
clusters are short-ranged dipoles. The nature of this strong
correlation is of a purely lattice origin as the recent data
shows~\cite{ref:vitaly}. Indeed, it was found in
Ref.~\cite{ref:vitaly} that the density of the UV monopoles
strongly depends on the UV-properties of the gluon action. The
density of the IR monopole clusters are also sensitive to the
lattice details of the gluon action, since the artificial UV
monopoles may randomly connect to the physical IR clusters and be
counted by a lattice algorithm as a part of the physical IR
cluster.

Thus the $a^{-1}$ and $a^{-2}$ scaling of the physical densities
of the scalar/axial and, respectively, invariant embedded
monopoles may be a result of the lattice procedure(s) which may be
sensitive to the UV-scale. On the other hand one can not exclude a
possibility that the embedded QCD monopoles may be strongly
correlated with objects which have surface-like and $3D$
volume-like world trajectories. This property is supported by the
observation~\cite{Gubarev:2005jm} that the low-lying fermion modes
show unusual localization properties being sensitive both to the
physical scale $\Lambda_{\mathrm{QCD}}$ and to the ultraviolet
cut-off, $a^{-1}$. If this suggestion is correct, then the scaling
of the ``slave'' monopoles may manifest the scaling of the
``master'' objects. In this case the scalar/axial and invariant
monopoles should be correlated with (or, as one can also say,
``lie on'') strings and membranes, respectively. In this paper we
are not performing a detailed scaling analysis of the monopole
clusters concentrating on simplest properties only. A review of
the lattice data on many-dimensional vacuum objects in
four-dimensional Yang-Mills theory can be found in
Ref.~\cite{ref:lower:dimensional:Zakharov}.

Finally, let us note that the embedded monopoles are likely not
lattice artifacts because their densities scale as non-zero
powers of the lattice spacing $a$. Still, the effect of the lattice
artifacts on the density may be noticeable since the gauge
fields are not improved at all contrary to the Dirac eigenmodes
which are greatly improved by using the overlap fermions.

\section{Embedded monopoles and fermion spectral density}
\label{sec:spectral}

One of the most essential characteristics of the fermion modes in the gauge theory
is the fermion spectral density $\rho_F$ which is formally defined as the expectation value
\beqn
\rho_F(\lambda) = \frac{1}{V} \langle \sum_{\bar \lambda}\delta(\lambda - \bar \lambda) \rangle\,,
\eeqn
where the sum goes over all Dirac eigenvalues $\bar \lambda = \bar \lambda(A)$ corresponding to
gauge fields configurations $A$ which enter the partition function. The low-lying part of
the fermion spectrum is important for the chiral symmetry breaking due to the Banks-Casher
formula~\cite{ref:Banks:Casher},
\beqn
\langle \bar \psi \psi\rangle = - \lim_{\lambda\to 0} \pi \rho_F(\lambda)\,,
\label{eq:Banks:Casher}
\eeqn
which relates the chiral condensate $\langle \bar \psi \psi\rangle$ with the spectral density.

\begin{figure}[!htb]
\includegraphics[width=90mm,clip=true]{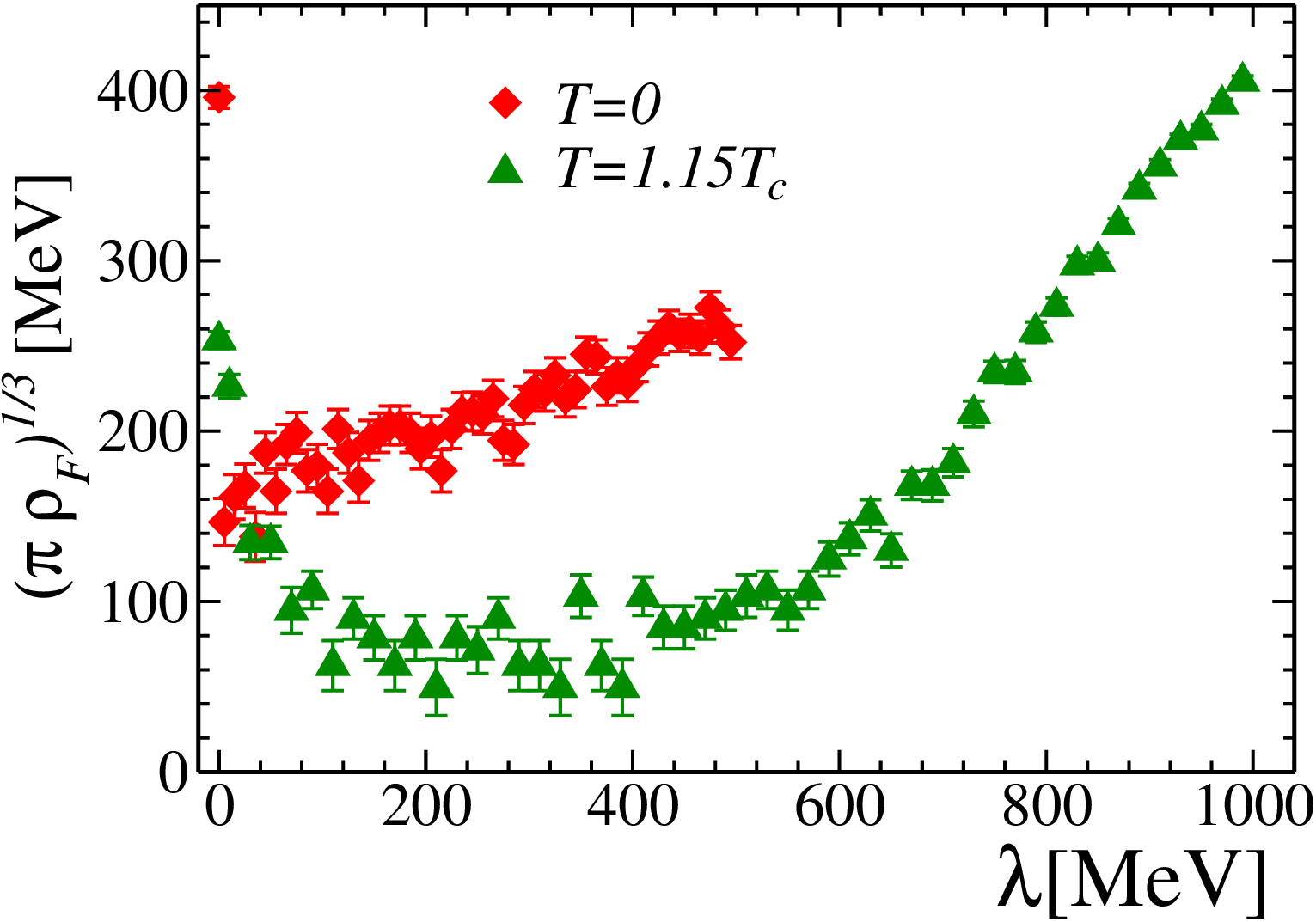}
\caption{(Color online) Spectral fermion density in the confinement
($T=0$, $\beta = 2.3493$) and in the deconfinement ($T=1.15 T_c$,
$\beta = 2.35$) phases.}
\label{fig:spectrum}
\end{figure}

Examples of the spectral density $\rho_F$ as functions of
$\lambda$ are shown in Fig.~\ref{fig:spectrum} for the confinement
($T=0$, $\beta = 2.3493$) and for the deconfinement ($T=1.15 T_c$,
$\beta = 2.35$) phases. In the zero temperature (confinement) case
the spectrum is a gradually increasing function of the eigenvalue
$\lambda$, and the low-energy part of the spectrum has a finite
$\lambda \to 0$ limit. In the deconfinement phase the low-energy
part of the spectrum is suppressed compared to the confinement
phase but is, however, non-zero.
This feature as well as the peak of the spectral density at
$\lambda = 0$ may be an artifact related to an effect of the finite volume. We expect
that in the limit of an infinite volume the spectrum above the deconfinement
temperature should vanish below some critical value
$\lambda_c(T)$. This property could imply vanishing of the chiral
condensate~\eq{eq:Banks:Casher} in the deconfinement phase,
$\langle \bar \psi \psi\rangle(T > T_c) = 0$.
In our case the critical value is $\lambda_c(T=1.15\,T_c) \approx 400\,\mbox{MeV}$.

The embedded QCD monopoles are suggested~\cite{ref:embedded:QCD}
to be agents the chiral symmetry restoration, since in the cores
of these monopoles the chiral invariance should be unbroken.
According to the proposed scenario, one can expect that at low
Dirac eigenvalues -- which are relevant to the chiral symmetry
breaking due to the Banks-Casher relation~\eq{eq:Banks:Casher} --
the density of the embedded monopoles should be high in the
chirally invariant (high temperature) phase and the density should
be relatively low in the chirally broken (low temperature) phase.
Thus suggestion implies, in turn, that the density of the embedded
monopoles should be anti-correlated with the fermion spectral
function: the lower value of the spectral function the higher
monopole density is expected to be. In particular, in the high
temperature phase the vanishing spectral function at $\lambda <
\lambda_c$ implies the high density of the quark monopoles for
$\lambda < \lambda_c$, and vice versa.

The anti-correlation of the quark monopole density and the fermion
spectral function is indeed observed in deconfinement phase as it
is shown in Figure~\ref{fig:densities:lambda:coarsed}(d). Indeed,
as one can see from comparison of
Figure~\ref{fig:densities:lambda:coarsed}(d) and the $T=1.15 T_c$
spectral function shown in Figure~\ref{fig:spectrum}, at high
$\lambda$ the fermion spectral function is high corresponding to
low embedded monopole density, while at low $\lambda$ the fermion
spectral density is suppressed in accordance with the observed
large valued of the monopole density.

In the confinement case the qualitative relation between the
spectral density and the embedded monopole density is true as well
according to Figures~\ref{fig:densities:lambda:coarsed}(a),(b),(c) and
Figure~\ref{fig:spectrum}: the fermion spectrum is an increasing
function of the Dirac eigenvalue $\lambda$ while the monopole
densities are generally decreasing function of $\lambda$.

\section{Excess of gluon action on quark monopoles}
\label{sec:energy}

If the embedded QCD monopoles are physical objects then we would
expect that these objects are locally correlated with the
action density and, presumably, with the topological charge.
Following Refs.~\cite{ref:Bakker,ref:Monopole:Structure} we
calculate numerically the excess of the SU(2) gauge action at the position of
the embedded monopole currents $j_{x,\mu}$,
\beqn
f_S = \frac{\langle |j_{x,\mu}| \, S_{c_{x,\mu}}\rangle - 6 \langle |j_{x,\mu}|\rangle
\langle S_P\rangle}{\langle |j_{x,\mu}| \rangle}\,,\,,
\label{eq:monopole:action:correlations}
\eeqn
where $S_c$ is the sum over the elementary plaquette actions,
\beqn
S_c = \sum_{P \in \partial c} S_P\,, \qquad S_P = 1 - \frac{1}{2} \, {\mathrm{Tr}} U_P\,, \nonumber
\eeqn
belonging to the six faces $P$ of the cubes $c \equiv c_{x,\mu}$ with non-zero monopole
charge, $j_{x,\mu} \neq 0$. Here $U_P$ is the $SU(2)$ plaquette constructed from the lattice
links $U_{x\mu}$ in the standard way
$U_{P_{x,\mu\nu}} = U_{x\mu} U_{x+\hat\mu,\nu} U^\dagger_{x+\hat\nu,\mu} U^\dagger_{x\nu}$.
The second term in Eq.~\eq{eq:monopole:action:correlations}
subtracts the vacuum average of the action from the value of the gluon action at the monopole.

Equation~\eq{eq:monopole:action:correlations} can be understood as
the (average) excess of the Yang-Mills action calculated at the
(average) distance $r=a/2$ from the center of the monopole. In
fact, in any given lattice configuration the position of the
monopole center can not be determined exactly within the lattice
the cube possessing a nonzero monopole charge. However on average
the monopole center is located as the cube center which resides at
the distance $a/2$ from any face (plaquette) of the cube. It is
worth noticing that Eq.~\eq{eq:monopole:action:correlations}
defines the excess of the chromomagnetic part of the action since
by construction the action around the monopole is calculated on
the plaquettes $P$ perpendicular to the corresponding link
$l=\{x,\mu\}$ of the monopole trajectory.

In the naive continuum limit the elementary plaquette, say
$P_{x,12}$, is expanded in powers of the lattice spacing as
follows
\beqn
S_{P_{x,12}}(a) = a^4\, \frac{g^2}{8} \, {\bigl[F^a_{12}(x)\bigr]}^2 + O(a^6)
\eeqn
where $g$ is the (bare) coupling and $U_{x\mu} = e^{i a g A_\mu(x+\hat\mu/2)}$. Thus
the excess of the action~\eq{eq:monopole:action:correlations}
can be written as (we omit $O(a^6)$ corrections starting from here)
\beqn
f_S(a) & = & a^4\, \frac{\pi}{2}\, \mathcal{B}_{\mathrm{mon}}(a/2)\,,
\label{eq:fS:B}
\eeqn
where
\beqn
\mathcal{B}_{\mathrm{mon}}(a/2) & = &
{\bigl\langle \alpha_s\, {({\mathbf{B}}^c)}^2 \,\bigr\rangle}_{\mathrm{mon}}{\Bigr{|}}_{r=a/2}
- {\bigl\langle \alpha_s \, {({\mathbf{B}}^c)}^2 \,\bigr\rangle}\,,
\eeqn
is the excess of the chromomagnetic condensate at the distance $r=a/2$ from the quark monopole
and $\alpha_s = g^2/(4 \pi)$. The chromomagnetic field at the segment of the monopole current $j_\nu$
is defined as $B^c_{\mu}(j_\nu) =
- \epsilon_{\mu\nu\alpha\beta} F^c_{\alpha\beta}/2$. This definition
reduces to the standard one for a static ($\nu=4$) monopole: $B^c_{i} = \epsilon_{ijk} F^c_{jk}/2$ with $i,j,k=1,2,3$.
In the Euclidean space-time at zero temperature the chromomagnetic and the standard gluon condensates are
related as ${\bigl\langle \alpha_s \, {({\mathbf{B}}^c)}^2 \bigr\rangle} =
{\bigl\langle \alpha_s \, {(F^c_{\mu\nu})}^2 \,\bigr\rangle}/2$.

Before proceeding with analysis of the numerical data it would be
instructing to discuss the expected behavior of the chromomagnetic
fields inside the embedded monopoles at least on a qualitative
level. In Ref.~\cite{ref:embedded:QCD} the embedded QCD monopole
is associated with the Nambu monopole in the Electroweak model.
The Nambu monopole is essentially the
't~Hooft-Polyakov~\cite{ref:thooft,ref:polyakov} (HP) monopole
configuration embedded into the EW model. Therefore one {\it
naively} can expect that the behavior of the chromomagnetic fields
inside the embedded monopole in QCD is qualitatively similar to
that of the HP monopole in the Georgi-Glashow model.

As an illustrative example let us consider the
Bogomol'ny-Prassad-Sommerfeld (BPS) limit~\cite{ref:BPS} of
the Georgi-Glashow model,
\beqn
\cL_{\mathrm{GG}} = \frac{1}{4} (F^a_{\mu\nu})^2 + \frac{1}{2}  \left(D^{\mathrm{ad}}_\mu \Phi\right)^2
+ \frac{\lambda}{4} \left((\Phi^a)^2 - \eta^2\right)^2\,.
\label{eq:GG}
\eeqn
This model describes the dynamics of the $SU(2)$ gauge field $A^a_\mu$
interacting with the triplet (adjoint) Higgs field $\Phi^a$, $a=1,2,3$. The
adjoint covariant derivative is given in Eq.~\eq{eq:D:adjoint}.
The scalar coupling $\lambda$ describes self-interaction of the Higgs field.
The condensate of the Higgs field is $|\langle \vec \Phi \rangle| = \eta$.
The masses of the gauge and Higgs fields in the Georgi-Glashow
model are, respectively, $m_A = g \, v$ and $m_\Phi = \sqrt{2
\lambda} \, \eta$.

The BPS limit is defined by the condition $\lambda = 0$, which sets
the mass of the Higgs particle to zero, $m_\Phi=0$. Due to the absence of
the quartic Higgs self-interaction the classical static
't~Hooft-Polyakov solution can be found explicitly~\cite{ref:BPS}:
\beqn
\Phi^a =\frac{r^a}{g\, r^2} H(\eta g\, r)\,, \qquad
A^a_i = \epsilon_{aij} \frac{r^j}{g\, r^2} [1 - K(\eta g\, r)]\,, \qquad
A^a_0 = 0\,,
\label{eq:Phi}
\eeqn
where
\beqn
K(\xi) = \frac{\xi}{\sinh \xi}\,,\qquad H(\xi) = \xi \coth\xi - 1\,,
\label{eq:K}
\eeqn
The chromoelectric field of the HP monopole is zero, $F^a_{0i} \equiv 0$,
and the chromomagnetic field $B^c_{i} = \epsilon_{ijk} F^c_{jk}/2$, is:
\beqn
B^c_i = \frac{r^c \, r^i}{g\, r^4} \Bigl( 1 - K^2 - H K\Bigr) + \frac{\delta^{ci}}{g\, r^2} \, H K\,,
\eeqn
The corresponding ``condensate'' of the chromomagnetic field tends to a finite value in the monopole
center, $r\to 0$:
\beqn
\mathcal{B}_{\mathrm{HP}}(r) & \equiv & \frac{g^2}{4\pi} {[{\mathbf{B}}^c(r)]}^2 =
\frac{1}{4\pi r^4}\, \Bigl[\Bigl(1 - K^2(\eta g\, r)\Bigr)^2
+ 2 H^2(\eta g\, r) \, K^2(\eta g\, r) \Bigr]\,,
\label{eq:B}\\
\mathcal{B}_{\mathrm{HP}}(r) & = &
\mathcal{B}_{\mathrm{HP}}^{(0)} \cdot \Bigl[1 + O((\eta g r)^2)\Bigr]\,, \qquad
\mathcal{B}_{\mathrm{HP}}^{(0)} = \frac{\eta^4 g^4 }{12\, \pi} \qquad
\mbox{for} \quad r \to 0\,. \label{eq:B0}
\eeqn

Let us take for a moment this illustrative example of the hedgehog
configuration seriously. The QCD counterpart of the field $\Phi^a$
is an octet quark-antiquark composite field $\xi^a$.
We do {\it not} expect the presence of the octet condensates
$\langle \xi^a\rangle$ in vacuum of the Yang-Mills theory because
such a condensate must inevitably break the color
symmetry (see, however, a discussion in Ref.~\cite{ref:wetterich}). On the other hand one may
expect~\cite{ref:embedded:QCD} that the non-perturbative
color-invariant four-quark condensates~\cite{ref:four:quark} of
the form $\langle (\xi^a)^2 \rangle$ should stabilize the
hedgehog-like configurations made of the composite ``Higgs'' field
$\xi^a$ in the confinement phase.

One can expect that the value of the condensate $\eta$ in
Eqs.~(\ref{eq:B}), (\ref{eq:B0}) should be of the
order of a typical
dimensional quantity describing the chiral condensate, $\eta \sim
|\langle \bar \psi \psi\rangle|^{1/3} \sim 0.2 \dots
0.3\,\mbox{GeV}$, which, in turn, is of the order of the QCD scale
parameter $\Lambda_{\mathrm{QCD}}$.
One can think of $\eta$ as of the condensate outside the core.
The gauge coupling $g$ can
be associated with the QCD running coupling.
Then Eq.~(\ref{eq:B0})
predicts that the chromomagnetic field inside the embedded
monopole should be ``soft'',
$$
\mathcal{B}_{\mathrm{mon}}(r) \sim g^4(r \Lambda_{\mathrm{QCD}}) \cdot
\Lambda^4_{\mathrm{QCD}}\,.
$$
In particular, this relation implies the absence of the hard ultraviolet divergences
of the energy density inside the monopole cores contrary to the case of the
Abelian (Dirac) monopoles~\cite{ref:fine:tuning,ref:Monopole:Structure}.

We perform the fit of the numerical data for the correlation
function~\eq{eq:monopole:action:correlations} by
\beqn
f^{\mathrm{fit,HP}}_S = 2\, \Bigl[\Bigl(1 - K^2(x)\Bigr)^2 + 2 K^2(x) H^2(x) \Bigr]\,,\quad
x = g(\Lambda_{\mathrm{HP}} \, a/2)\cdot \eta  a/2\,,
\label{eq:fHP}
\eeqn
which can be obtained from Eqs.~(\ref{eq:fS:B}), (\ref{eq:B}) by identification $r=a/2$. The function $K$
is given in Eq.\eq{eq:K}, and instead of the Georgi-Glashow coupling $g$
we take the one-loop running coupling constant of the SU(2) Yang--Mills theory,
\beqn
g^{-2}(\Lambda a) = \frac{11}{12 \pi^2} \log \frac{1}{\Lambda a}\,.
\label{eq:g:run}
\eeqn
The fitting parameters are the HP scale parameter $\Lambda_{\mathrm{HP}}$ and the ``condensate'' parameter $\eta$.

\begin{figure}[!htb]
\begin{tabular}{cc}
\includegraphics[width=70mm,clip=true]{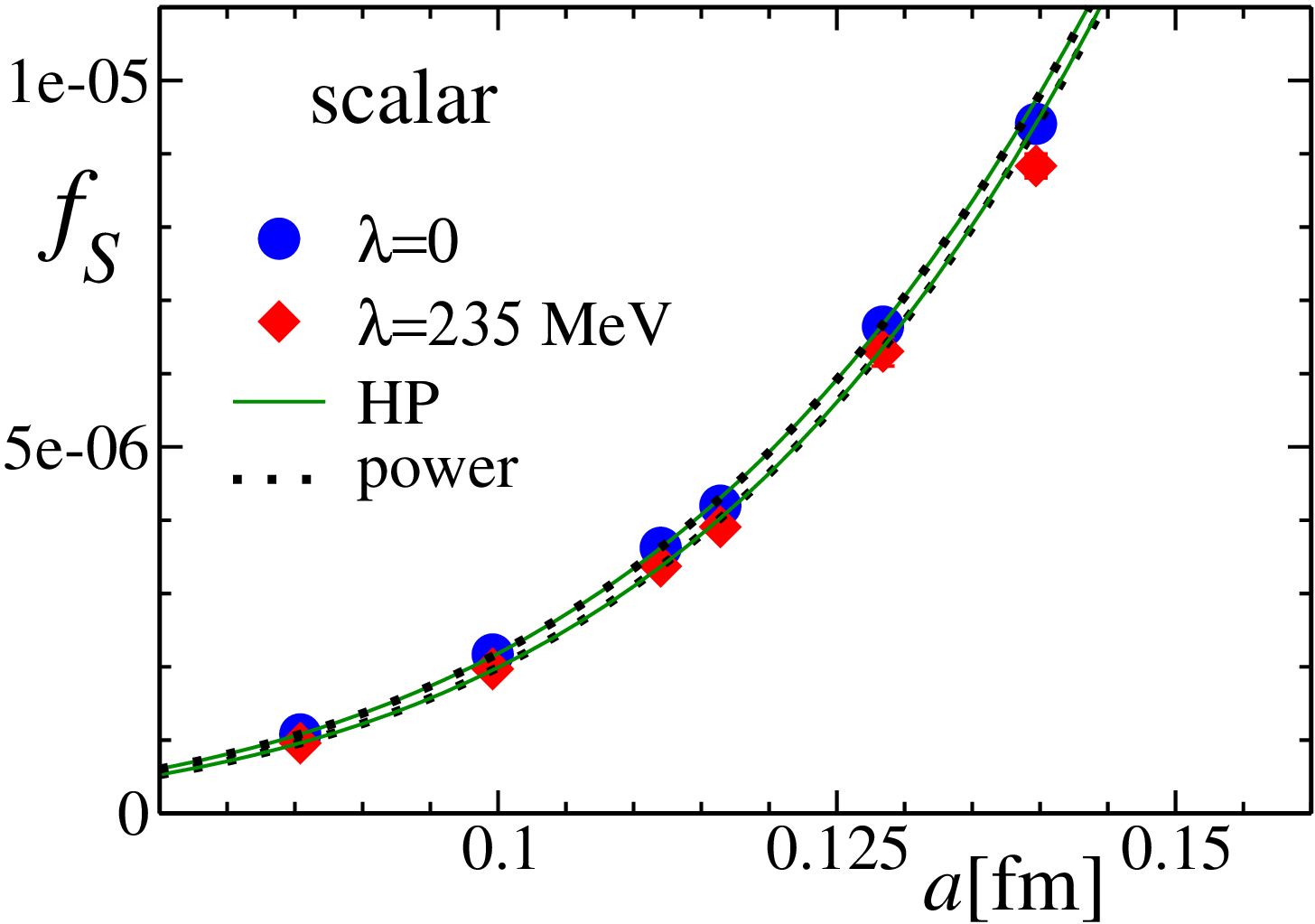} & \hspace{5mm}
\includegraphics[width=70mm,clip=true]{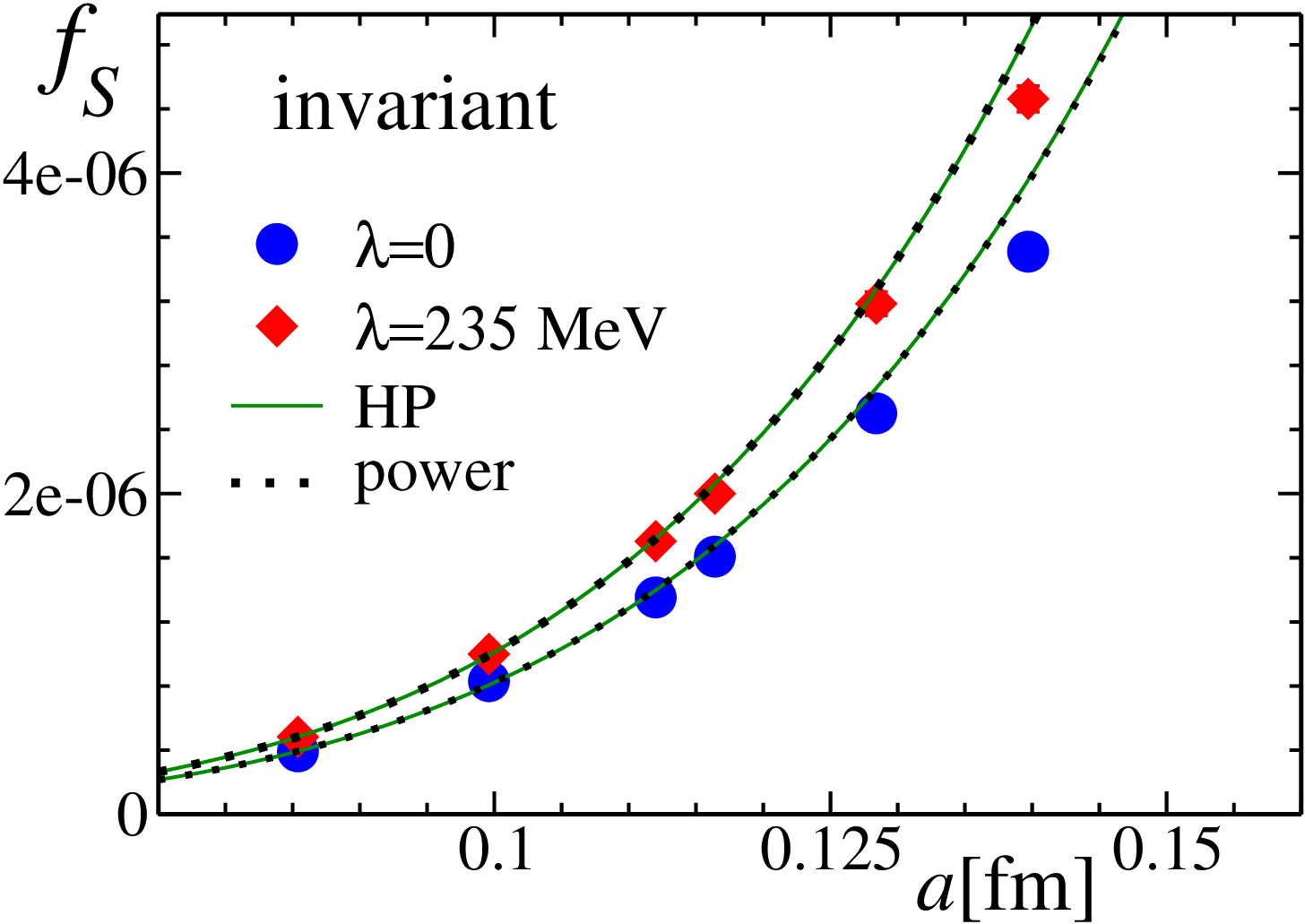} \\
(a) &  \hspace{5mm} (b)
\end{tabular}
\caption{(Color online) The excess of the chromomagnetic
action~\eq{eq:monopole:action:correlations} at (a) the scalar and
(b) the chirally invariant monopoles at $\lambda=0$ and
$\lambda=235\,\mbox{MeV}$. The solid and the dotted lines refer to
the fits Eq.~(\ref{eq:fHP}) and \eq{eq:power:fit}, respectively.}
\label{fig:fitting:energy}
\end{figure}

Examples of the correlation function $f_S$ at $\lambda=0$ and
$\lambda=235\,\mbox{MeV}$ are shown in
Figures~\ref{fig:fitting:energy}(a) and (b) for scalar and
chirally invariant monopoles, respectively. The excess of the
chromomagnetic energy for the scalar and axial monopoles coincide
with each other within error bars. The examples of the fits of the
excess energy by the function~\eq{eq:fHP} are shown by the solid
lines. One can see that the fitting
formula~(\ref{eq:K}), (\ref{eq:fHP}), (\ref{eq:g:run}) -- which is
resembling the 't~Hooft-Polyakov monopole configuration in the
Bogomol'ny limit~(\ref{eq:K}), (\ref{eq:Phi}), (\ref{eq:K}) -- emulates
our numerical data relatively well. The fitting is better near the
continuum limit, $a \to 0$. However at relatively large distances,
$a \gtrsim 0.13\,\mbox{fm}$, the data and the best fit function
show some noticeable difference which makes $\chi^2/d.o.f \sim 3
\dots 5$ for these fits.

The best fit parameters $\Lambda_{\mathrm{HP}}$ and $\eta$
obtained in our fits by the function~\eq{eq:fHP} come with
relatively large errors. For example, for scalar monopoles at
$\lambda=0$ we have $\Lambda_{\mathrm{HP}} = 17(26)\,\mbox{MeV}$
and $\eta=136(14)\,\mbox{MeV}$, while at $\lambda=235\,\mbox{MeV}$
we get $\Lambda_{\mathrm{HP}} = 28(31)\,\mbox{MeV}$ and
$\eta=98(12)\,\mbox{MeV}$. The corresponding numbers for the
invariant monopoles are: in the $\lambda=0$ case we obtain
$\Lambda_{\mathrm{HP}} = 30(36)\,\mbox{MeV}$ and
$\eta=72(17)\,\mbox{MeV}$, while for $\lambda=235\,\mbox{MeV}$ we
get $\Lambda_{\mathrm{HP}} = 24(20)\,\mbox{MeV}$ and
$\eta=75(10)\,\mbox{MeV}$. Thus the values of the parameter
$\Lambda_{\mathrm{HP}}$ can not be defined well due to quite weak
dependence of the logarithm function~\eq{eq:g:run} on the value of
its argument. Moreover, the value of $\Lambda_{\mathrm{HP}}$ is
very small what makes $g \sim 1$ in all our fits. On the other
hand the fit quantitatively confirms that the values for the
``condensate'' $\eta$ is of the order of the QCD scale, $\eta \sim
\Lambda_{\mathrm{QCD}}$. The effective size of the ``HP monopole''
core, $1/m_A = 1/(g \eta)$, can be estimated to be of the order of
1~fm for all values of $\lambda$. Since this value is
unrealistically large we conclude that the fact that the HP
fitting formula~\eq{eq:fHP} works relatively well is only a
manifestation of the ``softness'' of the gluonic action inside the
core of the monopole.
Quantum corrections to the HP monopole fields may be important.

In order to get prescription-independent result on the scaling of the
average action excess we fit the available data by the power-like
fitting function
\beqn
f^{\mathrm{fit,power}}_S(a) = {\left(a/h\right)}^{4 + \delta}\,,
\label{eq:power:fit}
\eeqn
where the scale $h$ and the ``anomalous'' exponent
$\delta$ are the fitting parameters. If the action density is
independent on the distance to the monopole center (or, in other
words, if the core of the embedded monopoles is structure-less)
then we would naively expect the vanishing anomalous exponent,
$\delta = 0$. If $\delta \neq 0$ then one can expect some
structure of the monopole core.

The example of the fits~\eq{eq:power:fit} at
$\lambda=0$ and $\lambda=235\,\mbox{MeV}$ are shown in
Figures~\ref{fig:fitting:energy}(a) and (b) by the dotted lines.
As one can see from these Figures, the HP monopole
fit~\eq{eq:fHP} and the simple power fit~(\ref{eq:power:fit}) are
practically indistinguishable from each other. Note that both fits
are two-parametric ones.

\begin{figure}[!htb]
\begin{tabular}{cc}
\includegraphics[width=70mm,clip=true]{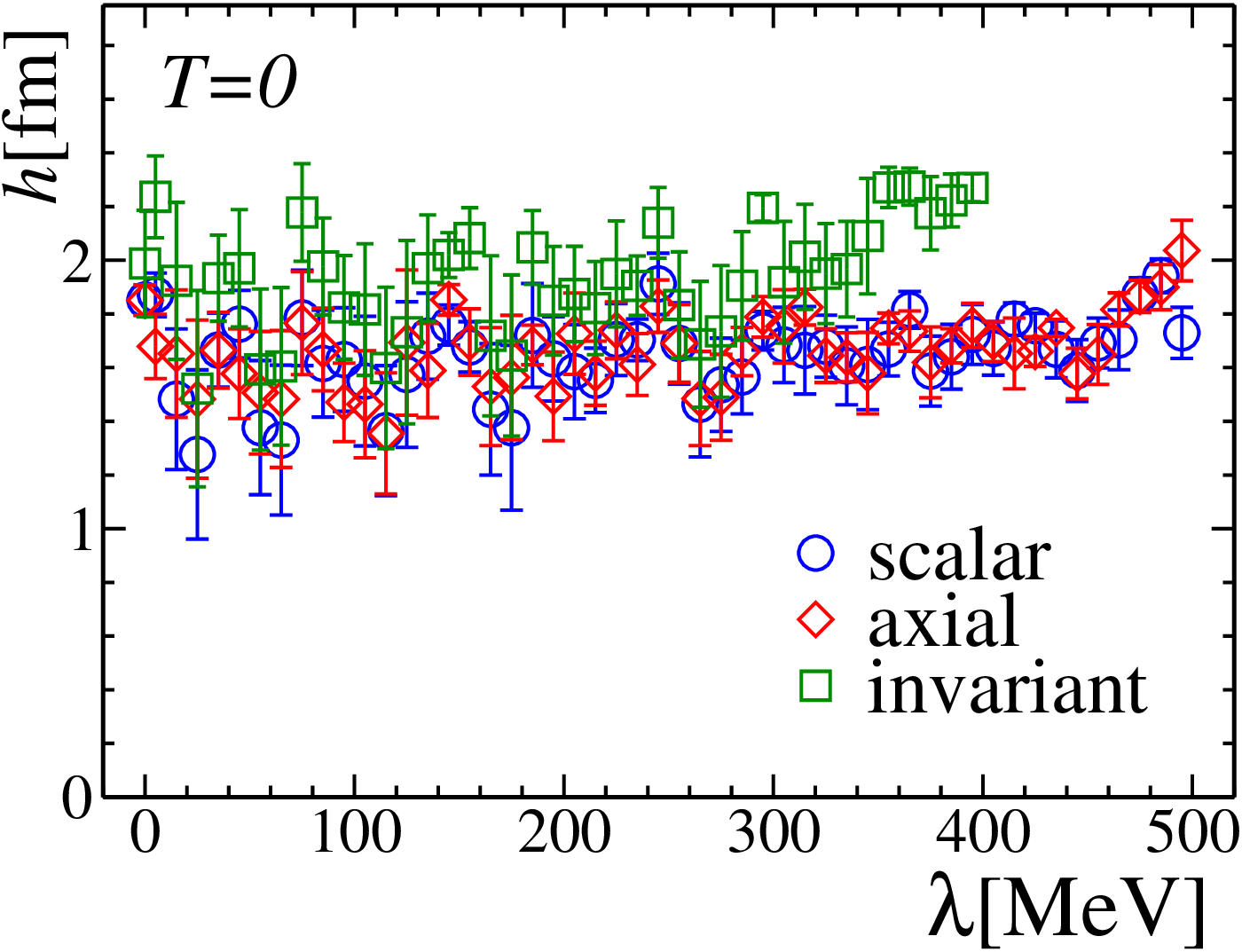} & \hspace{5mm}
\includegraphics[width=70mm,clip=true]{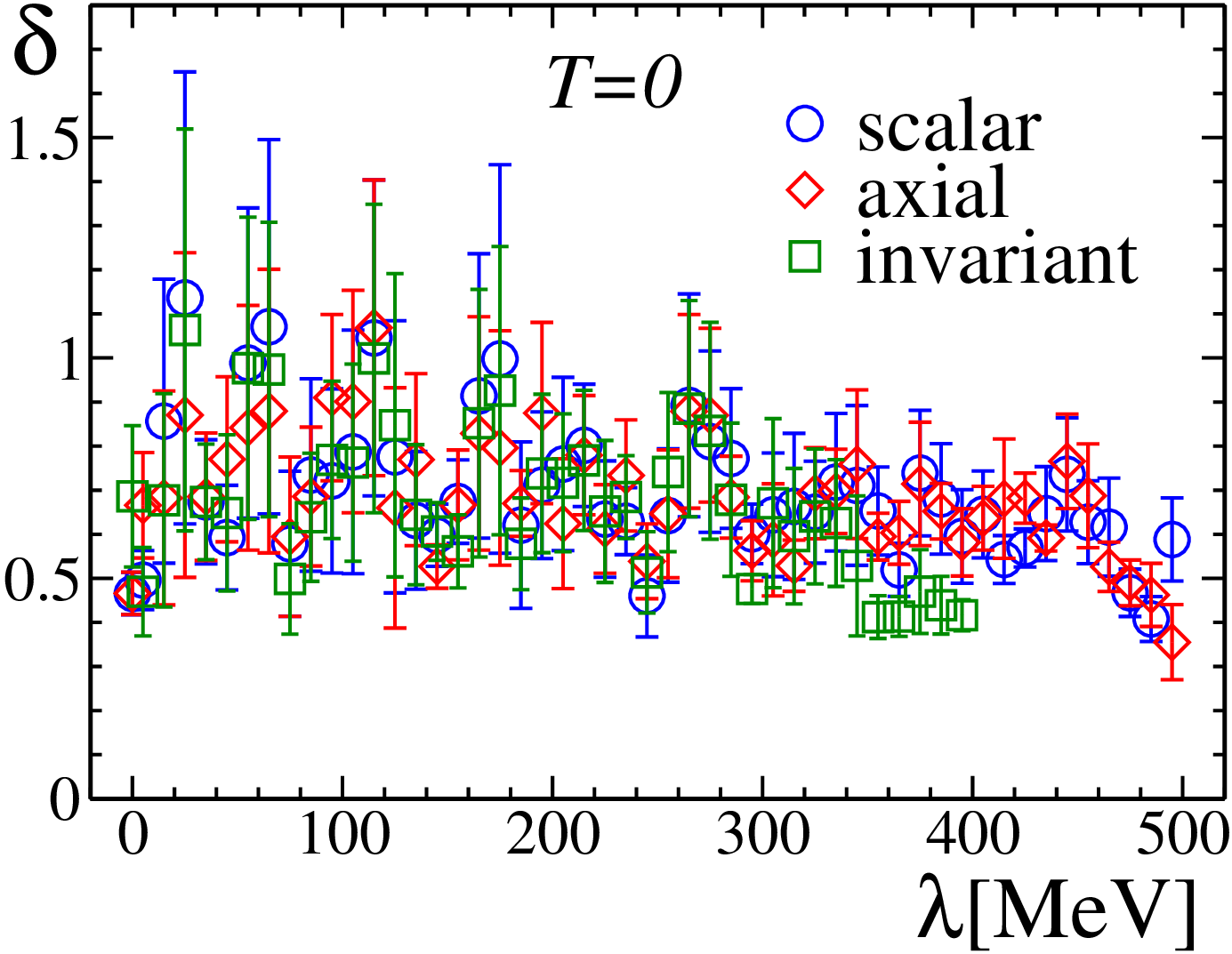} \\
(a) &  \hspace{5mm} (b)
\end{tabular}
\caption{(Color online) The best fit parameters (a) $h$ and (b) $\delta$ of the fit function~\eq{eq:power:fit}
{\it vs.} the Dirac eigenvalue $\lambda$ for the scalar, axial and chirally invariant embedded monopoles.}
\label{fig:power:coefficients}
\end{figure}
The best fit parameters for the power function~\eq{eq:power:fit}
are shown in Figures~\ref{fig:power:coefficients}(a) and (b) as
functions of $\lambda$ for all studied types of the quark
monopoles. We find that both $h$ and $\delta$ parameters are
almost independent on the Dirac eigenvalue $\lambda$. Moreover,
these parameters for different types of the quark monopole are
quite close to each other. This result suggests that the gluonic
structure of the embedded monopoles seems to be independent on the
value of $\lambda$. The typical values of the fitting parameters
are concentrating around central values $h_{S,A} \approx 1.6\,
\mbox{fm}$, $h_{I} \approx 2\, \mbox{fm}$ and $\delta_{S,A,I}
\approx 0.7$ with, however, relatively large error bars.

\begin{figure}[!htb]
\begin{tabular}{cc}
\includegraphics[width=74mm,clip=true]{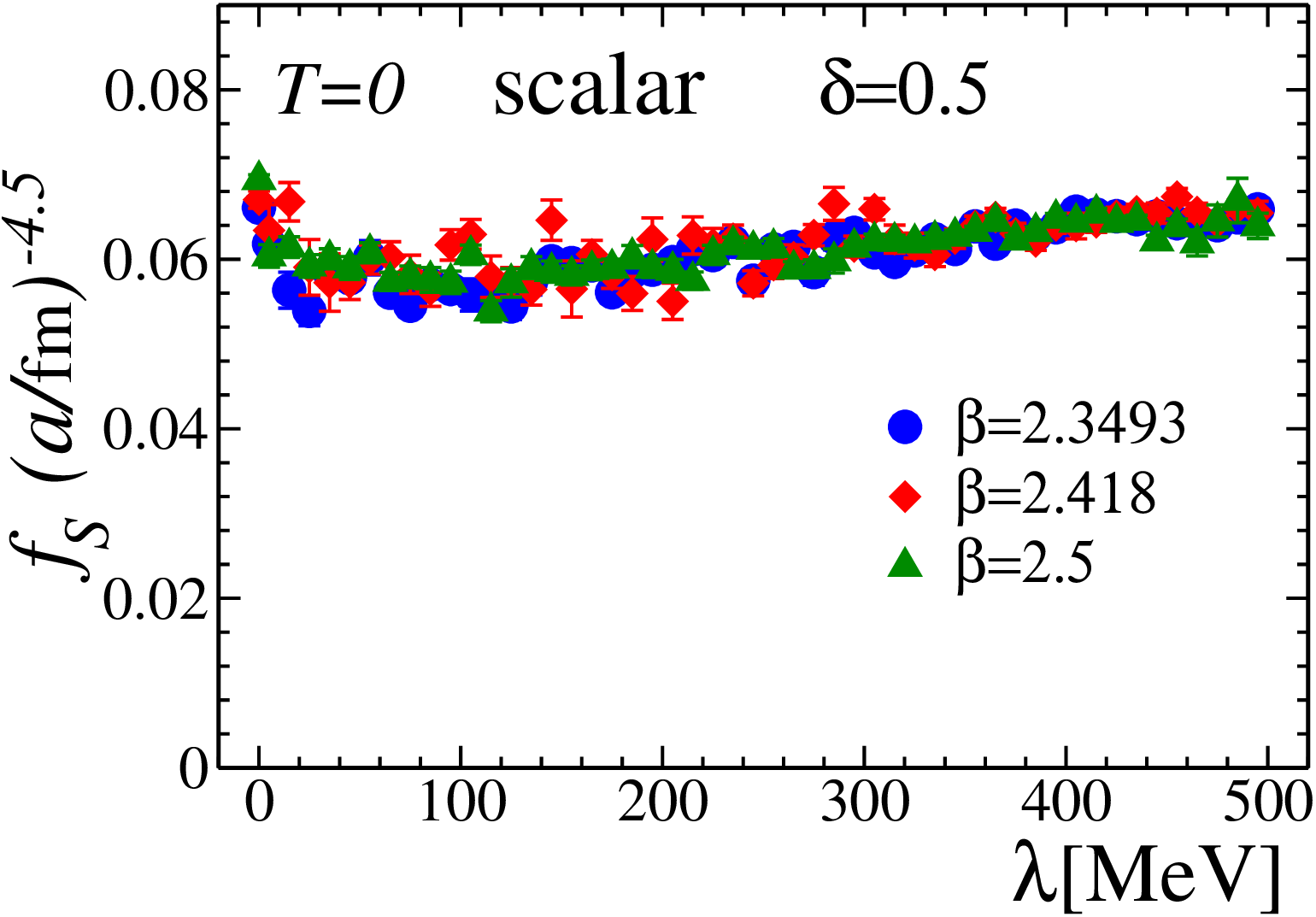} & \hspace{5mm}
\includegraphics[width=74mm,clip=true]{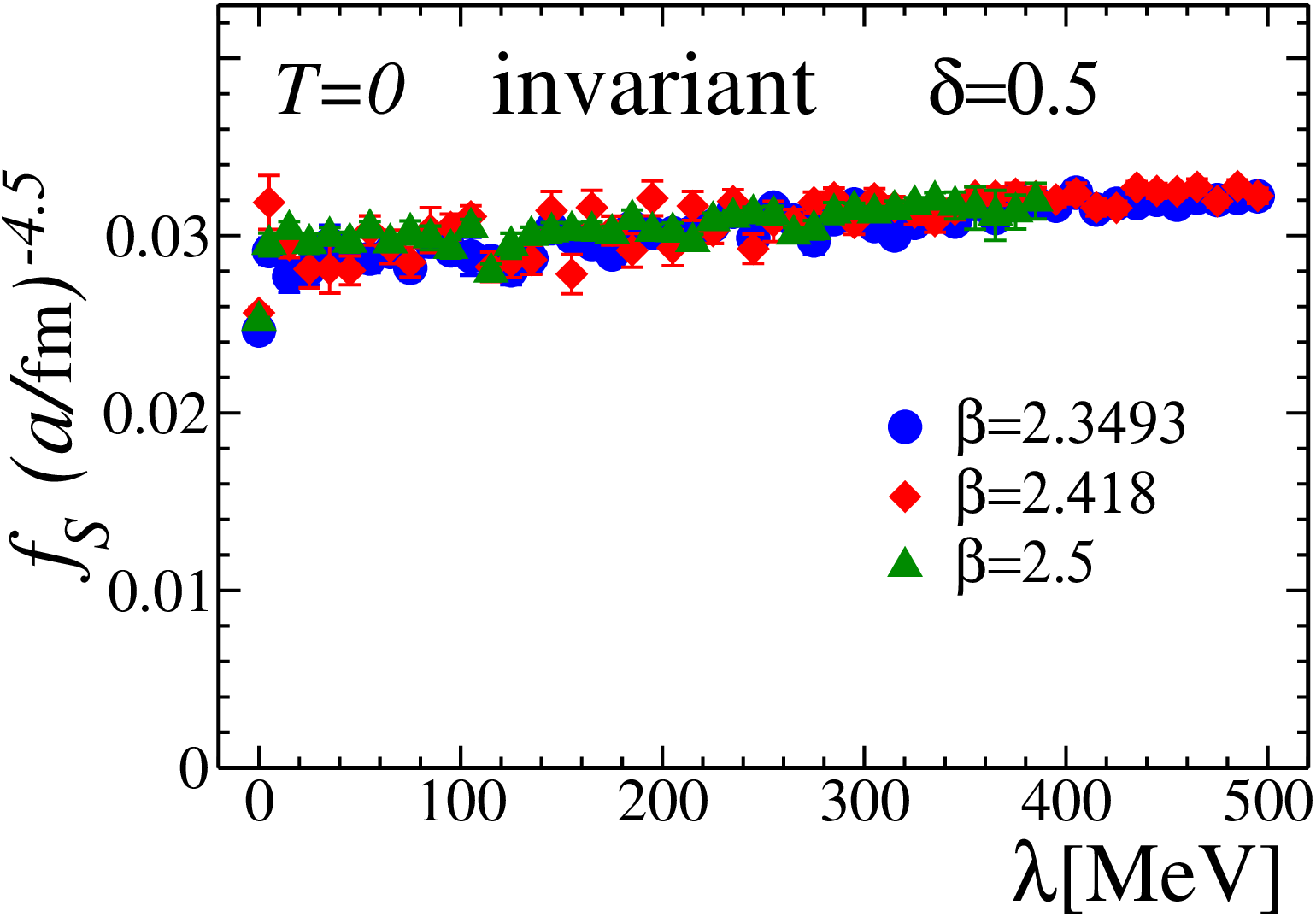} \\
(a) &  \hspace{5mm} (b)
\end{tabular}
\caption{(Color online) The local excess~\eq{eq:monopole:action:correlations} of the Yang-Mills action at the
position of the (a) scalar and (b) invariant monopoles, scaled by the power function,
$f_S(a)\,a^{-(4 + \delta)}$, with the anomalous scaling exponent $\delta=1/2$. The three
values of the lattice coupling $\beta$ are shown.}
\label{fig:action}
\end{figure}
In order to get an impression on the quality of the power-like
scaling~\eq{eq:power:fit} we plot in Figures~\ref{fig:action} (a)
and (b) the scaling function $f_S$ multiplied by the factor
$a^{-(4 + \delta)}$ with $\delta=1/2$ for the three different
values of the lattice coupling constant $\beta$.
Figures~\ref{fig:action}(a) and (b) clearly show that the quantity
$f_S(a)\,a^{-4.5}$ is independent of the lattice spacing $a$ for
the embedded monopoles of all three types. Since the fits of by
the power~\eq{eq:power:fit} and the HP-inspired~\eq{eq:fHP}
functions are practically indistinguishable at our data sets (as
one can see from examples plotted in
Figures~\ref{fig:fitting:energy}(a) and (b)), the scaling of our
data with the HP-type energy excess~\eq{eq:fHP} should be as
remarkable as it is plotted in Figures~\eq{fig:action} for the
power-profile excess~\eq{eq:power:fit}.

It is interesting to point out that despite noticeably different
scaling~\eq{eq:extrapolation:Lneq0} of the scalar/axial and
chirally invariant monopole densities towards continuum limit, the excess
of the action density at the positions of these monopoles scales
essentially in the same way.

Since the monopole cores are ``soft'' we do not expect a
fine-tuning~\cite{ref:fine:tuning} between the energy and entropy
of these monopoles (at least, in the studied case of the pure
SU(2) Yang-Mills theory). On the other hand, the embedded monopoles does not scale as
particle like objects, therefore the investigation of the
energy-entropy balance in the case of the embedded monopoles in
the quenched case may be a complicated issue.

A cautionary remark here is that the monopoles are detected with
the help of the operators~\eq{eq:j} which implicitly depend on the
lattice spacing $a$. If the monopole has a core-like structure of
the size of the lattice spacing then the ability of the numerical
procedure to ``detect'' a lattice monopole within the lattice cube
should be very sensitive to the size of the ``detector'' ({\it
i.e.} to the lattice spacing). In order to get physically reliable
results on the monopole density and the monopole correlations one
should probably study the scaling of the extended (blocked)
monopoles~\cite{ref:Extended:Monopoles}.

Summarizing, both scalar, axial and invariant embedded monopoles
are locally correlated with the magnetic part of the gluonic
action. The excess of the action at fixed lattice spacing $a$ (or,
equivalently, at fixed $\beta$) is a slowly increasing function of
the Dirac eigenvalue $\lambda$. The average action excess on the
scalar and axial monopoles coincide with each other and is
approximately two times higher than the excess of the action on
the invariant monopoles. The positive value of $f_S(a)$ indicates
that the action density near the monopole is increased compared to
the average density. Thus the embedded monopole has a
chromomagnetic core. On the other hand, the positive value of
the anomalous scaling exponent
$\delta$ indicates that the excess of
the chromomagnetic action decreases as one approaches the center
of the monopole core. Concretely, the chromomagnetic condensate
vanishes in the center of the embedded monopoles as \beqn
\mathcal{B}_{\mathrm{mon}}(r) \sim r^\delta\,, \qquad \delta \sim
0.5 \dots 1\,. \eeqn Thus, the embedded monopoles possess
``chromomagnetically'' empty cores.

\section{Conclusions}

We study basic properties of the embedded QCD monopoles in the
pure SU(2) Yang-Mills theory. The monopole trajectories are
found with the help of the low-lying eigenmodes of the overlap
Dirac operator. These modes are then treated as $c$-valued quark
fields, the behavior of which emulates the basic chiral properties
of the QCD vacuum.

The embedded monopoles are explicitly gauge-invariant, and the
magnetic charge of the monopole is quantized and conserved.
Basically, the embedded QCD monopoles are the gauge-invariant
hedgehogs in the quark-antiquark condensates (therefore we also
call these monopoles as ``quark monopoles'').

We give the lattice definitions of the embedded monopoles of
scalar, axial and chirally invariant types. We find that the
scaling of the scalar and axial monopole densities towards the
continuum limit is the same as the scaling of the string-like
objects. The scaling of the chirally invariant monopoles
corresponds to the one of the membrane-like objects. This result
may indicate that the monopole trajectories are correlated with
higher-dimensional (string-like and membrane-like) objects in
the $SU(2)$ Yang-Mills theory. The ``scalar/axial string'', for
example, may be a border of the ``chirally invariant membrane''.
We also observe a difference in the scaling properties of the
monopoles corresponding to the non-zero and to the zero Dirac
eigenvalues.

The embedded QCD monopoles were suggested~\cite{ref:embedded:QCD}
to be related to the restoration of the chiral symmetry in the
high-temperature phase since their cores should contain the
chirally symmetric vacuum. Our numerical study supports this
suggestion since the monopole density is anti-correlated with the
density of the Dirac eigenmodes. In particular, at low Dirac
eigenvalues -- which are relevant to the chiral symmetry breaking
due to the Banks-Casher relation -- the density of the embedded
monopoles is high in the chirally invariant (high temperature)
phase and is relatively low in the chirally broken (low
temperature) phase.

We find that the embedded monopoles have gluonic cores, which are
more pronounced for the chirally invariant monopoles compared to
the scalar/axial monopoles. On average, the chromomagnetic energy
near the monopole trajectories is higher compared to the
chromomagnetic energy far from the monopole core. However, our
scaling analysis suggests that at the very center of the embedded
QCD monopole, the excess of the chromomagnetic energy reduces back
to the vacuum expectation value. Therefore a typical monopole core
is a bump in the chromomagnetic energy which takes its maximum
value at a certain finite distance from the monopole center.
Outside this bump -- towards the monopole center and/or far from
the monopole core -- the energy density diminishes to its vacuum
expectation value. This structure is similar to the structure of
the `t~Hooft-Polyakov monopoles if one attributes  to the
asymptotic freedom the suppression of the chromomagnetic gluon
condensate in the monopole center.

Finally, we would like to remark that one can not exclude a
possibility that the properties of the embedded monopoles in the
full QCD may drastically be different from the quenched case.
On the other hand the quenched theory mimics chiral instability of the
full QCD to develop a chiral condensate at low temperatures. Therefore
our results support the suggestion that the quark monopoles are
tightly related to the chiral symmetry restoration also in the
case of the real QCD.

\section*{Note added}
When numerical calculations were completed we became aware that a
similar procedure has been applied earlier in $SU(3)$ lattice
gauge theory~\cite{ref:Gerrit}.

\begin{acknowledgments}
S.M.M. acknowledges an initial assistance with overlap fermions
given to him by the DESY group lead by Prof. G.~Schierholz.
M.N.Ch. is thankful to the members of Department of Theoretical
Physics of Uppsala University for kind hospitality and stimulating
environment. The authors are grateful to F.V.Gubarev for noticing
particularities of the Euclidean fermions. The work is supported
by a STINT Institutional grant IG2004-2 025, the grants RFBR
grants 04-02-16079, 05-02-16306a, 05-02-17642, and grants DFG 436
RUS 113/739/0, MK-4019.2004.2.
\end{acknowledgments}

\appendix

\section{Details of numerical simulations}
\label{app:details}

We simulate numerically the SU(2) lattice gauge theory with the standard Wilson action,
$S_P = \beta (1 - (1/2) {\mathrm{Tr}}\, U_P)$, where $\beta$ is the SU(2) gauge coupling
and $U_P$ is the SU(2) plaquette variable constructed from the lattice link fields $U(x,\mu)$.
We used various values of $\beta$ at different lattice sizes to check the scaling
of the numerically calculated observables towards the continuum limit.
The parameters of the numerical simulations are given in
Table~\ref{tbl:simulation:parameters}.

\begin{table}[htb]
\begin{tabular}{|c|c|c|c|c||c|c|c|c|c||c|c|c|c|c|}
\hline
\multicolumn{15}{|c|}{$T = 0$} \\
\hline
$\beta$ & $a$[fm] & $L_s$ & $L_t$ & $N_{conf}$ &
$\beta$ & $a$[fm] & $L_s$ & $L_t$ & $N_{conf}$ &
$\beta$ & $a$[fm] & $L_s$ & $L_t$ & $N_{conf}$ \\
\hline
2.3493 & 0.1397(15) & 10 & 10 & 300 &
2.3772 & 0.1284(15) & 10 & 14 &  90 &
2.4071 & 0.1164(15) & 12 & 12 & 180 \\
2.4180 & 0.1120(15) & 12 & 14 & 150 &
2.4500 & 0.0996(22) & 14 & 14 & 200 &
2.5000 & 0.0854(4)  & 16 & 18 & 200 \\
\hline
\hline
\multicolumn{15}{|c|}{$T = 1.15 T_c$} \\
\hline
\multicolumn{5}{|c||}{} & 2.3500 & 0.1394(8) & 16 & 4 & 200 & \multicolumn{5}{c|}{} \\
\hline
\end{tabular}
\caption{(Color online) Parameters of the simulations.}
\label{tbl:simulation:parameters}
\end{table}

The first 6 points in Table~\ref{tbl:simulation:parameters}
correspond to the zero temperature (confinement) phase. The
lattice geometries and values of the lattice coupling $\beta$ are
tuned in order to keep the lattice volume constant,
$V=3.8\,\mbox{fm}^4$. The point with $\beta=3.5$ has a little
bigger volume, $V = 3.92\,\mbox{fm}^4$.

In order to have an impression about the behavior of the quark
monopoles in the high temperature phase we study one point at
asymmetric lattice $16^3 \times 4$. At these lattices the system
is just above the finite temperature critical point with $T
= 1.15\,T_c$.

In order to define the quark monopoles one may use eigenmodes of
the Dirac operator in the background of the gauge field. In our
lattice simulations we use the overlap fermions which possess an
exact chiral symmetry, enjoy automatic $O(a)$ improvement, and are
not defaced by exceptional configurations~\cite{ref:overlap}.
The overlap Dirac operator is
\beqn
D = \frac{\rho}{a} \left(1 + D_w/\sqrt{D_w D_w^\dagger}\right) =
\frac{\rho}{a}(1+\gamma_5 \,{\mathrm{sign}}(H)),\ \ \ H = \gamma_5 D_w\,,
\eeqn
where $D_w$ is the Wilson Dirac operator with negative mass
term and $H$ is hermitian Wilson Dirac operator.
The value of $\rho$ parameter is equal to $1.4$. We have used the
minimax
polynomial approximation to compute the sign function.
In order to improve the accuracy and performance about one hundred lowest
eigenmodes of $H$ were
projected out. The eigenvalues of $D$, which lies on the circle in the complex
plain, were stereographically projected onto the imaginary axis in order to
relate them with continuous eigenvalues of the Dirac operator.



\begin{thebibliography}{99}

\bibitem{ref:general:reviews:QCD}
K. Rajagopal,  F. Wilczek, in {\it At the Frontier of Particle
Physics}, edited by M. Shifman (World Scientific,
Singapore, 2001), Vol. 3, p. 2061 [hep-ph/0011333].

\bibitem{ref:crossover:lattice:facts} S.~D.~Katz,
Nucl.\ Phys.\ Proc.\ Suppl.\  {\bf 129}, 60 (2004);
Z.~Fodor,  S.~D.~Katz,
JHEP {\bf 0404}, 050 (2004).

\bibitem{ref:embedded:QCD}
M.~N.~Chernodub, Phys.\ Rev.\ Lett. {\bf 95}, 252002 (2005).

\bibitem{ref:semilocal:review} For a review, see
A.~Achucarro,  T.~Vachaspati,
Phys.\ Rept.\  {\bf 327} (2000) 347.

\bibitem{ref:Nambu}
Y.~Nambu,
Nucl.\ Phys.\ B {\bf 130}, 505 (1977).

\bibitem{ref:Z:vortex}
N.~S.~Manton,
Phys.\ Rev.\ D {\bf 28}, 2019 (1983).

\bibitem{Va94} T.~Vachaspati,
in {\it Electroweak Physics and the Early Universe}, J.~C.~Rom\~ao
and F.~Freire (Eds.), Plenum, New York, 1995, p. 171, {\tt
hep-ph/9405286}.

\bibitem{ref:chernodub:nambu}
M.~N.~Chernodub,
JETP Lett.\  {\bf 66}, 605 (1997).

\bibitem{ref:hot:electroweak}
M.~N.~Chernodub, F.~V.~Gubarev, E.~M.~Ilgenfritz, A.~Schiller,
Phys.\ Lett.\ B {\bf 434}, 83 (1998).

\bibitem{ref:Z:percolation}
M.~N.~Chernodub, F.~V.~Gubarev, E.~M.~Ilgenfritz, A.~Schiller,
Phys.\ Lett.\ B {\bf 443}, 244 (1998).

\bibitem{ref:EW:phase}
M.~Reuter,  C.~Wetterich,
Nucl.\ Phys.\ B {\bf 408}, 91 (1993);
K.~Kajantie, M.~Laine, K.~Rummukainen,  M.~Shaposhnikov,
Phys.\ Rev.\ Lett.\  {\bf 77}, 2887 (1996);
Nucl.\ Phys.\ B {\bf 466}, 189 (1996);
M.~Gurtler, E.~M.~Ilgenfritz,  A.~Schiller,
Phys.\ Rev.\ D {\bf 56}, 3888 (1997).

\bibitem{ref:Kertesz}
J. Kert\'esz, Physica {\bf A161}, 58 (1989).

\bibitem{ref:Fortuin:Kasteleyn}
C.~M.~Fortuin,  P.~W.~Kasteleyn, Physica {\bf 57}, 536 (1972).

\bibitem{ref:Arwed}
M.~Baig,  J.~Clua,
Phys.\ Rev.\ D {\bf 57}, 3902 (1998);
S.~Wenzel, E.~Bittner, W.~Janke, A.~M.~J.~Schakel,  A.~Schiller,
Phys.\ Rev.\ Lett.\  {\bf 95}, 051601 (2005).

\bibitem{ref:Satz:theory}
G. Baym, Physica {\bf 96A}, 131 (1979); T.~Celik, F.~Karsch,
H.~Satz,
Phys.\ Lett.\ B {\bf 97}, 128 (1980);
H.~Satz,
Rept.\ Prog.\ Phys.\  {\bf 63}, 1511 (2000);

\bibitem{ref:Satz:percolation}
For a review, see H.~Satz,
  Nucl.\ Phys.\ A {\bf 681}, 3 (2001).

\bibitem{ref:Greensite:Faber}
R.~Bertle, M.~Faber, J.~Greensite,  S.~Olejnik,
Phys.\ Rev.\ D {\bf 69}, 014007 (2004).

\bibitem{ref:thooft}
G.~'t Hooft,
Nucl.\ Phys.\ B {\bf 79}, 276 (1974).

\bibitem{ref:sphaleron}
M. N. Chernodub, F. V. Gubarev, E. M. Ilgenfritz,
Phys.\ Lett.\ B {\bf 424}, 106 (1998).

\bibitem{ref:Yasha}
Yakov M. Shnir, ``Magnetic Monopoles'' (Springer, 2005);
Phys. Scr. {\bf 69}, 15 (2004);
hep-th/0508210.

\bibitem{ref:chernodub:inpreparation} M.~N.~Chernodub (unpublished).

\bibitem{ref:Weinberg}
E.~J.~Weinberg,
Phys.\ Rev.\ D {\bf 20}, 936 (1979);
Nucl.\ Phys.\ B {\bf 167}, 500 (1980);
{\it ibid.},  {\bf 203}, 445 (1982).

\bibitem{ref:Fujikawa}
K.~Fujikawa, Phys. Rev. Lett. {\bf 44}, 1733 (1980);  Phys. Rev. D {\bf 21}, 2848 (1980).

\bibitem{DGT}
T. A. DeGrand,  D. Toussaint, { Phys. Rev.} {\bf D22}, 2478
(1980).

\bibitem{Bornyakov:2001ux}
    V.~Bornyakov,  M.~Muller-Preussker,
    Nucl.\ Phys.\ Proc.\ Suppl.\  {\bf 106}, 646 (2002);
    V.~G.~Bornyakov, P.~Y.~Boyko, M.~I.~Polikarpov,  V.~I.~Zakharov,
    Nucl.\ Phys.\ B {\bf 672}, 222 (2003).

\bibitem{ref:clusters}
S.~i.~Kitahara, Y.~Matsubara,  T.~Suzuki,
  Prog.\ Theor.\ Phys.\  {\bf 93}, 1 (1995);
A.~Hart,  M.~Teper,
  Phys.\ Rev.\ D {\bf 58}, 014504 (1998);
M.~N.~Chernodub,  V.~I.~Zakharov,
  Nucl.\ Phys.\ B {\bf 669}, 233 (2003).

\bibitem{ref:vitaly}
V.~G.~Bornyakov, E.~M.~Ilgenfritz,  M.~Muller-Preussker,
  Phys.\ Rev.\ D {\bf 72}, 054511 (2005).

\bibitem{ref:lower:dimensional:Zakharov}
  V.~I.~Zakharov,
  Phys.\ Atom.\ Nucl.\  {\bf 68}, 573 (2005).

\bibitem{Gubarev:2005jm}
C.~Aubin {\it et al.}  [MILC Collaboration],
  Nucl.\ Phys.\ Proc.\ Suppl.\  {\bf 140}, 626 (2005);
  F.~V.~Gubarev, S.~M.~Morozov, M.~I.~Polikarpov,  V.~I.~Zakharov,
  JETP Lett. {\bf 82}, 343 (2005);
 Y.~Koma, E.~M.~Ilgenfritz, K.~Koller, G.~Schierholz, T.~Streuer,  V.~Weinberg,
  PoS {\bf LAT2005}, 300 (2005);
C.~Bernard {\it et al.},
  PoS {\bf LAT2005}, 299 (2005);
E.~M.~Ilgenfritz, K.~Koller, Y.~Koma, G.~Schierholz, T.~Streuer,  V.~Weinberg,
{\it ``Probing the topological structure of the QCD vacuum with overlap fermions''},
hep-lat/0512005;
A.~V.~Kovalenko, S.~M.~Morozov, M.~I.~Polikarpov, V.~I.~Zakharov,
{\it ``On topological properties of vacuum defects in lattice Yang-Mills
theories''}, hep-lat/0512036.

\bibitem{ref:Banks:Casher}
T.~Banks,  A.~Casher,
  Nucl.\ Phys.\ B {\bf 169}, 103 (1980).

\bibitem{ref:Bakker}
B.~L.~G.~Bakker, M.~N.~Chernodub,  M.~I.~Polikarpov,
Phys.\ Rev.\ Lett.\  {\bf 80}, 30 (1998).

\bibitem{ref:Monopole:Structure}
V.~G.~Bornyakov {\it et al.},
  Phys.\ Lett.\ B {\bf 537}, 291 (2002);
 V.~A.~Belavin, M.~I.~Polikarpov,  A.~I.~Veselov,
  JETP Lett.\  {\bf 74}, 453 (2001).

\bibitem{ref:polyakov}
A.~M.~Polyakov,
JETP Lett.\  {\bf 20}, 194 (1974).

\bibitem{ref:BPS}
E.~B.~Bogomolny,
  Sov.\ J.\ Nucl.\ Phys.\  {\bf 24}, 449 (1976);
M.~K.~Prasad,  C.~M.~Sommerfield,
  Phys.\ Rev.\ Lett.\  {\bf 35}, 760 (1975);
for a review see Yakov M. Shnir, ``Magnetic Monopoles'' (Springer, 2005).

\bibitem{ref:wetterich}
C.~Wetterich,
Phys.\ Lett.\ B {\bf 462}, 164 (1999);
Phys.\ Rev.\ D {\bf 64}, 036003 (2001);
J.~Berges, C.~Wetterich,
Phys.\ Lett.\ B {\bf 512}, 85 (2001),
where the idea of the spontaneous breaking of color by quark
condensates is discussed.

\bibitem{ref:four:quark}
M.~B.~Johnson,  L.~S.~Kisslinger,
  Phys.\ Rev.\ D {\bf 61}, 074014 (2000).

\bibitem{ref:fine:tuning}
V.~I.~Zakharov,
{\it ``Fine tuning in lattice SU(2) gluodynamics vs continuum-theory
constraints''},
hep-ph/0306262;
{\it ``Confining fields in lattice SU(2)''}
hep-ph/0312210.

\bibitem{ref:Extended:Monopoles}
T.~L.~Ivanenko, A.~V.~Pochinsky,  M.~I.~Polikarpov,
  Phys.\ Lett.\ B {\bf 252}, 631 (1990);
  H.~Shiba,  T.~Suzuki,
  Phys.\ Lett.\ B {\bf 351}, 519 (1995);
 S.~Kato, N.~Nakamura, T.~Suzuki,  S.~Kitahara,
  Nucl.\ Phys.\ B {\bf 520}, 323 (1998);
 M.~N.~Chernodub {\it et al.},
  Phys.\ Rev.\ D {\bf 62}, 094506 (2000).

\bibitem{ref:Gerrit}
V.~G.~Bornyakov, G.~Schierholz, T.~Streuer, unpublished (2005).

\bibitem{ref:overlap}
H.~Neuberger,
Phys.\ Lett.\ B {\bf 417}, 141 (1998);
{\it ibid.} {\bf 427}, 353 (1998)
%
T.~W.~Chiu, C.~W.~Wang,  S.~V.~Zenkin,
{\it ibid.} {\bf 438}, 321 (1998);
%
S.~Capitani, M.~Gockeler, R.~Horsley, P.~E.~L.~Rakow, G.~Schierholz,
%
{\it ibid.} {\bf 468}, 150 (1999);
%
T.~W.~Chiu,  S.~V.~Zenkin,
Phys.\ Rev.\ D {\bf 59}, 074501(R) (1999);
%
L.~Giusti, C.~Hoelbling, M.~Luscher,  H.~Wittig,
Comput.\ Phys.\ Commun.\  {\bf 153}, 31 (2003).

\end{thebibliography}
\end{document}